\documentclass[aip,jmp,amsmath,amssymb,preprint,reprint,author-year,author-numerical,]{revtex4-1}

\usepackage{graphicx}
\usepackage{dcolumn}
\usepackage{bm}
\usepackage{url}
\usepackage[export]{adjustbox}
\usepackage{multirow}

\begin{document}


\title[Lid Driven Cavities]{Proper orthogonal decomposition analysis of square lid driven cavity flows containing particle suspensions}

\author{J. Shelton}
\email{jshelton4@niu.edu.}
\author{N. Katiki}
\author{M. Adesemowo}
\affiliation{ 
Department of Mechanical Engineering, Northern Illinois University, DeKalb, IL, 60115, USA
}%


\begin{abstract}
Continual perturbations to the flow field in the square lid driven cavity can significantly impact both its stability and the dynamic characteristics of the system. In this study, the effect of these perturbations are analyzed via computational fluid dynamics simulations of a square, two-dimensional, lid driven cavity containing varying area fractions of suspended particles. The evolution over time of these particle-particle and particle-fluid interactions generate two-dimensional disturbance velocity changes in the flow field. In order to better understand the characteristics and effects of these changes, proper orthogonal decomposition analysis is performed on the velocity difference flow field to determine the dominant flow structures over the simulation time domain. The resulting eigenvectors and eigenvalues represent the varying dominant flow structures and their respective dominance over the simulation time, respectively. This is carried out at particle suspension area fractions from 10\% - 50\% and for Reynolds numbers of 100, 400, 700, and 1000. Critical flow regimes and particle suspension area fractions are discussed in order to characterize the stability of this particle-in-fluid system.
\end{abstract}

\keywords{Lid driven cavity, stream function, proper orthogonal decomposition, particle laden flow, granular flow}
\maketitle

\section{Introduction}

The lid driven cavity problem has proven to be incredibly powerful in understanding fundamental fluid dynamics behavior of a fluid-filled square two-dimensional (2D) domain of arbitrary side length and a moving top boundary. The location and strength of both the central primary vortex and the secondary corner eddies that develop within the domain and their relationship to Reynolds number provide insight into flow stability and base flow structures within the fluid domain \cite{Shankar:2000,GHIA1982387}. Yet, despite being one of the most studied problems in fluid dynamics, there are still interesting observations that can be obtained from investigating underlying stability behavior given the addition of a perturbation field \cite{Theofilis:2011}. Many investigations have used a three-dimensional (3D) mathematical perturbation field on both 2D \cite{Ramanan:1994,Albensoeder:2001} and 3D lid driven cavity flow fields \cite{Ding:1999,giannetti2009linear} and have used linear stability analysis to determine the Hopf bifurcation transition point from a steady 2D flow towards either a periodic, quasiperiodic, or purely unsteady 3D flow. Even an applied 2D perturbation field onto a 2D flow field can yield characteristics of the steady-to-unsteady flow transition behavior \cite{AUTERI20021,Sahin:2003}. The key to these linear stability analyses involves the assumption that velocity $\textbf{v}'(\textbf{x},t)$ and pressure $p'(\textbf{x},t)$ perturbations to the base flow steady-state velocity $\textbf{U}(\textbf{x})$ and pressure $P(\textbf{x})$ are sufficiently small to satisfy the criteria $\textbf{v}'\ll\textbf{U}$ and $p'\ll P$. This assumption allows the nonlinear Navier Stokes equations to be linearized, which can then be used to predict the onset of instability in this dynamical system. As a result, these mathematical perturbations can now be described mathematically as 
\begin{equation}
\begin{bmatrix}
\textbf{v}'(\textbf{x},t)\\
p'(\textbf{x},t)
\end{bmatrix}=\begin{bmatrix}
\hat{\textbf{v}}'(\textbf{x})\\
\hat{p}'(\textbf{x})
\end{bmatrix}\exp(i\kappa z+\sigma t )
\label{eq:one}
\end{equation}

\noindent where $\sigma$ is the complex growth rate which predicts the stability of the system, and for the cases when the span wise direction is nonzero and less than infinity, $\kappa$ is the spanwise wave number. This approach can be used to predict regimes of stability (steady, periodic, quasi-periodic, chaotic) and the corresponding Hopf bifuraction points and reconstruct the flow structures within these regimes.

Particle suspensions interacting both with each other and with the base fluid they are suspended in possibly can be viewed as physical representations of the mathematical perturbations to the base flow within a lid driven cavity described above. In this case, however, the velocity $\textbf{v}'(\textbf{x},t)$ and pressure $p'(\textbf{x},t)$ perturbations generated by these interactions of the particle suspensions are significantly larger than their respective mathematical counterparts. As a result, these perturbations do not satisfy the criteria that allows for linear stability analysis to be employed, i.e. $\textbf{v}'\ll\textbf{U}$ and $p'\ll P$. So, a different approach has to be employed. The velocity perturbation field $\textbf{v}'(\textbf{x},t)$ generated by particle suspensions can be described mathematically as the difference between the steady-state base flow velocity $\textbf{U}(\textbf{x})$ and the fluid velocity field due to particle-particle $\textbf{U}_{pp}(\textbf{x},t)$ and particle-fluid $\textbf{U}_{pf}(\textbf{x},t)$ interactions 
\begin{equation}
\textbf{v}'(\textbf{x},t)=[\textbf{U}_{pp}(\textbf{x},t) + \textbf{U}_{pf}(\textbf{x},t)] - \textbf{U}(\textbf{x})
\label{eq:two}
\end{equation}

One approach towards analyzing this perturbed flow field is through the use of proper orthogonal decomposition (POD). In this mathematical framework, dominant patterns are isolated from large data sets and then used to approximate the overall dataset as a combination of their contributions.  The use of POD analysis on lid driven cavity flows have been performed previously by Ahlman, et al.\cite{ahlman:2002}, Cazemier, et al.\cite{cazemier:1998}, and Lestandi, et al.\cite{lestandi:2018} and they were able to a) test the effectiveness of the mathematical approach to predict the onset of instability in flow field, b) obtain Hopf bifurcation points and determine stability regimes, and c) use the coherent structures obtained by POD to characterize the flow structures within each regime. 

In this study, characterization of the fundamental fluid dynamics behavior of the lid driven cavity perturbed by two-dimensional suspended particles is performed. The perturbations to the base flow during both transient and steady-state flow conditions are characterized by means of either particle-fluid drag interaction or particle-particle/particle-wall collisions. Proper orthogonal decomposition analysis isolates the dominant eigenmodes/flow structures of the field and quantifies the contribution each perturbation type has to the resulting flow field. This procedure is carried out for Reynolds numbers of 100, 400, and 1000 for area fractions from 10\% to 50\% to ascertain how these parameters affect the dominant flow structures.

\section{Theory}
\subsection{Base Flow Model}
We consider a two-dimensional lid-driven cavity with a size of $\textit{L}\times\textit{L}$ with its entire domain $\Omega(x,y)$ being filled with a fluid that is incompressible, isothermal, and Newtonian (Fig. 1). The fluid is driven by the top boundary of the cavity ($\textit{y}=L$) moving at a constant speed of $U_\infty$. 

\begin{figure}[!t]
	\vspace{1.3cm}
	\setlength{\unitlength}{0.14in} 
	\centering 
	\begin{picture}(32,13) 
	\put(6,2){\framebox(12,12){$\Omega(x,y)$}}
	\put(9,15){\vector(1,0){6}}
	\put(3,0){\vector(0,1){2}}
	\put(3,0){\vector(1,0){2}}
	\put(9.5,15.5) {$u(x,L) = U_{\infty}$} 
	\put(5.5,-0.3) {$(x,u)$} 
	\put(2,2.3) {$(y,v)$} 
	\end{picture}
	\caption{Lid Driven Cavity} 
	\label{fig:ldc_picture} 
\end{figure}
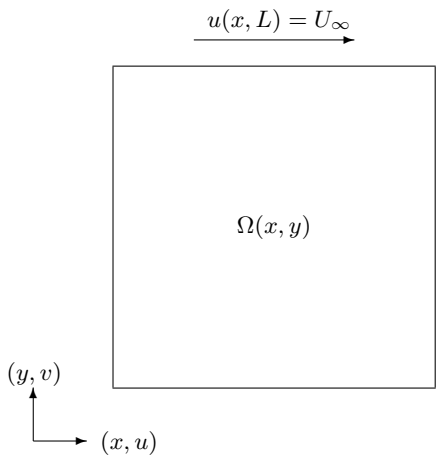

In the case of fluid flow with no particle suspensions, the equations that govern fluid flow within the domain in the $u_x$ and $u_y$ directions are 
\begin{equation}
\nabla_*\cdot \textbf{u}=0
\label{eq:three}
\end{equation}
\begin{equation}
\rho\left(\frac{\partial \textbf{u}}{\partial t} + (\textbf{u} \cdot \nabla_*)\textbf{u}\right)=-\nabla_* p+\mu\nabla^{2}_*\textbf{u}+\rho g
\label{eq:four}
\end{equation}

\noindent where $\nabla_*$ is the two-dimensional gradient operator ($\nabla_*=\textbf{i}\partial_{,x}+\textbf{j}\partial_{,y}$). No-slip conditions are imposed on all boundaries within the cavity and all non-moving boundaries have a velocity ($u_x,u_y$)=(0,0). The only forces acting on the fluid are due to pressure, viscosity, and gravity.
	
\subsection{Particle Laden Flow Model}
Unlike the mathematically generated perturbations utilized in previous investigations, the perturbations employed here are generated from particle suspensions and their interactions with each other and the surrounding fluid. These types of flows are generally called $\textit{particle-laden}$ and are characterized by its two phases: one continuous phase that serves as a carrier for a secondary phase that is dispersed throughout it. In this investigation, both the continuous and secondary phases are considered to be incompressible, with the carrier characterized as a liquid and the particulate phase as a solid of a single diameter. In order to investigate the interactions between this particulate secondary phase and the continuous fluid phase, the governing equations for fluid flow have to modified.  Starting with the continuity equation, the physical presence of the suspended particles in the base flow is accounted for with the addition of the fluid volume fraction ($\epsilon$) term as shown below
\begin{equation}
\nabla_* \cdot \left(\epsilon\textbf{u}\right)=0
\label{eq:five}
\end{equation}
This fluid volume fraction is equal to 1 for a fluid domain with no particle suspensions and approaches 0 as both the number of particles occupying the fluid domain increases towards a closed packed system and the diameter of the particles reduces to 0. This fluid volume fraction term is also added to the momentum equation and is expressed in the form  
\begin{equation}
\rho\left(\frac{\partial (\epsilon \textbf{u})}{\partial t} + \left(\epsilon\textbf{u}\cdot\nabla_*\right)\textbf{u}\right) = -\epsilon\nabla_* p +\mu\nabla^2_* (\epsilon\textbf{u}) + \rho\epsilon\textbf{g}+\textbf{F}_{d}
\label{eq:six}
\end{equation}
The interaction between the carrier fluid and the particles suspended within it has been characterized in this investigation as four-way coupling\cite{elghobashi:1994}. Momentum exchange due to a generated drag force between the suspended particle and the carrier fluid defines two of this four-way coupling. This drag force $\textbf{F}_{d}$ originates from the difference in velocity between the fluid $\textbf{u}_{f}$ and the granular suspension $\textbf{u}_{p}$ and can be expressed as
\begin{equation}
\textbf{F}_d = K_{pg}(\textbf{u}_{p} - \textbf{u}_{f})
\label{eq:seven}
\end{equation}
The term $K_{pg}$ represents an interphase momentum exchange due to the drag of particles present in the system. This momentum exchange coefficient can be obtained either through a variety of empirically-determined models, such as Syamlal-O'brien\cite{Syamlal87thederivation} and Gidaspow\cite{Gidaspow:1488821}, or computationally-determined models, such as  Hill-Koch-Ladd\cite{hill_koch_ladd_2001_1,hill_koch_ladd_2001_2,benyahia:2006}. In this investigation, the Gidaspow model is selected due to its ability to model two distinct flow regimes: 1) the dilute particle flow regime ($\epsilon > 0.8$), where viscous forces dominate fluid flow behavior over other internal forces, like particle-particle collisions, and can be modeled using the Ergun equation\cite{Ergun:1952}
\begin{equation}
K_{pg}=150\frac{\mu_f(1-\epsilon)^2}{\epsilon\left(d_p\phi\right)^2}+1.75\frac{\rho_{f}\left(1-\epsilon\right)}{d_p\phi}\vert \textbf{u}_f-\textbf{u}_p\vert
\label{eq:eight}
\end{equation}
where $\mu$ is the dynamic viscosity, $\rho$ is the density, $d_{p}$ is the particle diameter, $\phi$ is a constant shape factor, and $\vec{u}$ is velocity, with the subscripts $p$ and $f$ denoting either the particle phase or the fluid phase, respectively; and 2) the dense particle flow regime ($\epsilon < 0.8$), where particles are densely packed and each particle is assigned a value of drag related to the pressure drop across the densely packed system, and can be modeled using the empirically-derived Wen-Yu equation\cite{Wen1966a} 
\begin{equation}
K_{pg}=\frac{3}{4}\frac{\rho C_d\left(1-\epsilon\right)\epsilon^{-2.65}_f}{d_p}\vert \textbf{u}_f-\textbf{u}_p\vert
\label{eq:nine}
\end{equation}

\begin{equation}
C_d=\frac{24}{\epsilon Re_p}\left[1+0.15\left(\epsilon Re_p\right)^{0.687}\right] 
\label{eq:ten}
\end{equation}

\begin{equation}
Re_p=\frac{\rho_{f}d_p}{\mu_{f}}\vert \textbf{u}_f-\textbf{u}_p\vert
\label{eq:eleven}
\end{equation}
Originally used to describe the friction factor of packed bed in terms of the Reynolds number, this model also allows for a description of the resulting frictional drag forces acting on the fluid due to the presence of suspended particles.

\subsection{Granular Flow Model}
In addition to the Particle Laden Flow Model that characterizes generated perturbations as originating from drag forces due to particle-fluid interactions, particle-particle and particle-boundary interactions also generate additional velocity perturbations that can disturb the base flow. These interactions define the final two of the four-way coupling mentioned above\cite{elghobashi:1994}. These particle collision induced perturbations generate forces that locally disturb fluid velocities and increase in frequency and intensity with increasing particle area fraction (2D case). Tsuji, et al. \cite{TSUJI1992239} presented a theorectical model to describe the equations of motion for these particles and takes the form (with no particle rotation)  
\begin{equation}
\ddot{\textbf{r}}=\frac{\textbf{F}_{c} + \textbf{F}_{d}}{m} + \textbf{g}
\label{eq:twelve}
\end{equation}
The motion of these particles are governed by three forces: gravity, drag, and collisions. The drag force $\textbf{F}_{d}$ corresponds to the equal and opposite force that was described in Eq. 7 and uses the interphase momentum exchange coefficient mathematically described by either the Ergun or Wen-Yu equations. The collision force $\textbf{F}_{c}$ corresponds to the particle-particle and particle-boundary collisions that occur during the simulation and is governed by Hertzian contact theory which uses a spring and a dashpot to represent the deformation of the particles. The normal component of this collision force can be expressed mathematically in the form
\begin{equation}
\textbf{F}_{c,n} = (-K_n\delta_{nij}^{3/2}-\eta_{n}\textbf{v}_{ij}\textbf{n}_{ij})\textbf{n}_{ij}
\label{eq:thirteen}
\end{equation}
The first term in Eq. 13 contains a normal stiffness coefficient $K_n$ that is a function of the particle diameter, elastic modulus, and Poisson's ratio. For the case when two particles of the same diameter collide with each other, this normal stiffness coefficient can be expressed as
\begin{equation}
K_{n}=\frac{E_p\sqrt{d_p}}{3(1-\sigma^2_p)}
\label{eq:fourteen}
\end{equation}
For the case when a particle collides with a boundary such as a wall in the lid driven cavity, this normal stiffness coefficient can be expressed as
\begin{equation}
K_{n}=\cfrac{\cfrac{4\sqrt{\cfrac{d_p}{2}}}{3}}{\cfrac{1-\sigma_p^{2}}{E_p}+\cfrac{1-\sigma_w^{2}}{E_w}}
\label{eq:fifteen}
\end{equation}
In Eqs. 14 and 15, $\textit{E}$ is the elastic modulus and $\sigma$ is the Poisson's ratio, with the subscripts $\textit{p}$ and $\textit{w}$ specifying the particle suspension and wall boundary of the lid driven cavity, respectively.

The second term in Eq. 13 contains a normal damping coefficient $\eta_n$ that is a function of the particle mass and its normal stiffness coefficient and is expressed in the form
\begin{equation}
\eta_{n}=\alpha(mK_n)^{1/2}\delta_{nij}^{1/4}
\label{eq:sixteen}
\end{equation}
The constant $\alpha$ is an empirically defined value related to the coefficient of restitution that is both derived and explained in detail by Tsuji, et al.\cite{TSUJI1992239}

The tangential component of this collision force can be expressed mathematically in the form
\begin{equation}
\textbf{F}_{c,t} = K_{t}\delta_{nij}^{1/2}-\eta_{t}\textbf{v}_{ij}
\label{eq:seventeen}
\end{equation}
The first term in Eq. 17 contains a tangential stiffness coefficient $K_{t}$ that is a function of particle elastic modulus and Poisson's ratio and can either be describe in the form
\begin{equation}
K_{t} = \frac{2E_p\sqrt{d_p}}{2(2-\sigma_p)(1+\sigma_p)}
\label{eq:eighteen}
\end{equation}
for two colliding particles of equal diameter or in the form 
\begin{equation}
K_{t} = \cfrac{8E_p\sqrt{\cfrac{d_p}{2}}}{2(2-\sigma_p)(1+\sigma_p)}
\label{eq:nineteen}
\end{equation}
for one particle colliding with a boundary, such as a wall in the case of the lid driven cavity.

The second term in Eq. 17 contains a tangential damping coefficient $\eta_t$ that is also a function of the particle mass and its normal stiffness coefficient. As was done by Tsuji, et al., we have considered that $\eta_t$ is equal to $\eta_n$. 

\subsection{Proper Orthogonal Decomposition Theory}

Taira, et al.\cite{taira:2017} presented a theoretical overview of proper orthogonal decomposition to extract a set of dominant modes that can be used to sufficiently describes a dynamic fluid flow field. In this overview, a snapshot $\textbf{x}(t)$ of the fluctuation in the fluid velocity field variables ($u$,$v$) within the lid driven cavity is constructed by first arranging the fluid velocity field variables at time $\textit{t}$ into a single velocity vector $\textbf{q}(t)$ and subtracting from it the time-averaged velocity vector $\overline{\textbf{q}}$ 
\begin{equation}
\textbf{x}(t) = \textbf{q}(t) - \overline{\textbf{q}}
\label{eq:twenty}
\end{equation}
All fluid velocity fluctuation snapshots are then constructed into a single matrix $\textit{X}$ that represent the evolution of velocity fluctuations in the entire fluid domain over time
\begin{equation}
\textit{X} = \begin{bmatrix}\textbf{x}(t_1) && \textbf{x}(t_2) && \dots && \textbf{x}(t_m)\end{bmatrix}
\label{eq:twenty-one}
\end{equation}
An eigenvalue equation can now be constructed from the covariance of $\textit{X}$ 
\begin{equation}
XX^T\boldsymbol{\Phi}=\boldsymbol{\Lambda}\boldsymbol{\Phi}
\label{eq:twenty-two}
\end{equation}
The matrix of eigenvectors $\boldsymbol{\Phi}$ represent the dominant modes or flow structures that are present with the dynamic fluid flow field. The corresponding diagonal matrix of eigenvalues $\boldsymbol{\Lambda}$ can be interpreted as a percentage of the total observation time that each dominant mode is sustained within the system. 

\section{Numerical Methods}
\subsection{Base Flow Computational Methodology}
\begin{table}[t]
	\caption{Properties and Parameters for the Base, Particle Laden, and Granular Flows}
	\centering
	{\setlength{\extrarowheight}{5pt}
		\begin{tabular}{>{\raggedright\arraybackslash}p{3.5cm}  >{\centering\arraybackslash}p{2.25cm}
				>{\centering\arraybackslash}p{2.25cm}}   
			\hline\hline
			Property/Parameter & Units & Value  \\
			\hline
			\textit{Fluid}\\
			Density, $\rho$ & $\textrm{kg/m}^3$	 & 1.2   \\ 
			Dynamic viscosity, $\mu$ & $\textrm{Pa}\cdot\textrm{s}$ & 0.01  \\ 
			\textit{Particle}\\
			Density, $\rho_p$ & $\textrm{kg/m}^3$	 & 1200   \\ 
			Diameter, $d_p$ & m & 0.01\\
			Elastic Modulus, $\textit{E}_p$ & $\textrm{N/m}^2$ & 1.0E8\\
			Poisson's ratio, $\sigma_p$ & & 0.35\\
			Empirical constant, $\alpha$ & & 0.01\\
			\textit{Wall}\\
			Elastic Modulus, $\textit{E}_w$ & $\textrm{N/m}^2$ & 1.0E8\\
			Poisson's ratio, $\sigma_p$ & & 0.35\\
			\textit{Model Coefficients}\\
			$K_{n} (Particle-Particle)$ & $\textrm{N}/(\textrm{m}^{3/2})$ & 3.8E6\\
			$K_{n} (Particle-Wall)$ & $\textrm{N}/(\textrm{m}^{3/2})$ & 5.4E6\\
			$K_{t} (Particle-Particle)$ & $\textrm{N}/(\textrm{m}^{3/2})$ & 4.5E6\\
			$K_{t} (Particle-Wall)$ & $\textrm{N}/(\textrm{m}^{3/2})$ & 1.3E7\\
			\hline\hline
		\end{tabular}}
		\label{table:properties}
	\end{table}

The evaluation of the 2D flow characteristics within the lid-driven cavity involved a finite volume discretization and analysis of the fluid domain of Eqs. 3 and 4. The OpenFOAM (Open Field Operation and Manipulation) software package v4.0 was used to perform this task with the DPMFoam solver being used to solve the Navier Stokes equations. The Discrete Particle Modeling solver (DPMFoam) is a transient solver based on the PIMPLE algorithm, which is a combination of the pressure implicit with splitting operator (PISO) and the semi-implicit method for pressure-linked equations (SIMPLE) algorithms. The convective terms were discretized using the Gauss linear upwind scheme. The remaining terms were discretized using the Gauss linear scheme. We applied the preconditioned conjugate gradient (PCG) with the geometric-algebraic multigrid (GAMG) for the pressure, and the symmetric Gauss-Seidel smoother for the velocity and stress. The absolute tolerance at each time step was between 1.0 x $10^{-5}$ and 1.0 x $10^{-6}$ for the variables. The time step was 5 x $10^{-5}$s. All flow field fluid properties (dynamic viscosity and density) were held constant for all time and locations and their values are found in Table 1.  The Reynolds number for this study is defined and the Reynolds number is defined as $\textit{Re}=U_{\infty}L/\nu$, with $\nu$ as the fluid kinematic viscosity by the length of the square lid driven cavity, the lid velocity and the kinematic viscosity. The Reynolds numbers investigated in this study were 100, 400, 700, and 1000.

The computational domain that defines the square lid driven cavity was divided into \textit{M}x\textit{M} regular divisions. A grid independence study was performed for all Reynolds numbers investigated in order to determine the grid size that minimized the error in the minimum stream function value when compared to the values obtained by Ghia, et al. \cite{GHIA1982387}. Fig. 2 indicates that at a grid size of greater than 160 yielded an streamfunction error of less than 0.5\%.  

\begin{figure}[t]
	\centering
	\includegraphics[width=1.0\linewidth]{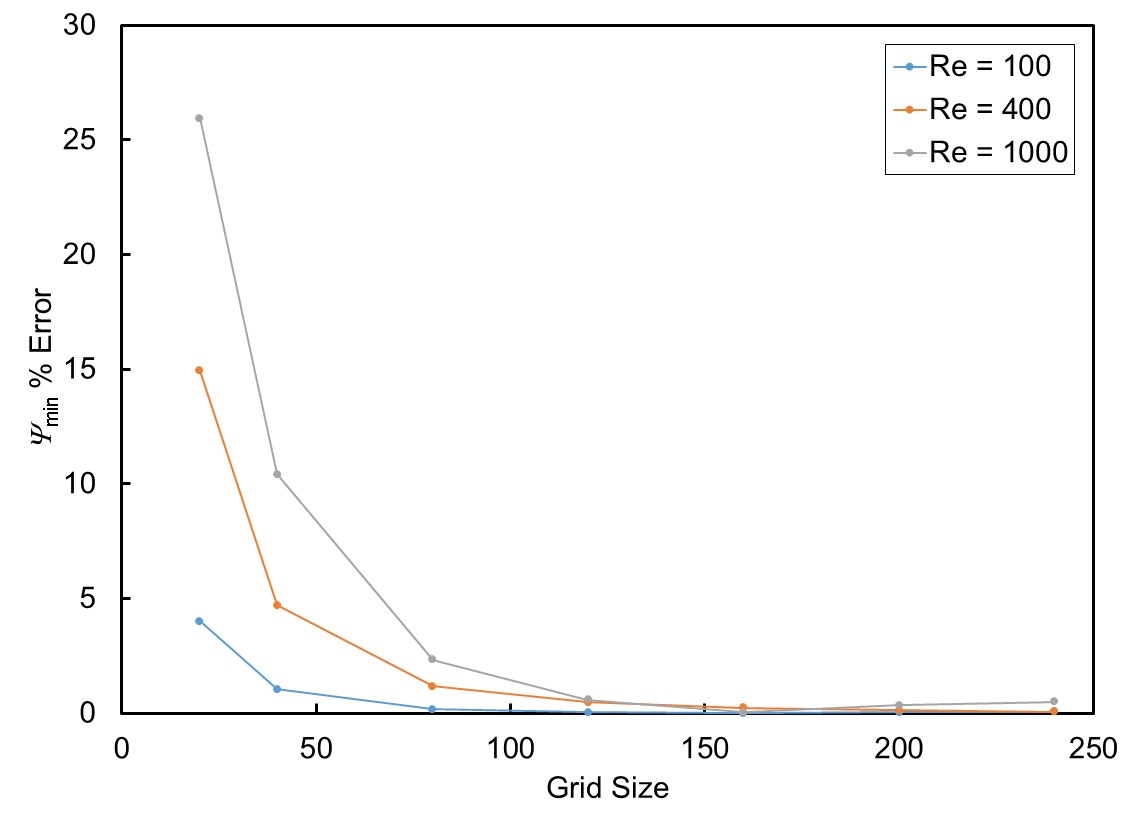}
	\caption{Grid Independence Study}
	\label{fig:grid_independence}
\end{figure}

\subsection{Particle-Related Computational Methodology}
The evaluation of the 2D flow characteristics within the lid-driven cavity involved a finite volume discretization and analysis of the fluid domain of Eqs. 7 and 8. Once again, the OpenFOAM (Open Field Operation and Manipulation) software package v4.0 was used to perform this task with the DPMFoam solver being used to solve the Navier Stokes equations using the same parameters described above for the base flow analysis. DPMFOAM employs the modified form the Navier-Stokes equation and continuity equation described equations 7 and 8. All flow field fluid properties (dynamic viscosity and density) were held constant for all time and locations and their values are found in Table 1.  The Reynolds number for this study is defined and the Reynolds number is defined as $\textit{Re}=U_{\infty}L/\nu$, with $\nu$ as the fluid kinematic viscosity by the length of the square lid driven cavity, the lid velocity and the kinematic viscosity. The Reynolds numbers investigated in this study were 100, 400, and 1000.

Granular flow characteristics were incorporated into the computational analysis via the DPMFOAM solver. The density of the particle was set to be 1000 times larger than the density of the surrounding fluid. Additional particle properties are presented in Table I. The elastic modulus was set at .97 throughout the domain for particle to particle and particle to wall. The spring constant determining the force experienced between the particles and the wall were calculated by means of Eqs. 14, 15, 18, and 19 and are also presented in Table I. The particle area fraction of the flow is defined as the total cross-sectional area of the particles over the cross-sectional area of the cavity
\begin{equation}
\phi=\frac{N\cdot\pi\frac{d_p^2}{4}}{L^2}
\label{eq:twenty-three}
\end{equation}

\noindent where N is the number of particles and $d_p$ is the diameter of the particle. The particle size chosen was 0.01 m and the particle area fractions under study were 10\%, 20\%, 30\%, 40\%, and 50\%, corresponding to a particle number of N = 13, 26, 38, 51, 64, respectively. 
\begin{table*}[t]
	\caption{Properties of the primary vortex and its equilibrium response for PL (Particle Laden Flow) and MX (Mixture of Particle Laden and Granular Flows) for all Reynolds numbers and particle area fractions.}
	\centering
	\begin{tabular}{>{\raggedright\arraybackslash}p{2.5cm} >{\raggedright\arraybackslash}p{3.5cm} >{\centering\arraybackslash}p{2.5cm} >{\centering\arraybackslash}p{2.5cm} >{\centering\arraybackslash}p{2.5cm} >{\centering\arraybackslash}p{2.5cm}} 
		\hline\hline
		&&\multicolumn{4}{>{\centering\arraybackslash}p{10cm}}{Re}  
		\tabularnewline
		\cline{3-6}
		Particle Area Fraction & Property & 100 & 400 & 700 & 1000 \\ [0.5ex] 
		\hline 
		\multirow{2}{*}{0\%} & Equilibrium Response & S & S & S & S\\[0.5ex]
		& $\psi_{min}$ & -0.10345 & -0.11364 & -0.11665 & -0.11797\\ 
		\multirow{2}{*}{10\%} & Equilibrium Response & PL (U-TI) & PL (U-TI) & PL (U-TI) & PL (U-TI)\\[0.5ex]
		& $\overline{\psi_{min}}$ & -0.08075 & -0.09082 & -0.09842 & -0.1027\\ 
		\multirow{2}{*}{20\%} & Equilibrium Response & PL (U-TI) & PL (U-TI) & PL (U-TI) & PL (U-TI)\\
		& $\overline{\psi_{min}}$ & -0.07035 & -0.07926 & -0.08929 & -0.09672\\ 
		\multirow{2}{*}{30\%} & Equilibrium Response & PL (U-TI) & PL (U-TI) & PL (U-TI) & MX (U-TD) \\
		& $\overline{\psi_{min}}$ & -0.0642 & -0.07883 & -0.08826 & -0.69643\\ 
		\multirow{2}{*}{40\%} & Equilibrium Response & PL (U-TI) & PL (U-TI) & MX (U-TI) & MX (U-TI)\\
		& $\overline{\psi_{min}}$ & -0.05868 & -0.07575 & -0.86196 & -0.75338\\ 
		\multirow{2}{*}{50\%} & Equilibrium Response & PL (U-TI) & PL (U-TI) & MX (U-TI) & MX (U-TI)\\ 
		& $\overline{\psi_{min}}$ & -0.05667 & -0.07674 & -0.90908 & -0.78299\\ 
		\hline\hline
	\end{tabular}
	\label{table:PL_vs_MX}
\end{table*}

\subsection{Proper Orthogonal Decomposition Computational Methodology}
The first step in solving the eigenvalue problem expressed in Eq. 22 involves constructing the snapshot of the fluctuations of the fluid velocity field variables. The size of this vector $\textbf{q}(t_i)$ ends up being [2*(\textit{M}x\textit{M}),1] and takes the form  
\begin{equation}
\textbf{q}(t_i)=\begin{bmatrix}
\begin{pmatrix}
u_{x_1,y_1}(t_i)\\u_{x_1,y_2}(t_i)\\\vdots\\u_{x_1,y_m}(t_i)
\end{pmatrix}\\
\begin{pmatrix}
u_{x_2,y_1}(t_i)\\u_{x_2,y_2}(t_i)\\\vdots\\u_{x_2,y_m}(t_i)
\end{pmatrix}\\
\vdots\\
\begin{pmatrix}
u_{x_m,y_1}(t_i)\\u_{x_m,y_2}(t_i)\\\vdots\\u_{x_m,y_m}(t_i)
\end{pmatrix}\\
\begin{pmatrix}
v_{x_1,y_1}(t_i)\\v_{x_1,y_2}(t_i)\\\vdots\\v_{x_1,y_m}(t_i)
\end{pmatrix}\\
\begin{pmatrix}
v_{x_2,y_1}(t_i)\\v_{x_2,y_2}(t_i)\\\vdots\\v_{x_2,y_m}(t_i)
\end{pmatrix}\\
\vdots\\
\begin{pmatrix}
v_{x_m,y_1}(t_i)\\v_{x_m,y_2}(t_i)\\\vdots\\v_{x_m,y_m}(t_i)
\end{pmatrix}
\end{bmatrix}
\label{eq:twenty-four}
\end{equation}
In order to be in line with previous stability analysis investigations and investigate how perturbations caused by particle-particle and particle-fluid interactions affect the base flow, the time-averaged velocity vector $\overline{\textbf{q}}$ is assumed to be the steady-state base flow velocity field in a lid driven cavity with no particle suspensions. The difference velocity field calculations were performed in ParaView. 

Using the resulting $\textbf{x}(t_i)$ to construct the $\textit{X}$ matrix results in a [2*(\textit{M}x\textit{M}),\textit{N}] matrix, where $\textit{N}$ is the number of timesteps required to achieve a steady-state condition in the perturbed fluid system. The resulting eigenvalue value problem with the covariance matrix of $\textit{X}$ of this size can be challenging, even with modern computational resources. Therefore, the $\textit{method of snapshots}$ methodological approach is employed and is described below. 

In the method of snapshots methodology, the spatial covariance matrix originally used in the eigenvalue equation above (Eq. 22) is replaced with a temporal covariance matrix that is constructed by premultiplying the matrix $\textit{X}$ by its transpose
\begin{equation}
X^TX\boldsymbol{\Psi}=\boldsymbol{\Lambda}\boldsymbol{\Psi}
\label{eq:twenty-five}
\end{equation}
In doing so, calculation of the eigenvalues is done on the much smaller matrix of size NxN, where N<<(2*MxM). The diagonal matrix $\lambda$ obtained from this new eigenvalue equation contains the same eigenvalues as those obtained from Eq. 22. Additionally, each of the spatial eigenvectors $\phi_j$ can be obtained from the temporal eigenvectors $\psi_j$ by
\begin{equation}
\begin{matrix}
\phi_j=X\psi_j\frac{1}{\sqrt{\lambda_j}} && j = 1,2,\dots,N
\end{matrix}
\label{eq:twenty-six}
\end{equation}

This $\textit{method of snapshots}$ approach to the POD analysis was executed utilizing MATLAB. 

\begin{figure*}[t]
	\centering
	\begin{tabular}{lcc}
		
		\adjincludegraphics[width=8.50cm]{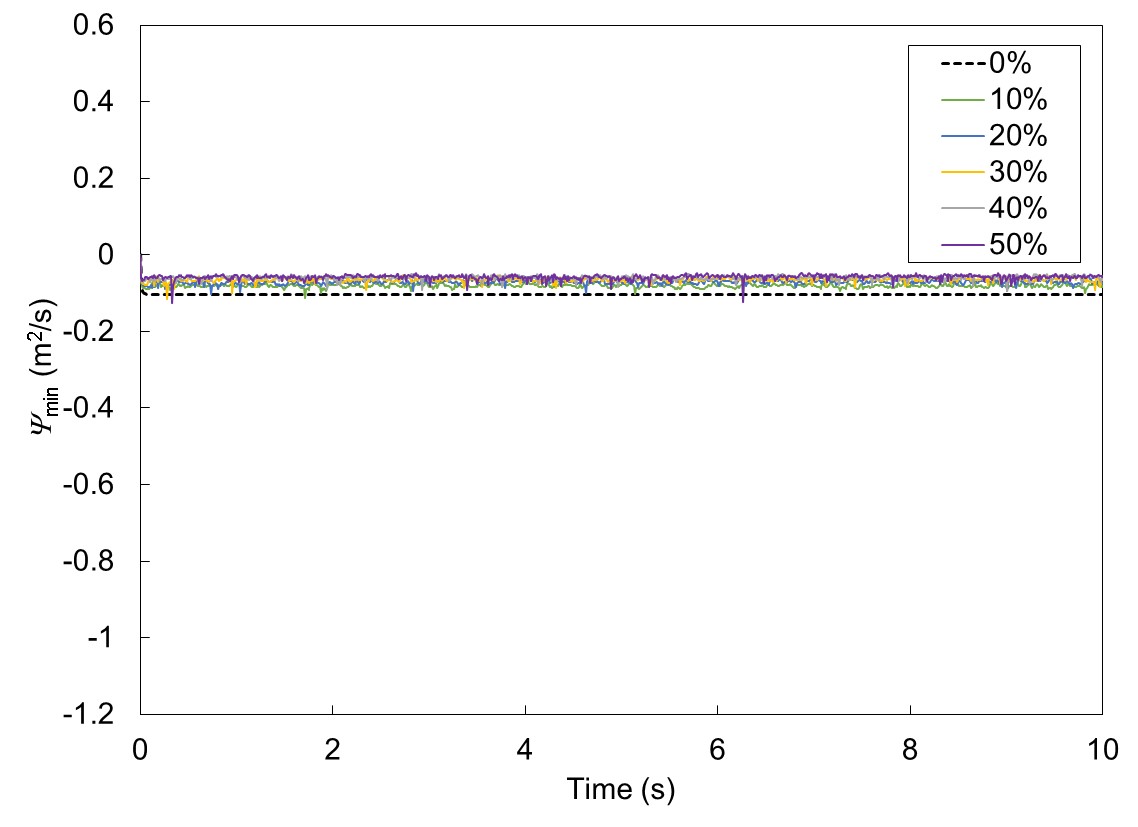}&
		\adjincludegraphics[width=8.50cm]{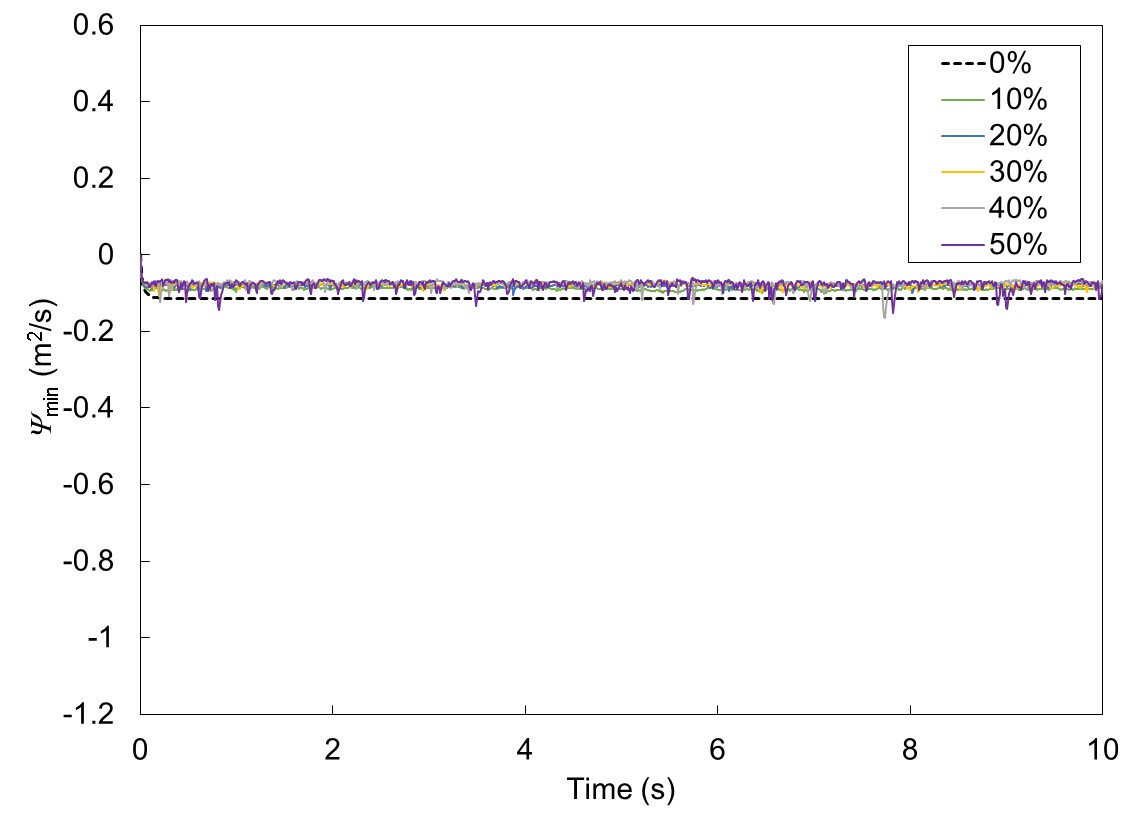}&
		\\
		\adjincludegraphics[width=8.50cm]{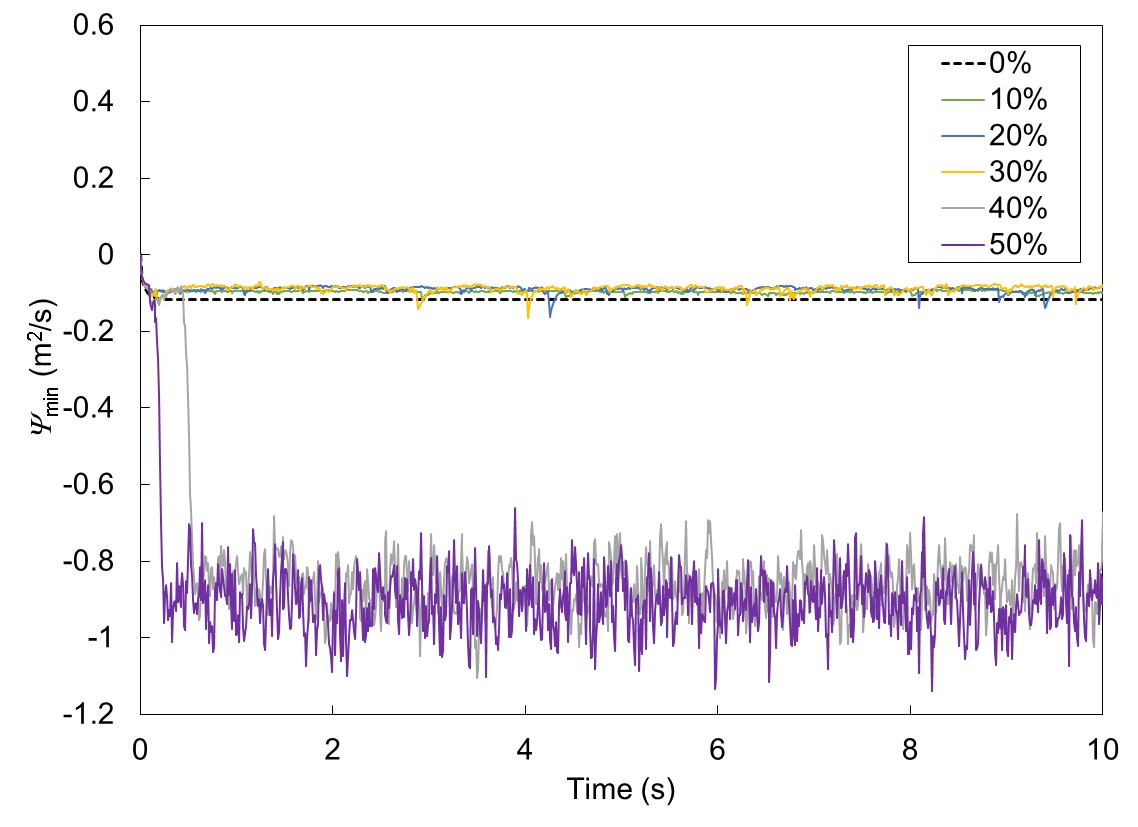}&
		\adjincludegraphics[width=8.50cm]{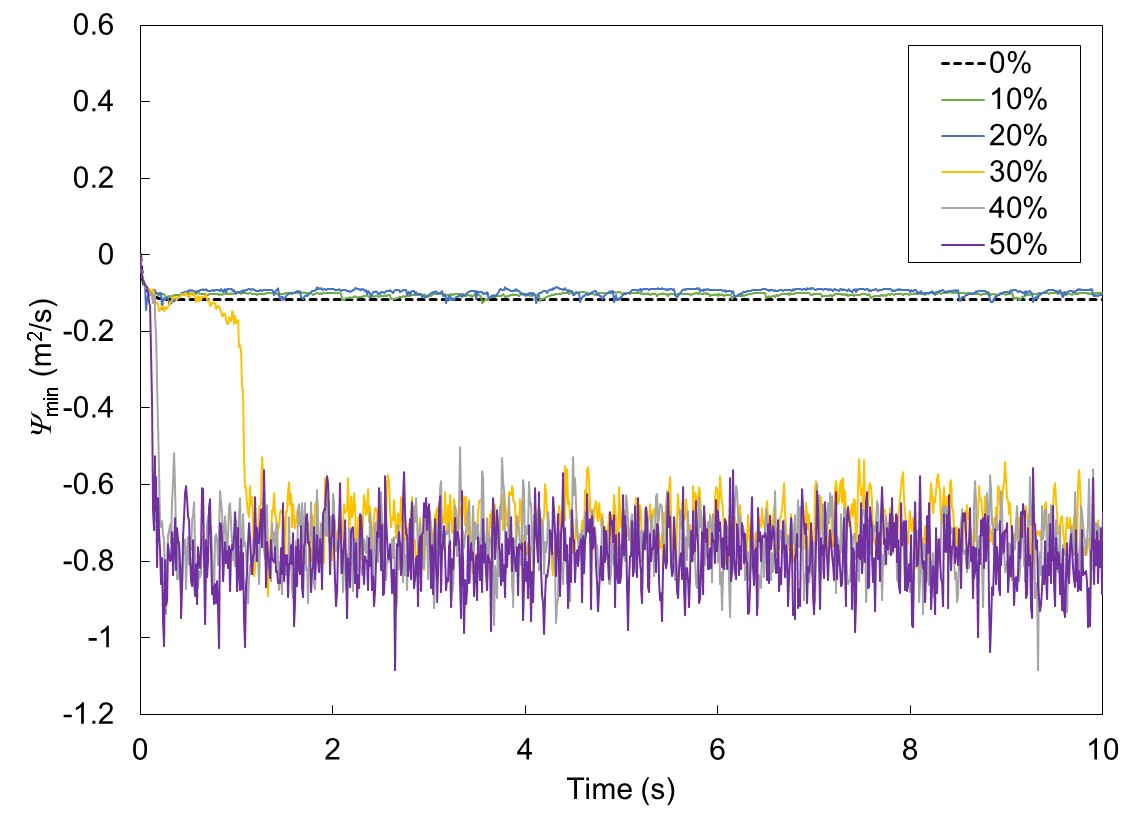}&				
		\\
	\end{tabular}
	
	\caption{Minimum stream function value vs. time for Reynolds numbers = 100 (Top Left), 400 (Top Right), 700 (Bottom Left), 1000 (Bottom Right).}
	\label{fig:minimum_sf_vs_time} 
\end{figure*}

\begin{figure*}[t]
	\centering
	\begin{tabular}{lcc}
		\adjincludegraphics[width=8.50cm]{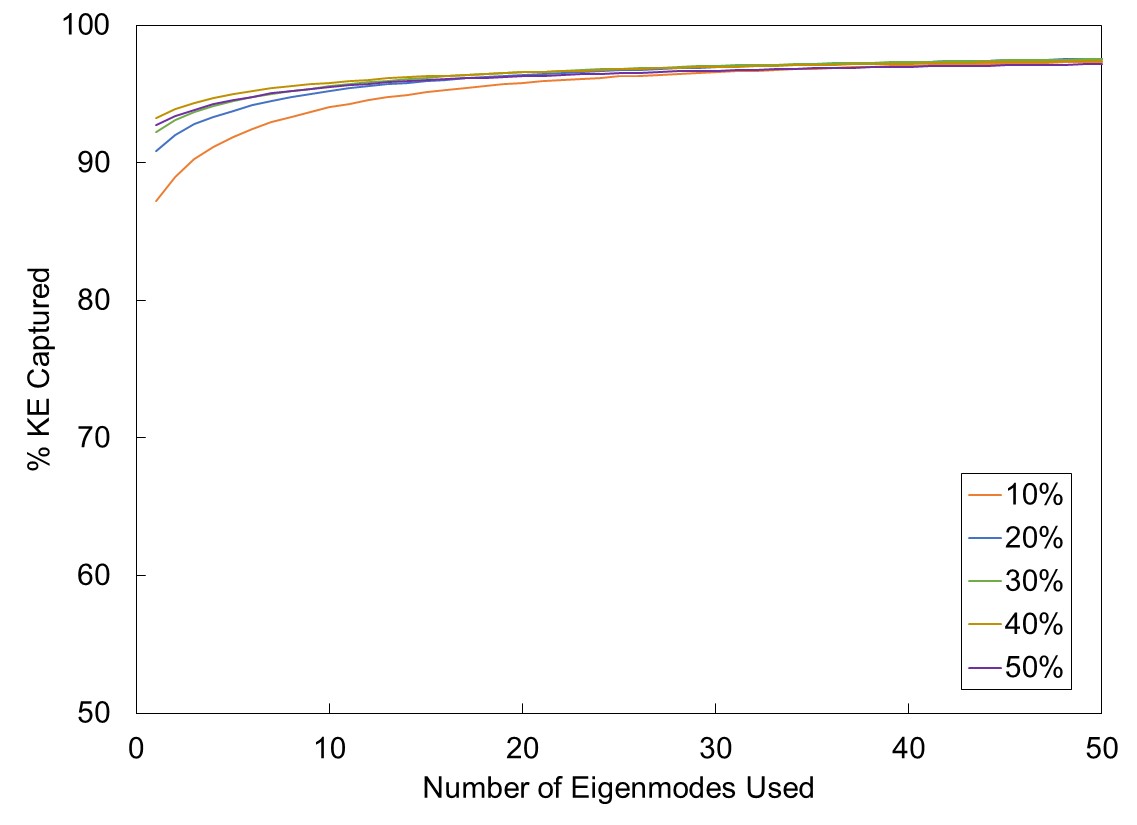}&
		\adjincludegraphics[width=8.50cm]{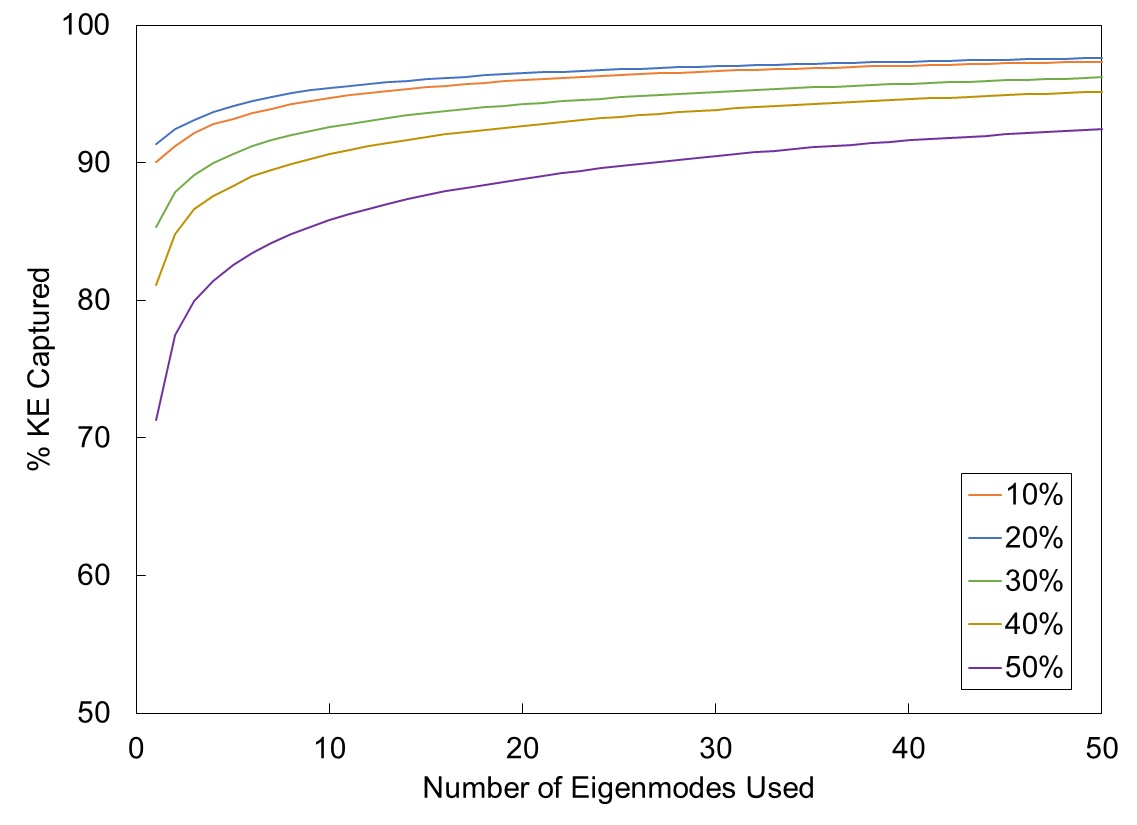}&
		\\
		\adjincludegraphics[width=8.50cm]{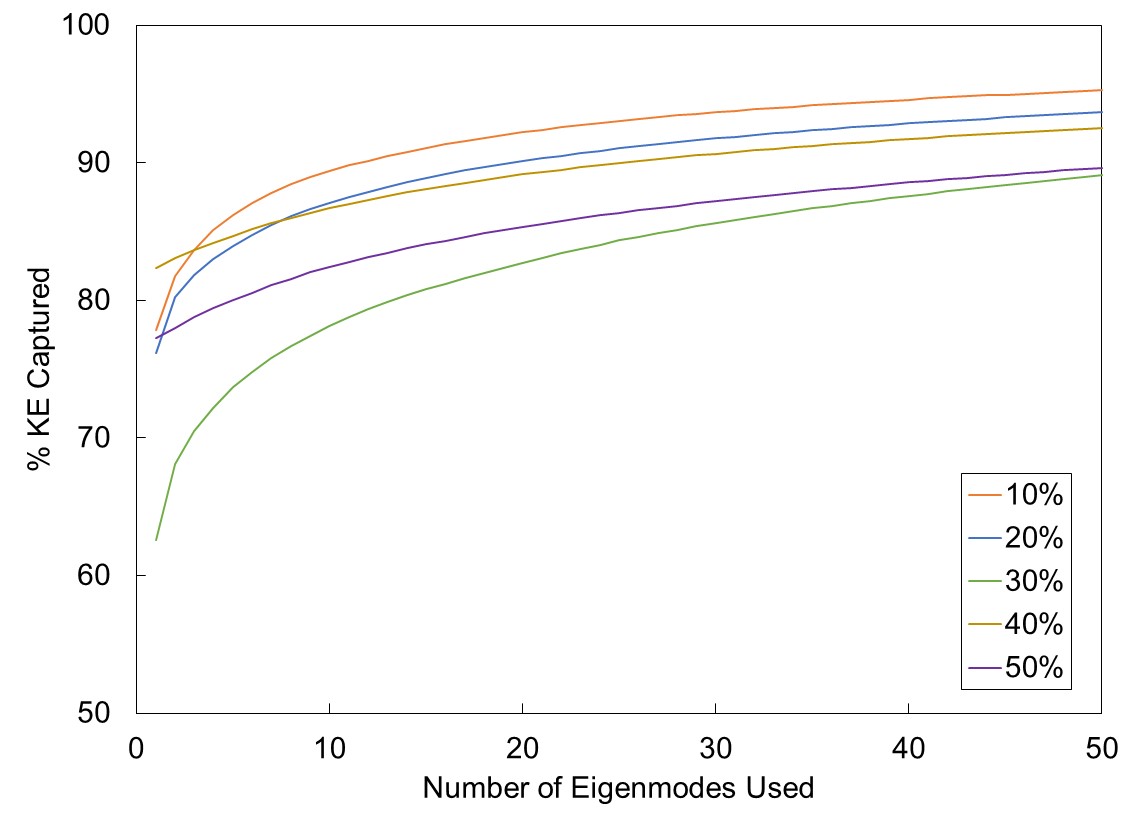}&
		\adjincludegraphics[width=8.50cm]{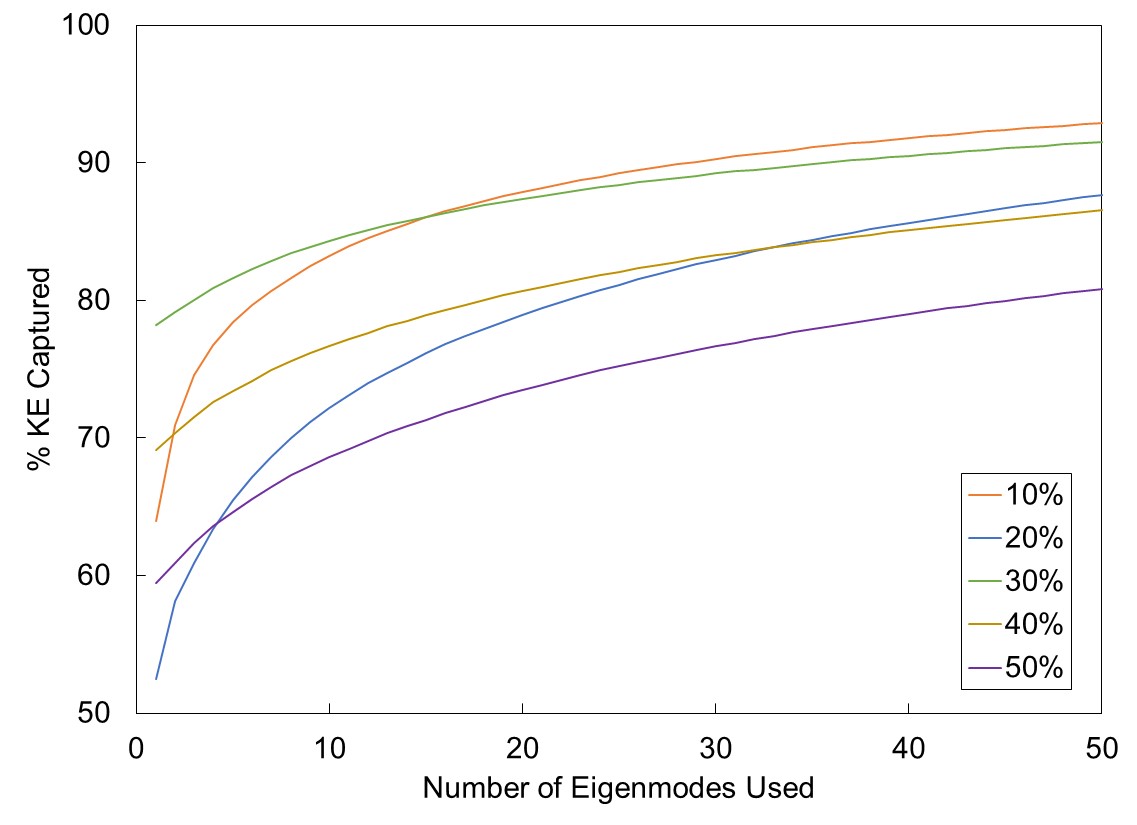}&				
		\\
	\end{tabular}
	
	\caption{Kinectic energy vs eigenmode for all Reynolds numbers.}
	\label{fig:ke_vs_eigenmode} 
\end{figure*}

\begin{figure*}
	\centering
	\begin{tabular}{lccc}
		\adjincludegraphics[width=5.7cm,trim={9cm 1cm 9cm 2cm} ,clip]{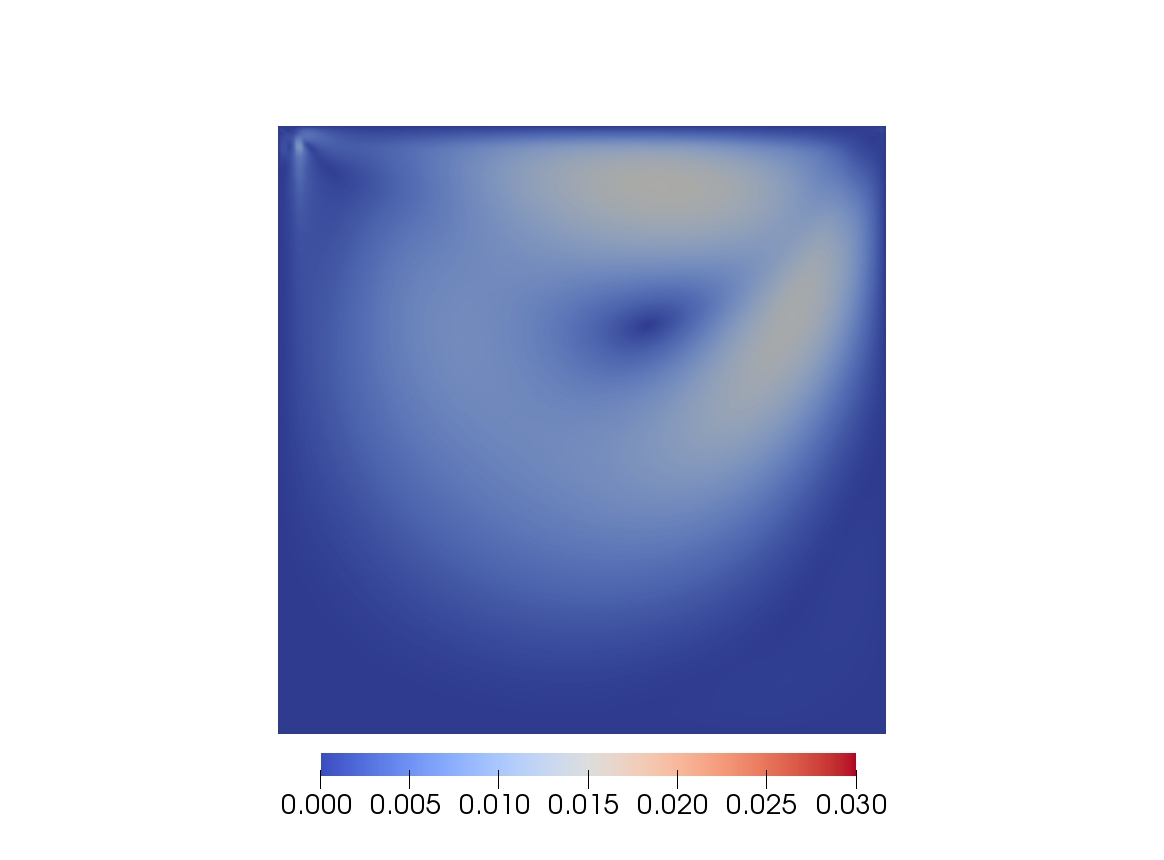}&				\adjincludegraphics[width=5.7cm,trim={9cm 1cm 9cm 2cm} ,clip]{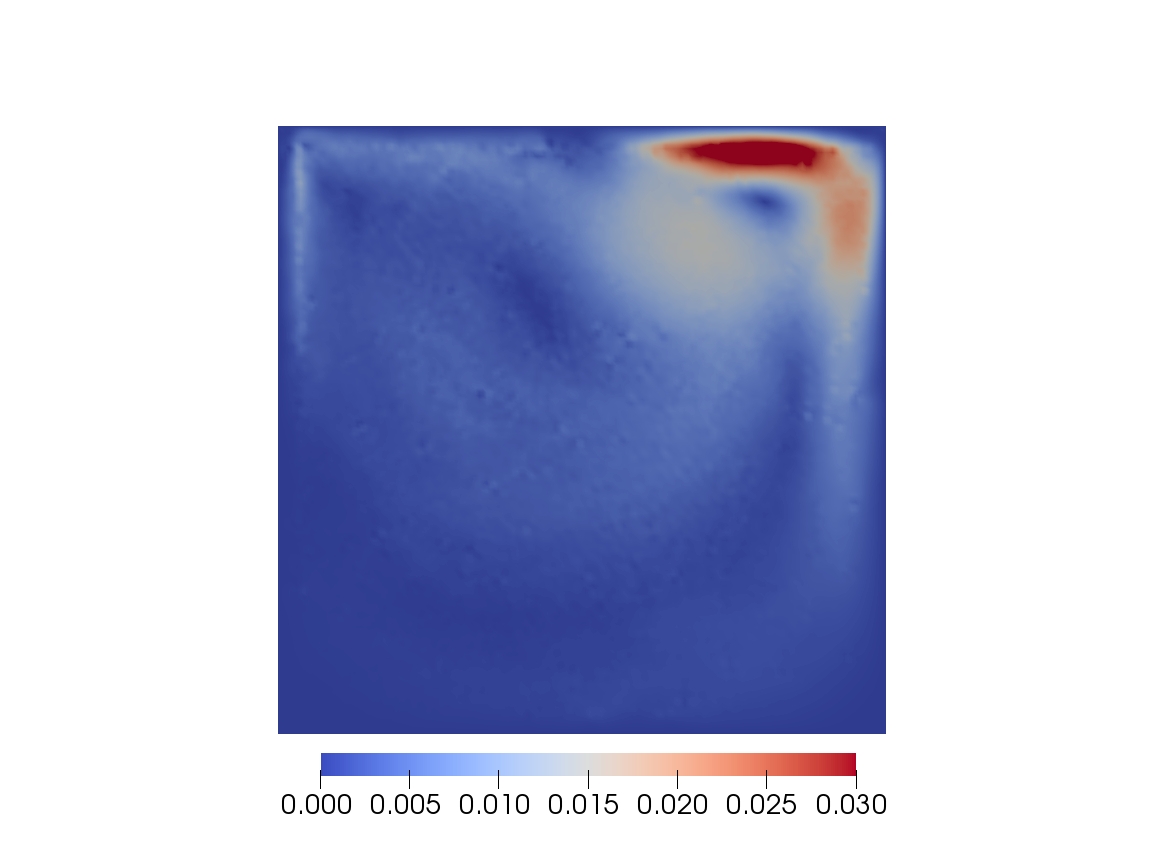}&
		\adjincludegraphics[width=5.7cm,trim={9cm 1cm 9cm 2cm} ,clip]{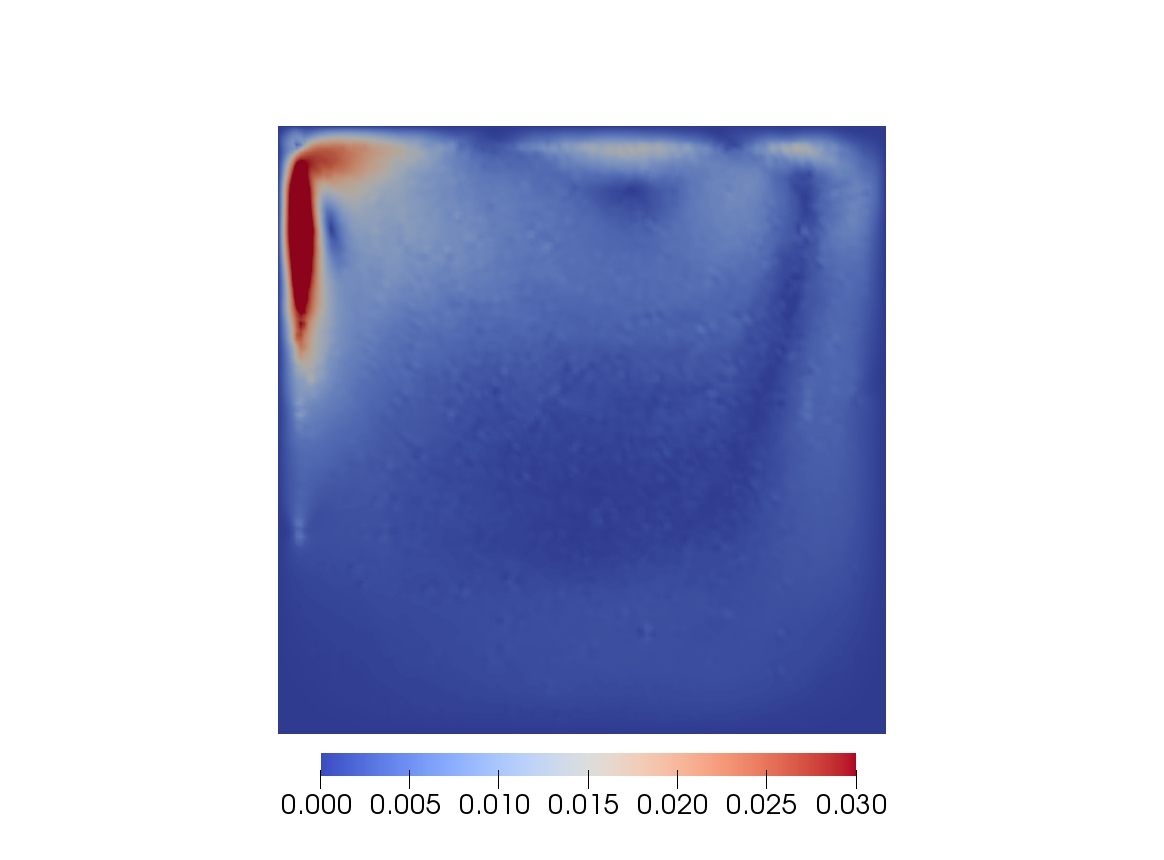}&
	\end{tabular}
	
	\caption{First (left), second (middle), and third (right) eigenmodes for $\phi={50\%}$ and for Reynolds number = 100.}
	\label{fig:eigenmodes_phi50_Re100} 
	
	\begin{tabular}{lccc}
		\adjincludegraphics[width=5.7cm,trim={9cm 1cm 9cm 2cm} ,clip]{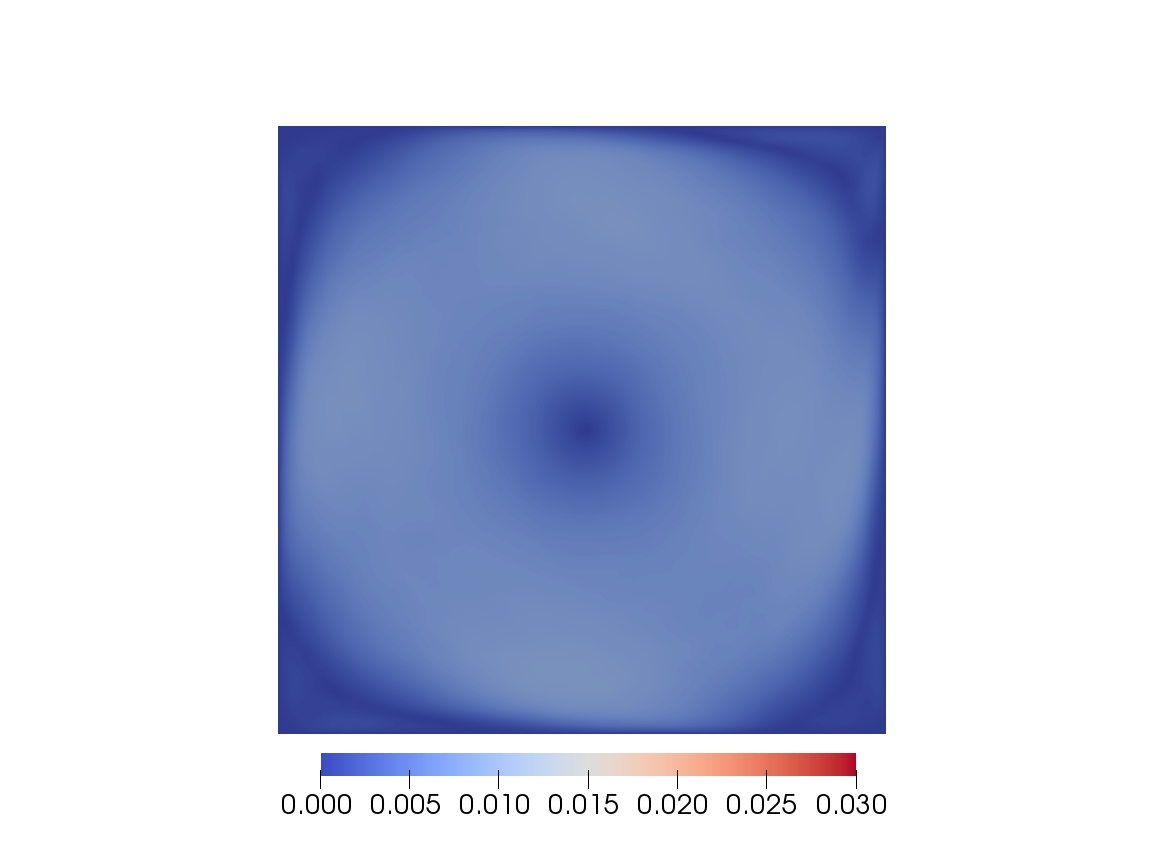}&				\adjincludegraphics[width=5.7cm,trim={9cm 1cm 9cm 2cm} ,clip]{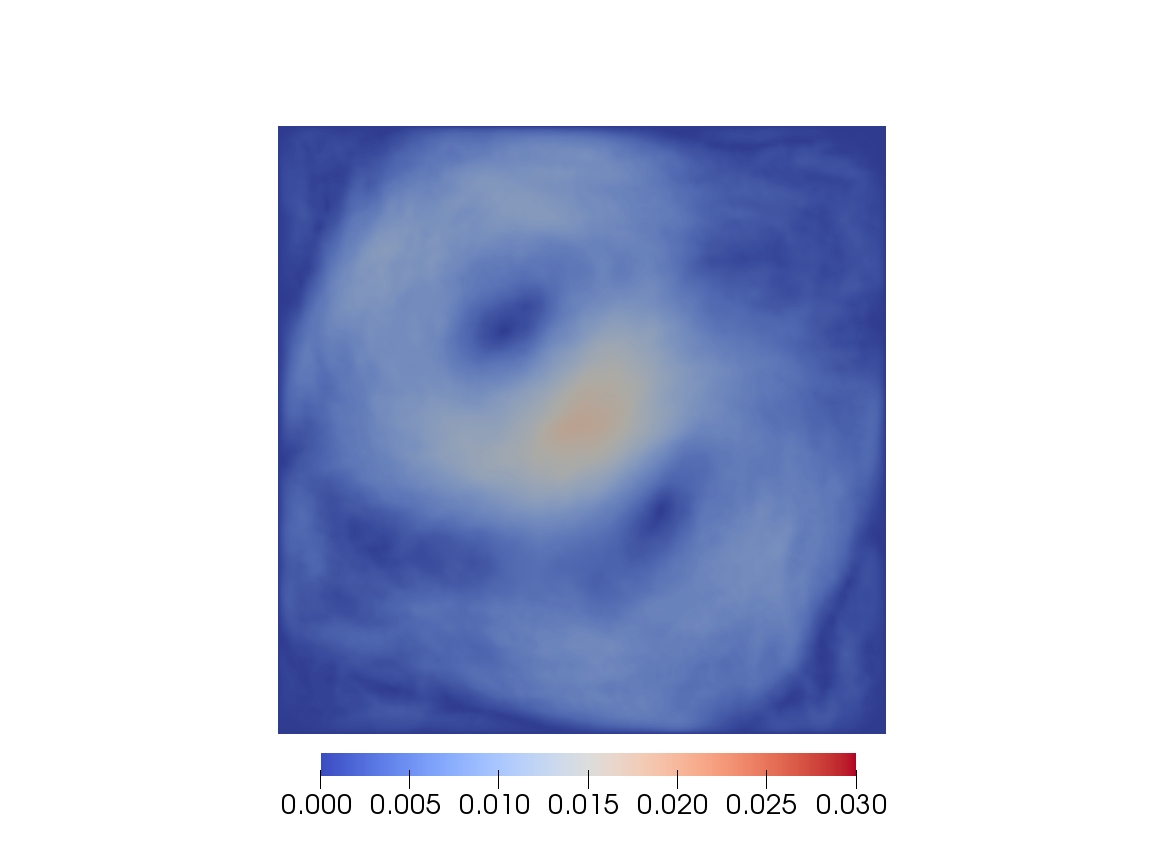}&
		\adjincludegraphics[width=5.7cm,trim={9cm 1cm 9cm 2cm} ,clip]{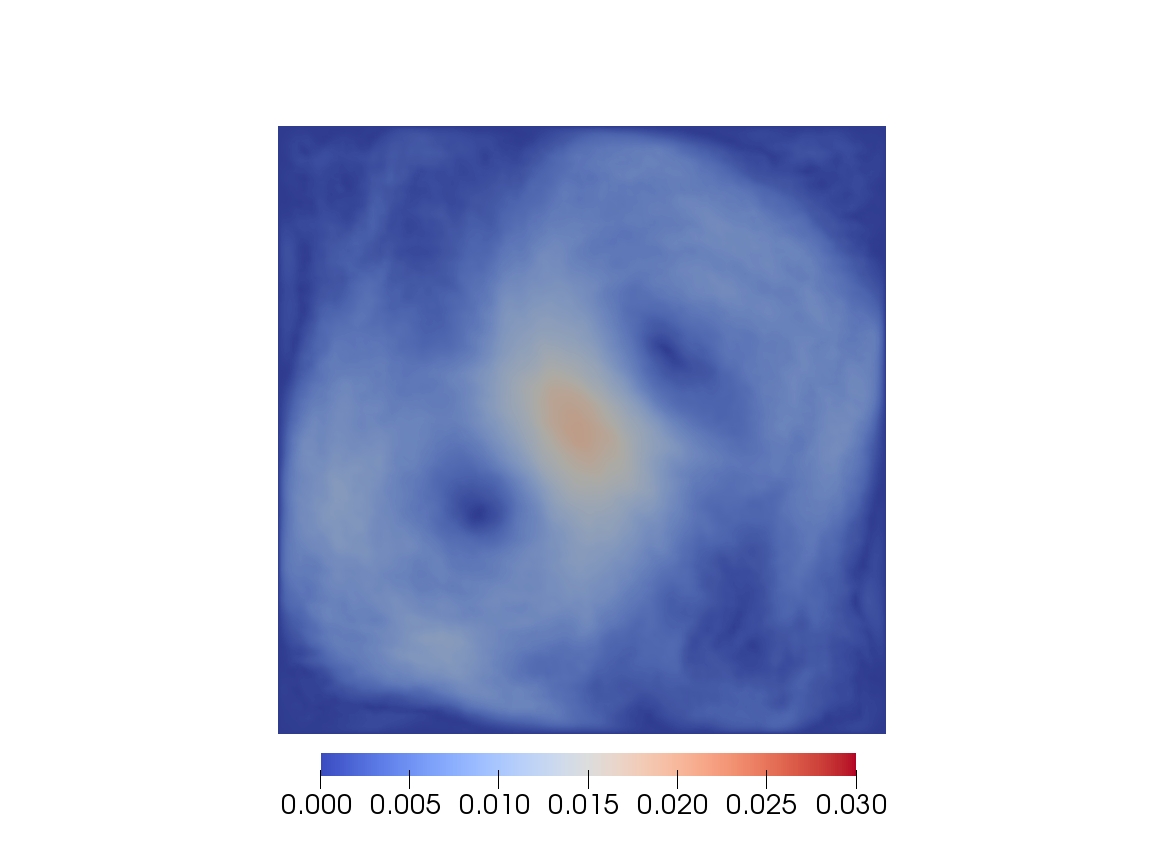}&
	\end{tabular}
	
	\caption{First (left), second (middle), and third (right) eigenmodes for $\phi={50\%}$ and for Reynolds number = 1000.}
	\label{fig:eigenmodes_phi50_Re1000} 
	
	\begin{tabular}{lccc}
		\adjincludegraphics[width=5.7cm,trim={9cm 1cm 9cm 2cm} ,clip]{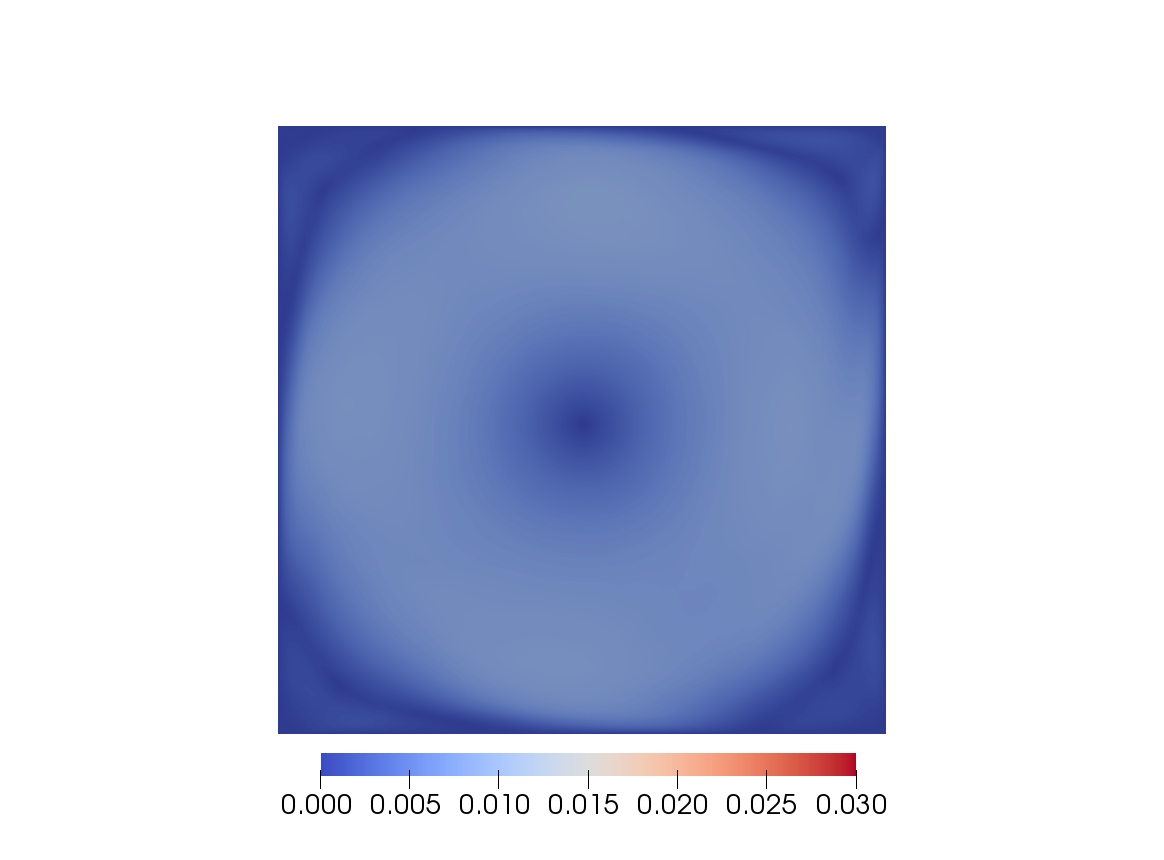}&				\adjincludegraphics[width=5.7cm,trim={9cm 1cm 9cm 2cm} ,clip]{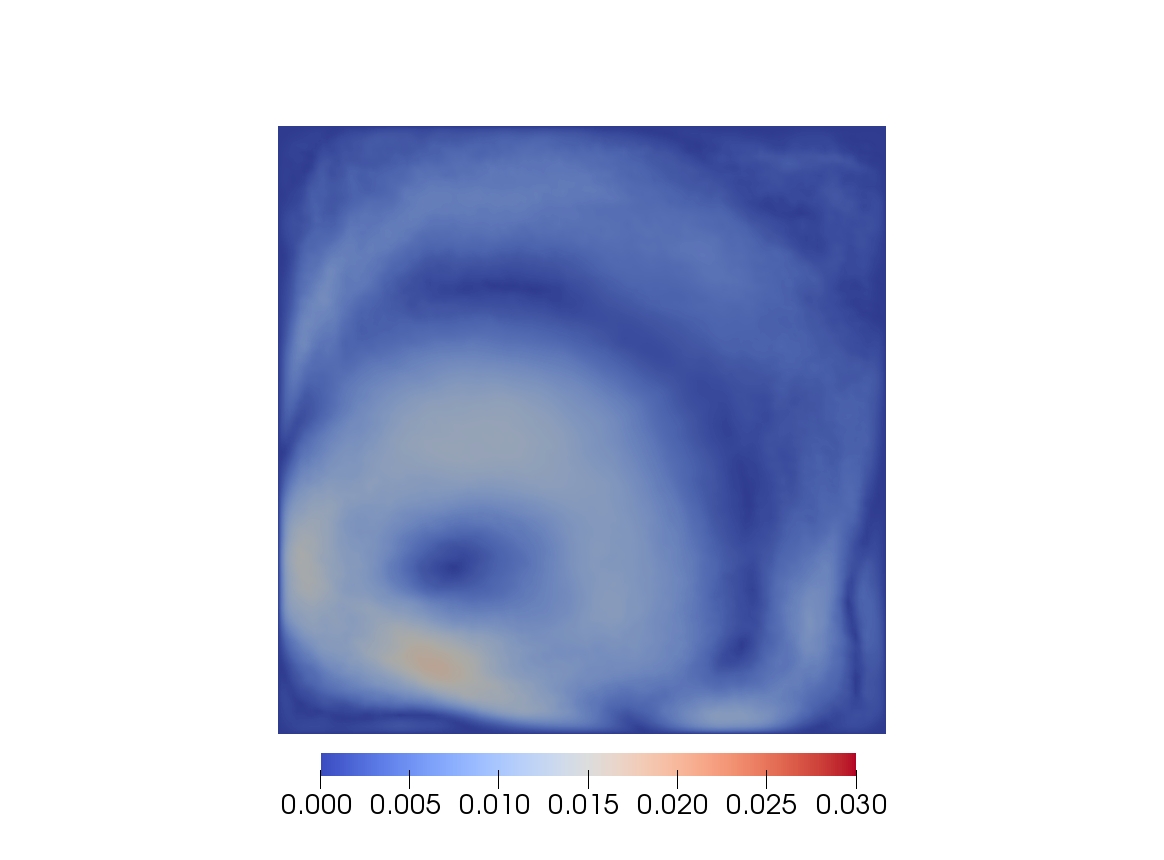}&
		\adjincludegraphics[width=5.7cm,trim={9cm 1cm 9cm 2cm} ,clip]{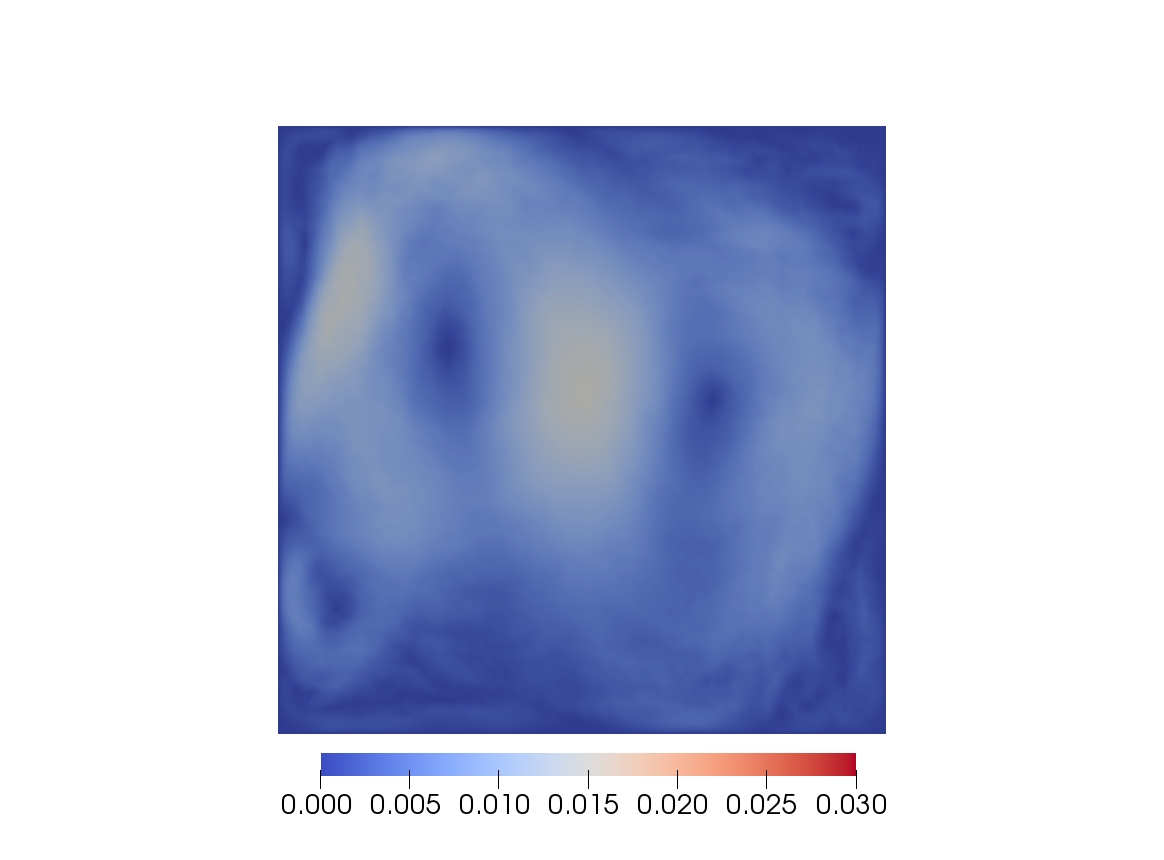}&
	\end{tabular}
	
	\caption{First (left), second (middle), and third (right) eigenmodes for $\phi={30\%}$ and for Reynolds number = 1000.}
	\label{fig:eigenmodes_phi30_Re1000} 
\end{figure*}

\section{Results and Discussion}

In order to understand the effect that particles have on the fundamental structure of the underlying flow field, the dynamics of the entire particle-fluid system within the cavity needs to be analyzed over time. This can be done by focusing on the minimum stream function value within the cavity which reflects the strength of the primary vortex at its center. Fig. 3 presents the minimum stream function value of each particle-fluid system over time simulated and two observations can be made. 

First, it can be seen that the overall flow within a cavity containing particle suspensions can be characterized as either being unsteady and time-independent (U-TI) or unsteady and time dependent (U-TD) for all particle area fractions and all Reynolds numbers investigated. This indicates that the transition from a regime of steady flow behavior (S) characteristically found in unperturbed systems towards a regime of unsteady flow behavior can occur at a Reynolds number less than 100 with a particle area fraction less than 10\%. This observation is in strong contrast to the steady-to-unsteady flow Hopf bifurcation points that were observed to occur at Reynolds numbers ranging from 7000-9000 in the two-dimensional lid driven cavity problem that had been observed in previous studies using either direct numerical simulations, linear stability analysis or proper orthogonal decomposition\cite{erturk:2004}. By the end of each numerical simulation performed, this unsteady flow behavior always ends up fluctuating about an equilibrium value which indicates that this unsteady flow can be characterized as a global attractors for this dynamical system. 

The second observation that can be made is that the presence of particles in the lid-driven cavity caused two distinct global attractors that we have characterized as either being a drag-dominated particle laden flows (PL) or a mixture of both particle laden and particle collision granular flow (MX). Drag-dominated equilibrium responses are those particle-fluid systems whose minimum stream function values have reached an equilibrium value near the minimum stream function value of the unperturbed base flow. In the case where there is a mixture of drag- and particle collision-dominated equilibrium responses, the minimum stream function value of these particle-fluid systems has shifted to a lower equilibrium value via a step function. Table I categorizes each particle-fluid system according to these two distinct responses. The characteristics of each of these equilibrium responses can be further analyzed using proper orthogonal decomposition where the dominant eigenvectors indicate where the effect of the perturbations occurred most often during the simulation and present the flow structures that resulted from them (See Supplementary Material for the complete set of eigenmodes for all Reynolds numbers and particle area fractions). According to Fig. 8, the most impact of these perturbations are captured in the first ten eigenmodes, where between 50\% and 90\% of the kinetic energy is captured. A closer examination of these eigenmodes within each type of equilibrium response is discussed below.

\subsection{Drag-Dominated Perturbed Flow}

The drag-dominated flows that we have identified in Fig. 3 are those whose average minimum stream function value is greater than the steady-state value for the 0\% particle case (see Table I). We can surmise that the strength of the primary vortex at the center of the cavity has been weakened due to localized drag force-induced perturbations from all particle suspended within the cavity. This weakening increases asymptotically with increasing particle area fraction within a given Reynolds number, with this asymptotic behavior reaching an equilibrium value at a much steeper rate with increasing Reynolds number. This equilibrium response does not appear to be dependent on the number of particle suspensions at low Reynolds numbers (Re = 100 and 400). These physical perturbations mathematically described by the interfacial momentum exchange coefficient $K_{pg}$ in Eq. 7 and the relative velocity of the particle to the fluid appears to generate the unsteady periodic response that was observed in the fluctuations in the minimum stream function value. The small amplitude of these fluctuations and the relative similarity in the minimum stream function value would suggest that overall fundamental flow structure within the lid driven cavity has not deviated significantly from unperturbed base flow system. This theory is supported by observing the eigenvectors obtained from proper orthogonal decomposition. In the example case shown in Fig. 5 for Reynolds number of 100 and an particle area fraction $\phi$ = 50\%, the first eigenmode shows that most of the perturbations occurred along the typical flow path circulating around the center of the primary vortex. This would indicate that the effect of the perturbations was most likely generated by the drag of the particles present in this flow path. The second and third eigenvectors also indicate that any additional effect of this perturbation type that occurred during the simulation were most likely drag-induced since the locations in the domain that are shaded red indicate where there happened to be most deviations from the time averaged flow regime for this system. 

\subsection{Particle Collision-Dominated Perturbed Flow}
The mixture of drag-dominated and particle collision-dominated flows that we have identified in Fig. 3 are those whose average minimum stream function value has shifted significantly lower than the steady-state equilibrium minimum stream function value for the 0\% particle case (see Table I). This shift would suggest that the strength of the primary vortex has increased by almost an order of magnitude. However, the strongest primary vortex that was calculated by Ghia, et al.\cite{GHIA1982387} over a range of Reynolds numbers of up to 10,000 was for a Reynolds number of 3200 with a value of -0.120377.  Therefore, it is possible that within this second equilibrium response type there is a flow structure that is significantly different than is typically found within a lid driven cavity. Additionally, the larger amplitudes of the fluctuations in the minimum stream function value indicates that the perturbations within this system are generating an unsteady periodic response in this new structure that may be significantly different than was observed for the purely drag-dominated cases. 

By using proper orthogonal decomposition to calculate the first eigenvector for the case of Reynolds number = 1000 and $\phi$ = 50\% (an MX case), it can be seen in Fig. 6 that the effect of the perturbations appear to reflect a drag-dominated flow structure where the perturbed flow follows around a central vortex. However, an examination of the second and third eigenmodes indicate that the effect of the perturbations do not follow this path around a central vortex and would suggest that drag is not the dominant perturbation in action in this equilibrium state. In fact, the effect of these perturbations appear to follow more clearly defined paths that are confined to the four quadrants of the cavity. Additionally, the region at the center of the cavity that is shaded red in the second and third eigenmodes indicate where the effect of the perturbation happened the most over the simulation time when compared to the time averaged flow for this system. Both of these effects could result from the motion of the particles generated by their collisions with themselves and the lid driven cavity boundaries. It also appears that neither the number of collisions due to an increase in particles present in the system nor the Reynolds number appear to affect the equilibrium response and shift towards a more unsteady chaotic equilibrium response. 

The equilibrium response for Reynolds number = 1000 and $\phi$ = 30\% (another MX case) presents an interesting case where two global attractors are present over the simulation time. For approximately one second of simulation time, the equilibrium response is characterized by a drag-dominated system: the minium stream function value is close to that of the unperturbed flow domain with small fluctuations about this equilibrium value. After one second has passed, an abrupt change in the equilibrium response is observed that is similar to the particle-collision dominated perturbed flow. This abrupt shift to this second global attractor appears to persist for the remainder of the simulation and does not either return back to the first global attractor or continue to a third global attractor. Once again, proper orthogonal decomposition reveals that first eigenvector in Fig. 7  indicates the effect of perturbations appear to reflect a drag-dominated flow structure where the perturbed flow follows around a central vortex. Additionally, the third eigenmode indicate a similar effect of particle collisions that was observed for the purely particle collision dominated perturbation case of  Reynolds number = 1000 and $\phi$ = 50\%. However, there is now a new second eigenmode type present in this system. This second eigenmode reflects a mixture of the two global attractors: 1) drag perturbations around a central vortex that has been shifted to the bottom left corner of the cavity and 2) a region of perturbations that follows a secondary path around this shifted central vortex. This second eigenmode possibly indicates the combined effect of drag and particle collision perturbations that occurs around the transition point observed at one second of simulation time. 

\section{CONCLUSION}
In this study, we examined the type of perturbations generated due to physical particles suspended within a lid driven cavity flow and their affect on the stability of the system. Proper orthogonal decomposition is used to probe the affect of these perturbations by comparing the perturbed flow to the steady, two-dimensional base flow. Our results indicate that the physical particles generated either an unsteady time independent or unsteady time dependent periodic response centered on two global attractors of this dynamical system. At low Reynolds numbers and low particle area fractions, an unsteady, time independent response was observed. This response was characterized as drag dominated, which led to a weakened primary vortex with slightly higher minimum stream function values for those cases. Proper orthogonal decomposition further supported the theory that drag dominated in this regime by indicated the effect of this perturbation type along the path surrounding the primary vortex in the first, second, and third eigenmodes. A second time independent response was observed that was characterized as being generated by a mixture of drag- and particle collision-dominated perturbations. The effect of the particle collision based perturbations in this second equilibrium response appeared in the second and third eigenmodes, while the effect of drag perturbations still appeared in the first eigenmode. There was also one case where an unsteady, time dependent periodic response occurred and both equilibrium responses were present. This particular case led to the observation of a second perturbation effect on the fluid flow. This third response type was considered to be the underlying transitory perturbation effect due to a mixture of drag and particle collision perturbations interacting with each other. These speculations require further experimentation and a more detailed analysis to determine the transitions between these regimes and their underlying flow structures. In the future, it will be also interesting to investigate the effect of the properties of the particles themselves on the characteristics of the equilibrium states. 

\section*{SUPPLEMENTARY MATERIAL}
See supplementary material for complete set of the first three eigenmodes for all Reynolds numbers and particle area fractions data sets.    

\section*{AUTHOR DECLARATIONS}
\subsection*{Conflict of Interest}
The authors have no conflicts to disclose.

\subsection*{Author Contributions}
\textbf{John Shelton}: Conceptualization (equal); Formal analysis (equal); Methodology (equal); Software (equal); Supervision (equal); Validation (equal); Visualization (equal), Writing - Original Draft (equal); Writing - review \& editing (equal). \textbf{Nitin Katiki}: Formal analysis (equal); Software (equal); Visualization (equal). \textbf{Morakinyo Adesemowo}: Formal analysis (equal); Software (equal); Visualization (equal).

\section*{DATA AVAILABILITY}
The data that support the findings of this study are available from the corresponding author upon reasonable request. 

\section*{REFERENCES}
\nocite{*}
\bibliographystyle{unsrt}
\bibliography{biblatex-Nitin}

\end{document}



\begin{titlepage}
	\begin{center}
	\huge\bfseries
	\vspace*{0.5cm}
	Supplementary Material:\\
	Proper orthogonal decomposition analysis of square lid driven cavity flows containing particle suspensions
	\vspace*{0.5cm}\\
	\mdseries
	\large 	J. Shelton$^1$, N. Katiki$^1$, M. Adesemowo$^1$ \\
	$^1$$\textit{Department of Mechanical Engineering, Northern Illinois University, DeKalb, IL 60115, USA}$\\
	\vspace*{\fill}
	\newpage	    
	\end{center}
\end{titlepage}

\pacs{Valid PACS appear here}

\onecolumngrid
\begin{flushleft}
\section{First Eigenmode: All Reynolds Numbers}
\subsection{10\% Particle Fraction}
\begin{figure*}[h]
	\centering
	\begin{tabular}{lcc}
		\adjincludegraphics[width=7cm,trim={9cm 1cm 9cm 2cm} ,clip]{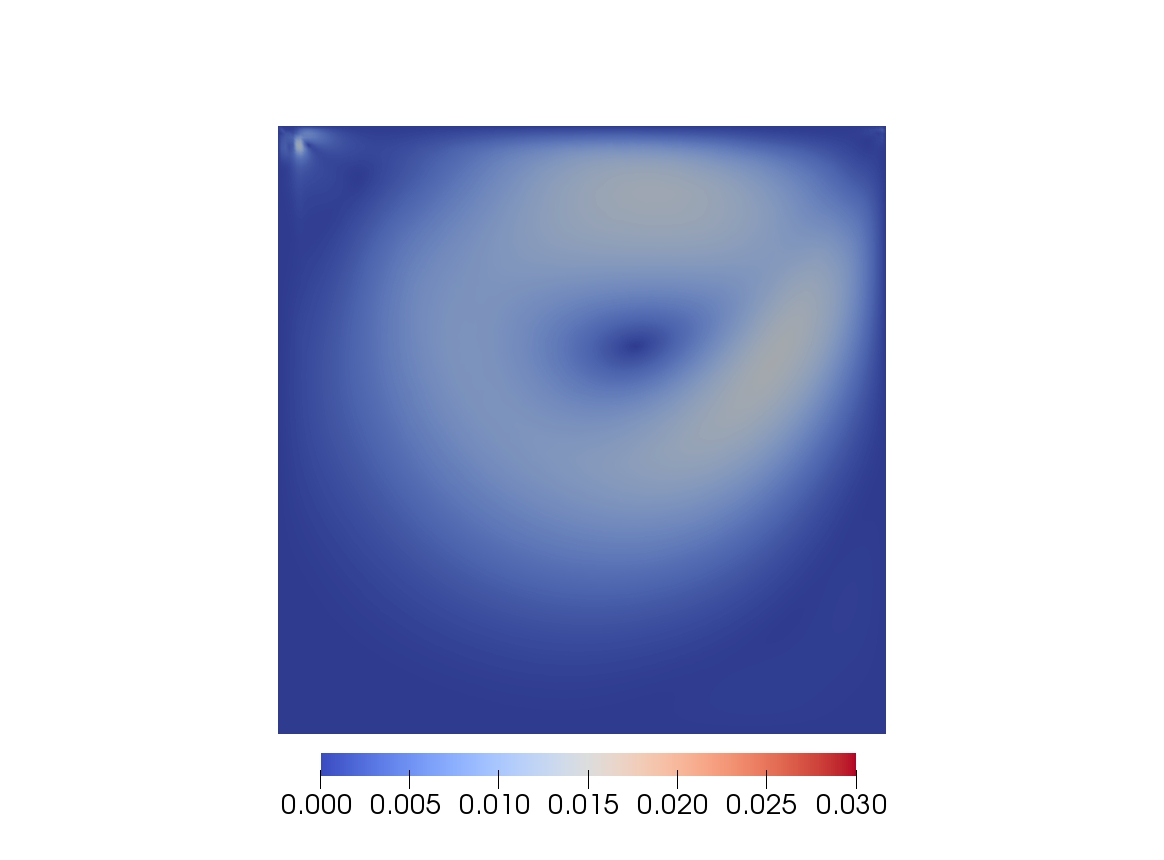}&
		\adjincludegraphics[width=7cm,trim={9cm 1cm 9cm 2cm} ,clip]{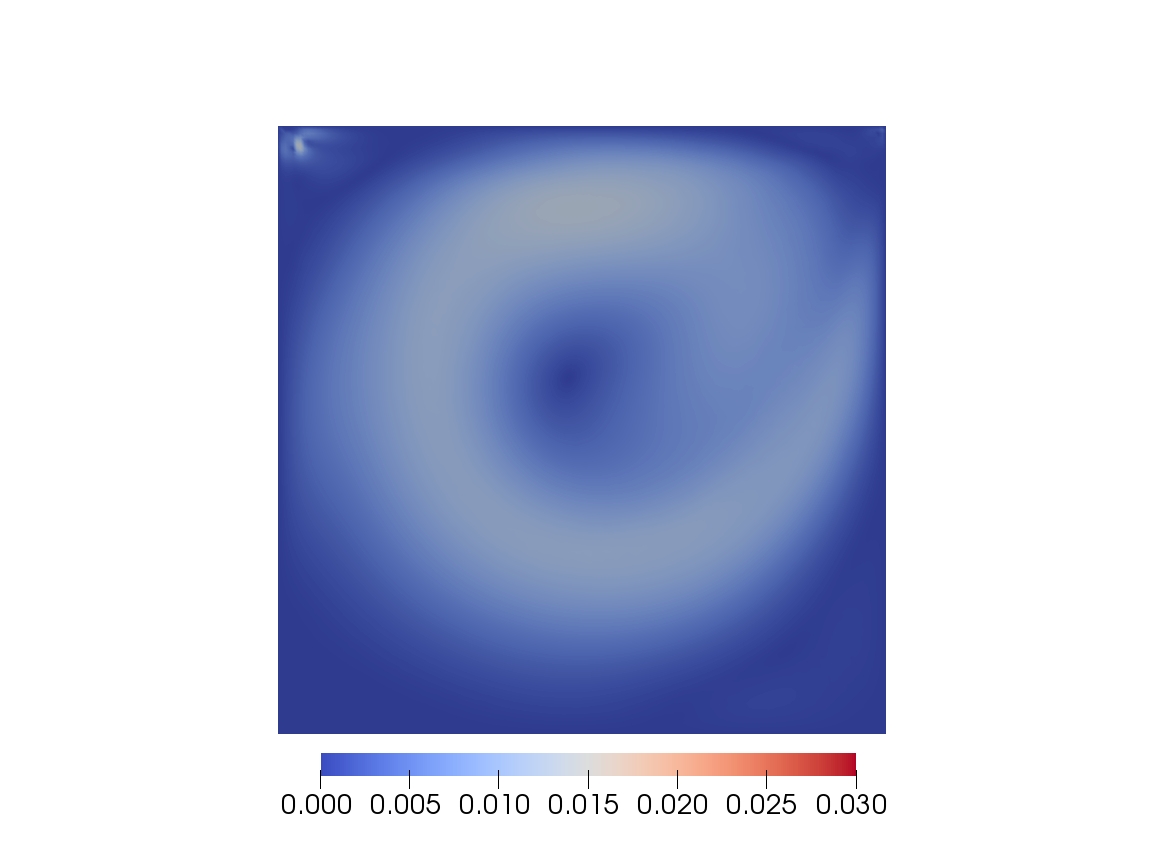}
		\\
		\adjincludegraphics[width=7cm,trim={9cm 1cm 9cm 2cm} ,clip]{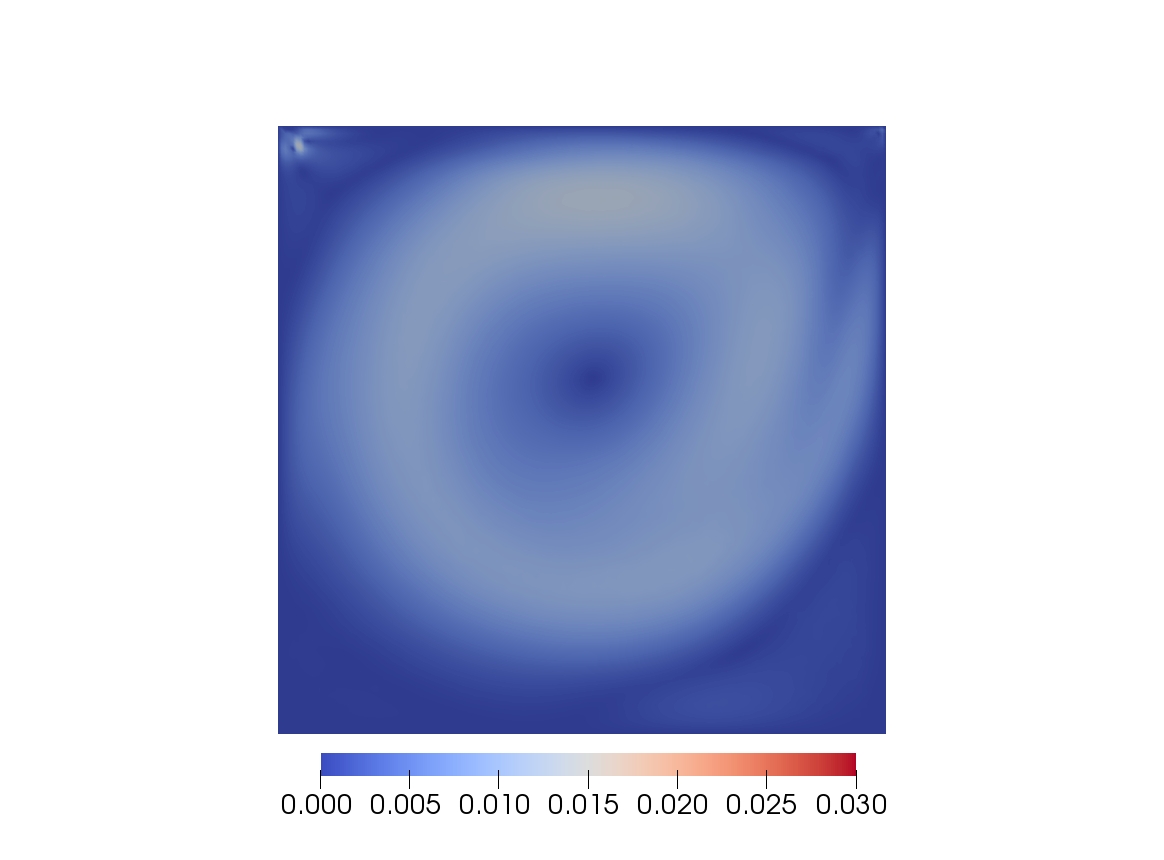}&
		\adjincludegraphics[width=7cm,trim={9cm 1cm 9cm 2cm} ,clip]{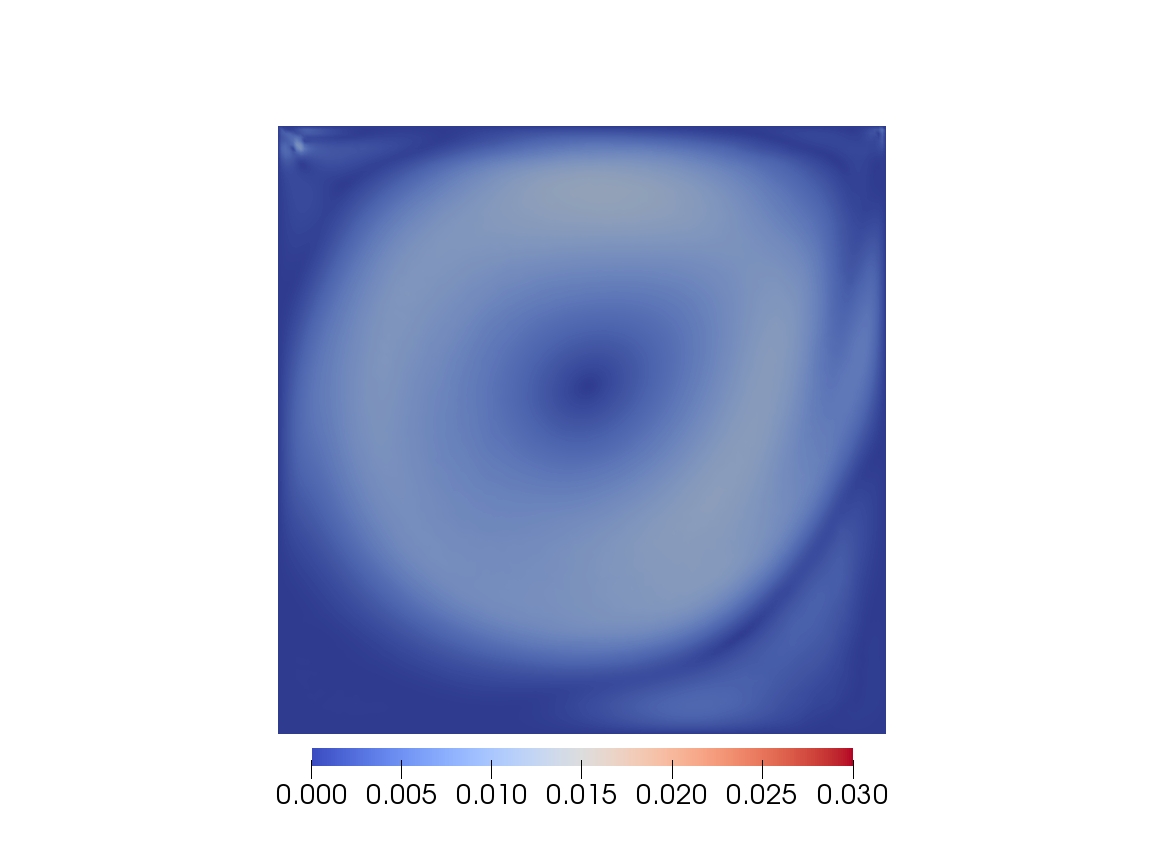}&
		\\
	\end{tabular}
	\caption{First eigenmode for Reynolds number = 100 (top left), 400 (top right), 700 (bottom left), 1000 (bottom right) and $\phi=10\%$.}
	\label{fig:fig4} 
\end{figure*}
\end{flushleft}
\newpage	    

\subsection{20\% Particle Fraction}
\begin{figure*}[h]
	\centering
	\begin{tabular}{lcc}
		\adjincludegraphics[width=7cm,trim={9cm 1cm 9cm 2cm} ,clip]{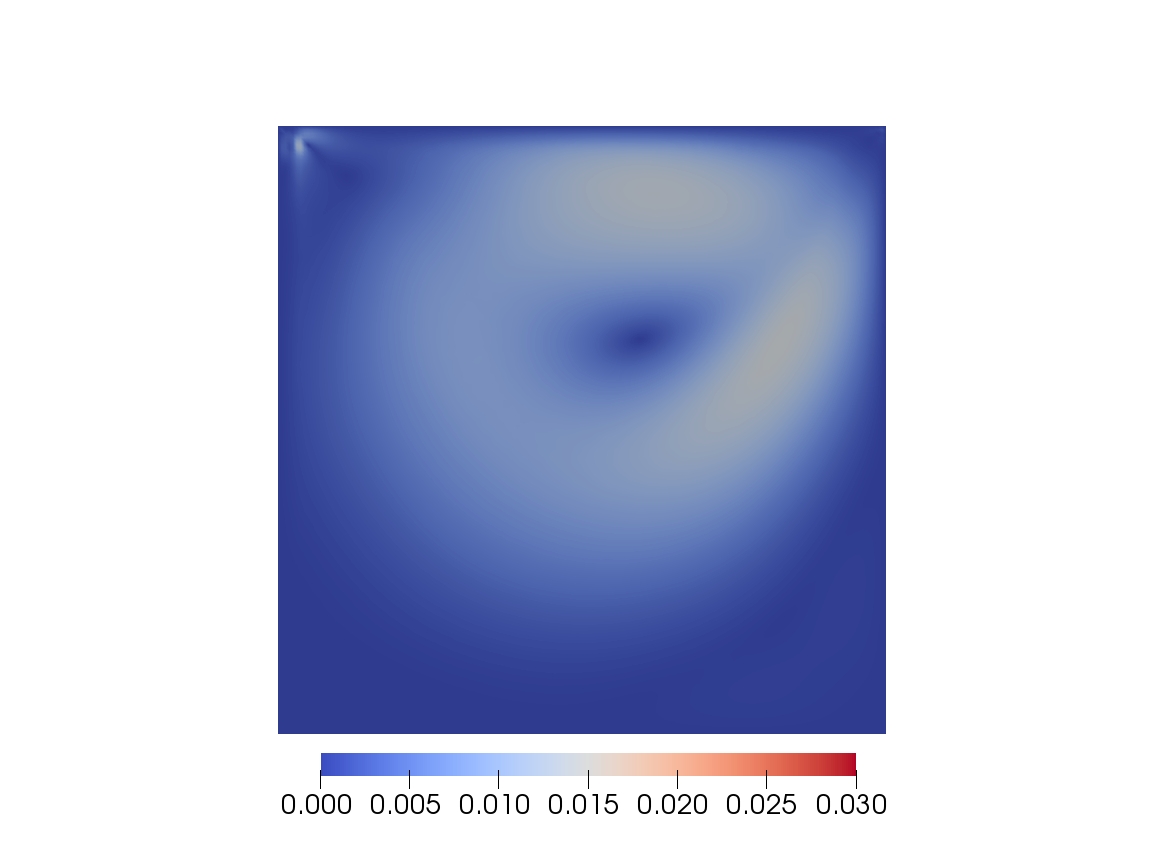}&
		\adjincludegraphics[width=7cm,trim={9cm 1cm 9cm 2cm} ,clip]{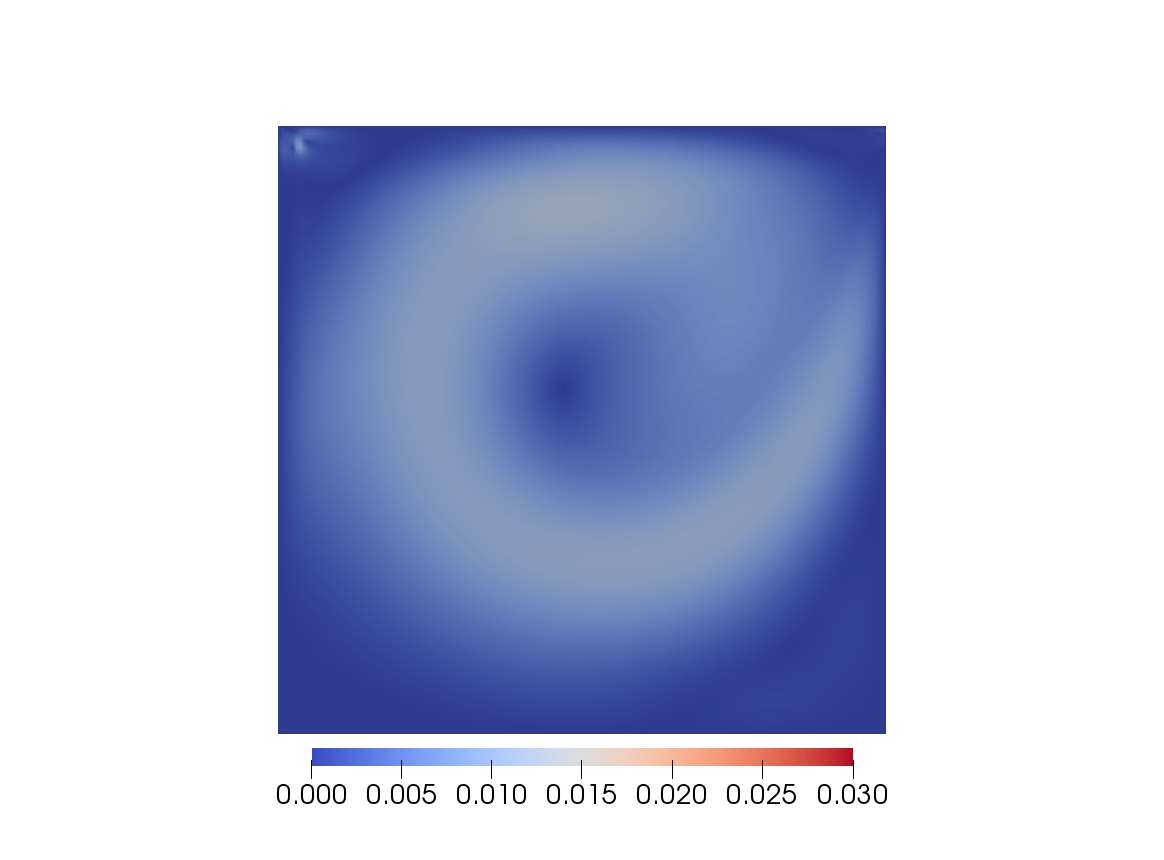}
		\\
		\adjincludegraphics[width=7cm,trim={9cm 1cm 9cm 2cm} ,clip]{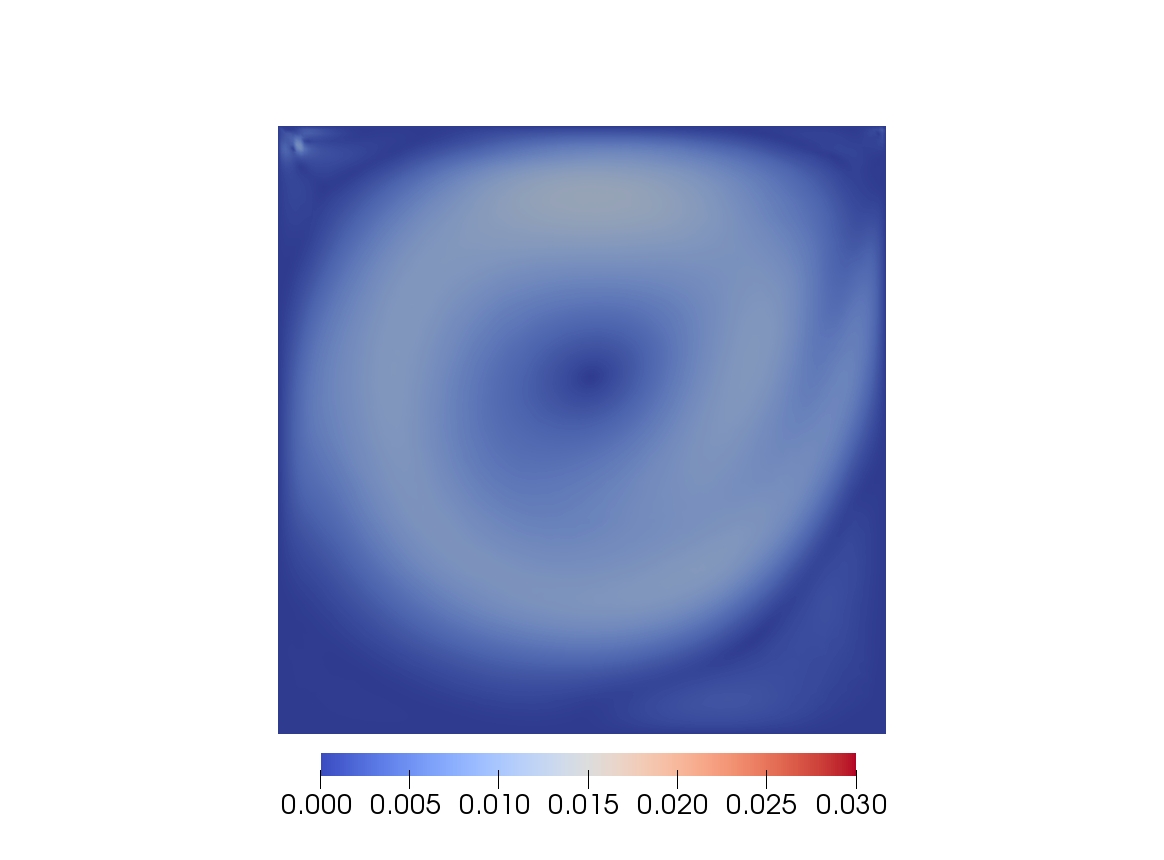}&
		\adjincludegraphics[width=7cm,trim={9cm 1cm 9cm 2cm} ,clip]{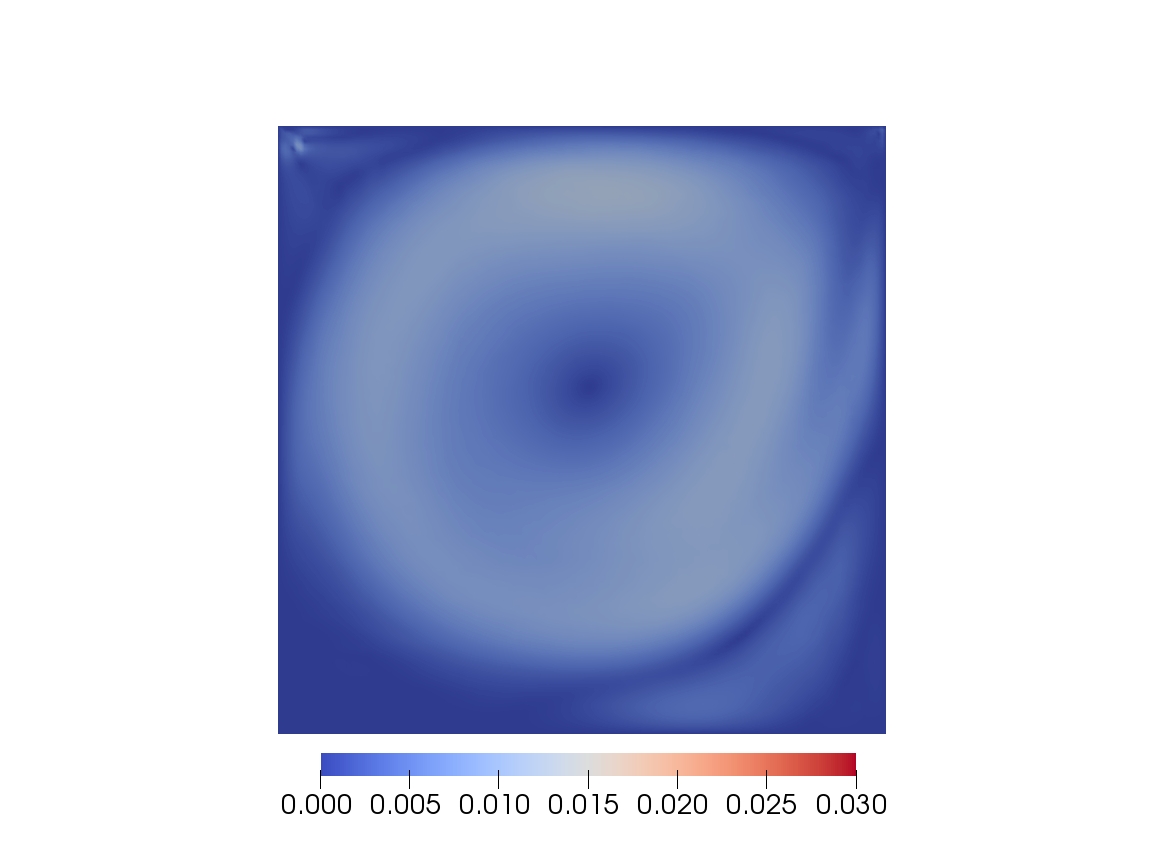}&
		\\
	\end{tabular}
	\caption{First eigenmode for Reynolds number = 100 (top left), 400 (top right), 700 (bottom left), 1000 (bottom right) and $\phi=20\%$.}
	\label{fig:fig4} 
\end{figure*}
\newpage	    

\subsection{30\% Particle Fraction}
\begin{figure*}[h]
	\centering
	\begin{tabular}{lcc}
		\adjincludegraphics[width=7cm,trim={9cm 1cm 9cm 2cm} ,clip]{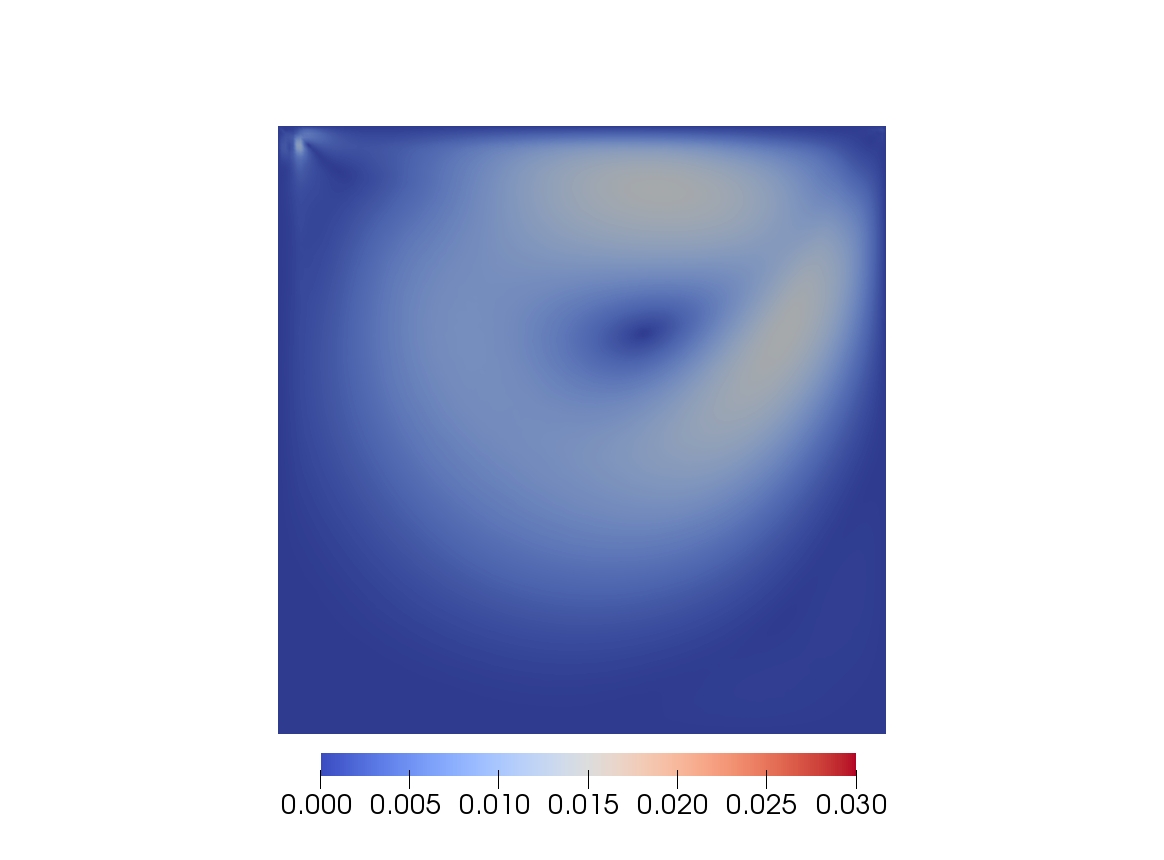}&
		\adjincludegraphics[width=7cm,trim={9cm 1cm 9cm 2cm} ,clip]{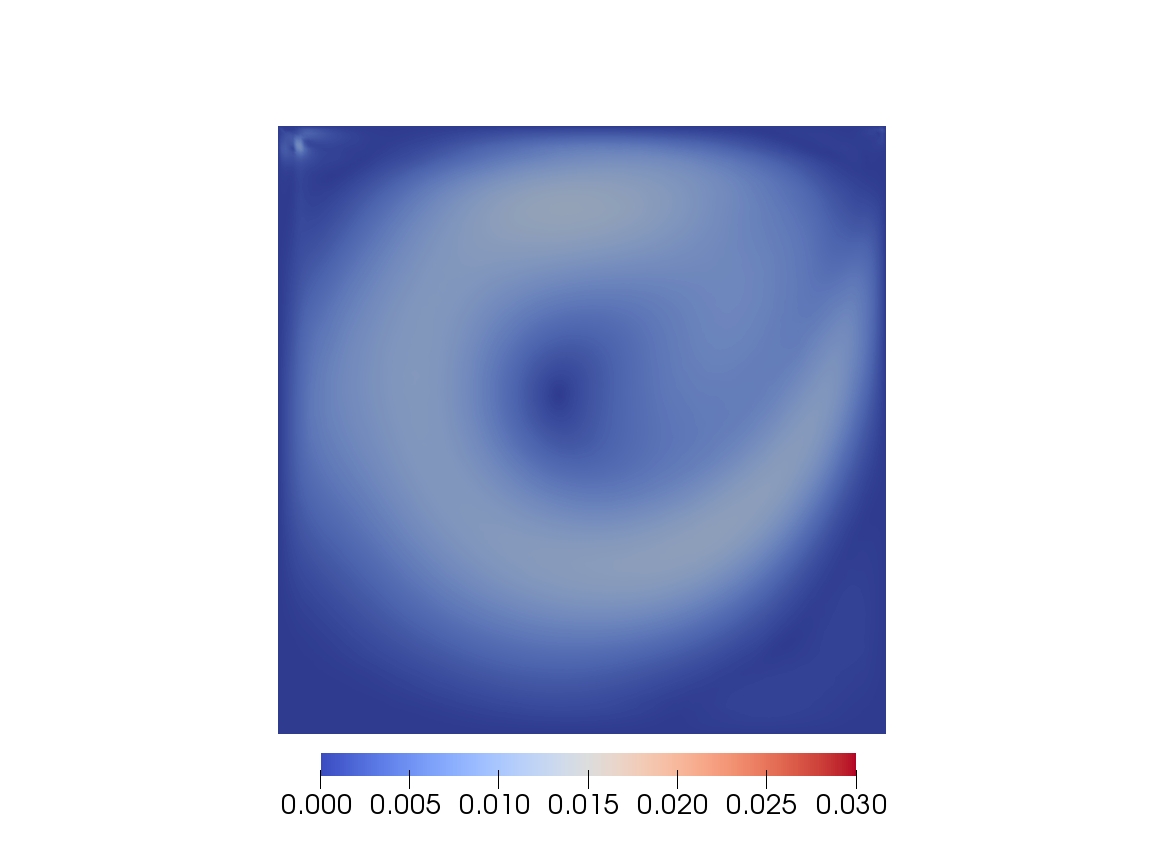}
		\\
		\adjincludegraphics[width=7cm,trim={9cm 1cm 9cm 2cm} ,clip]{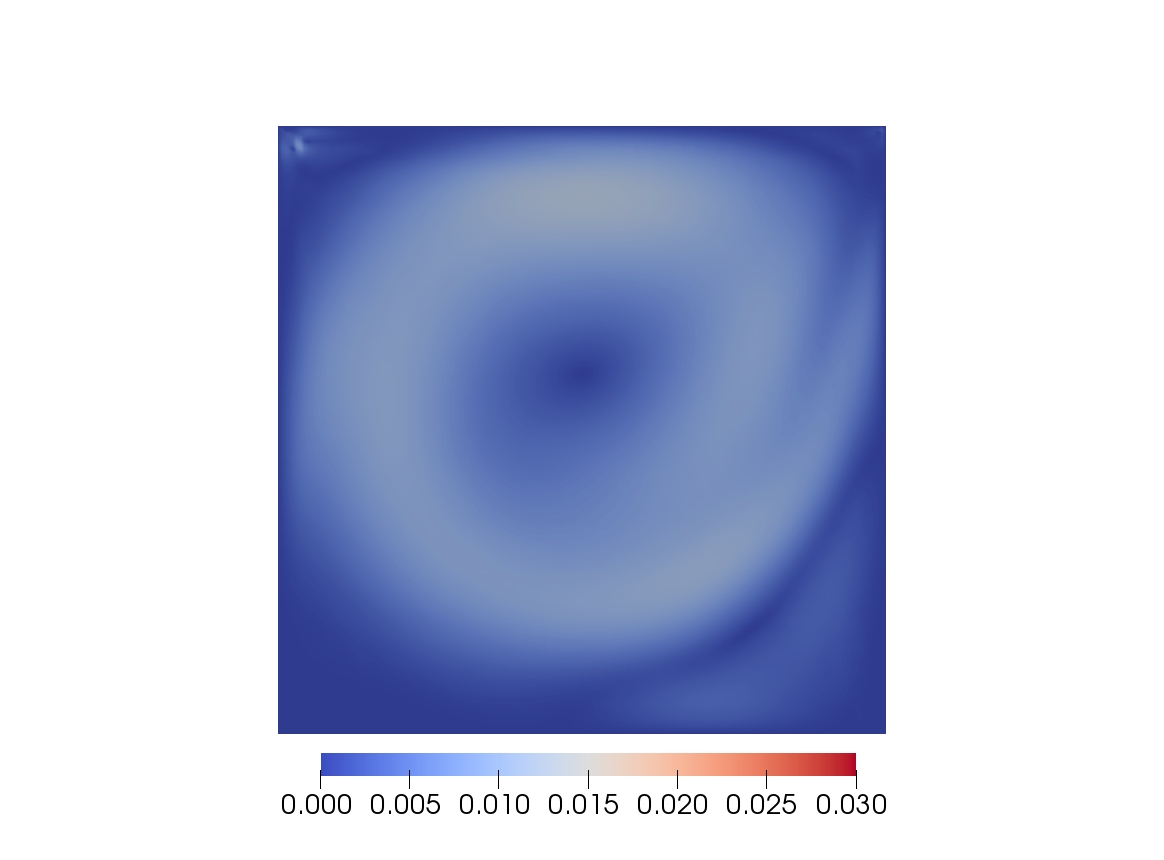}&
		\adjincludegraphics[width=7cm,trim={9cm 1cm 9cm 2cm} ,clip]{./images/Re1000/phi_1_30pct_rhorat_1000}&
		\\
	\end{tabular}
	\caption{First eigenmode for Reynolds number = 100 (top left), 400 (top right), 700 (bottom left), 1000 (bottom right) and $\phi=30\%$.}
	\label{fig:fig4} 
\end{figure*}
\newpage	    

\subsection{40\% Particle Fraction}
\begin{figure*}[h]
	\centering
	\begin{tabular}{lcc}
		\adjincludegraphics[width=7cm,trim={9cm 1cm 9cm 2cm} ,clip]{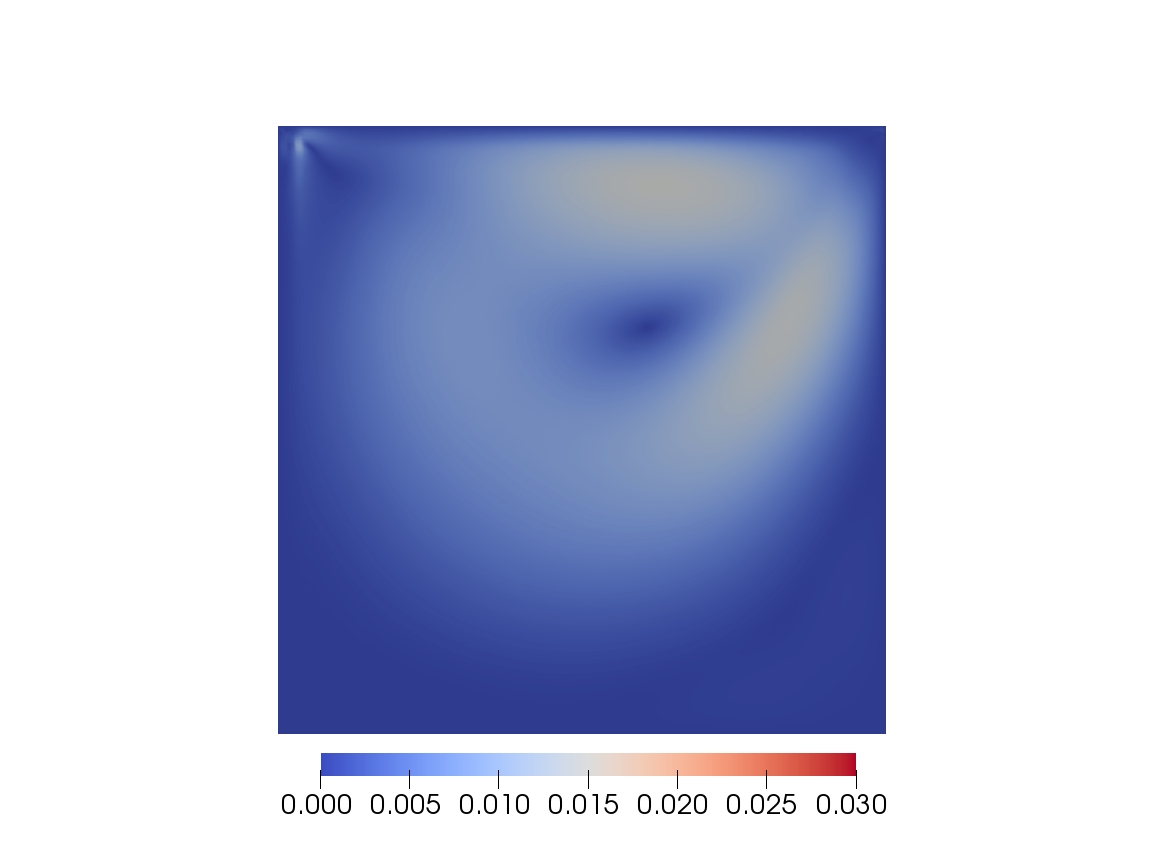}&
		\adjincludegraphics[width=7cm,trim={9cm 1cm 9cm 2cm} ,clip]{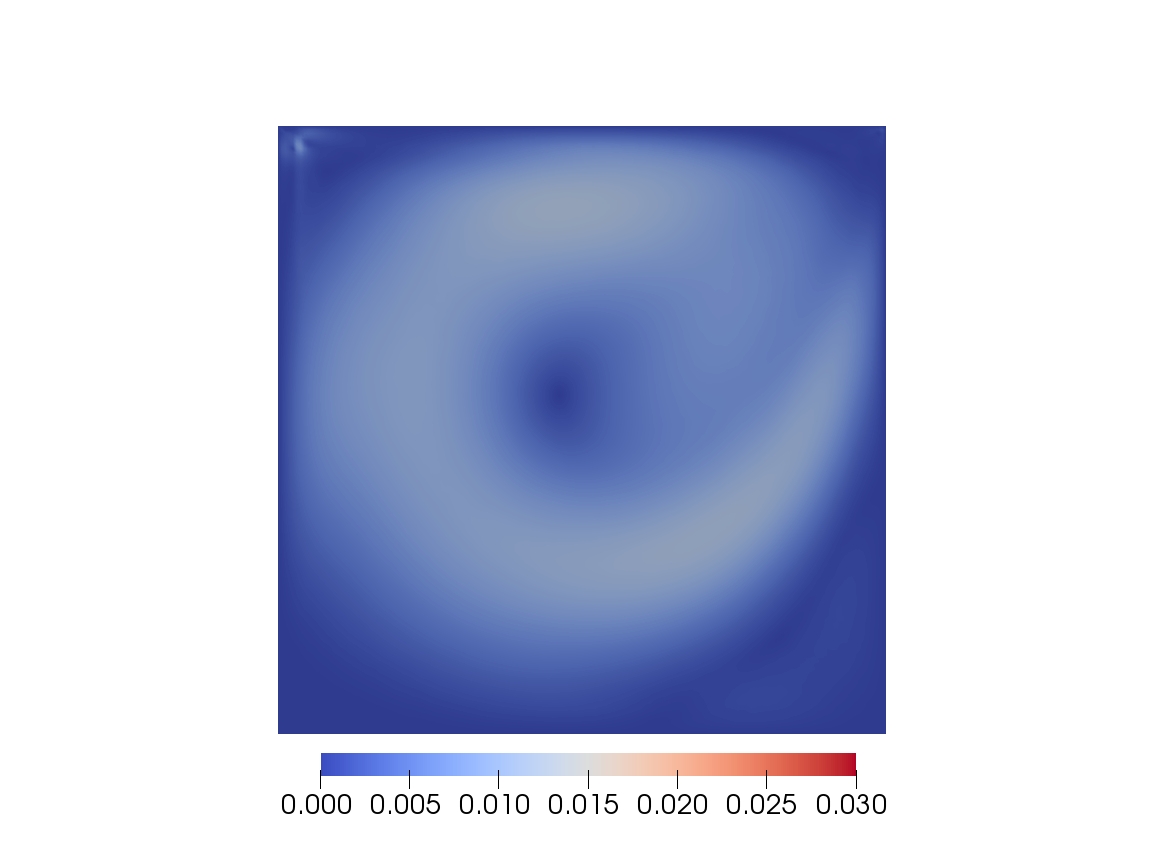}
		\\
		\adjincludegraphics[width=7cm,trim={9cm 1cm 9cm 2cm} ,clip]{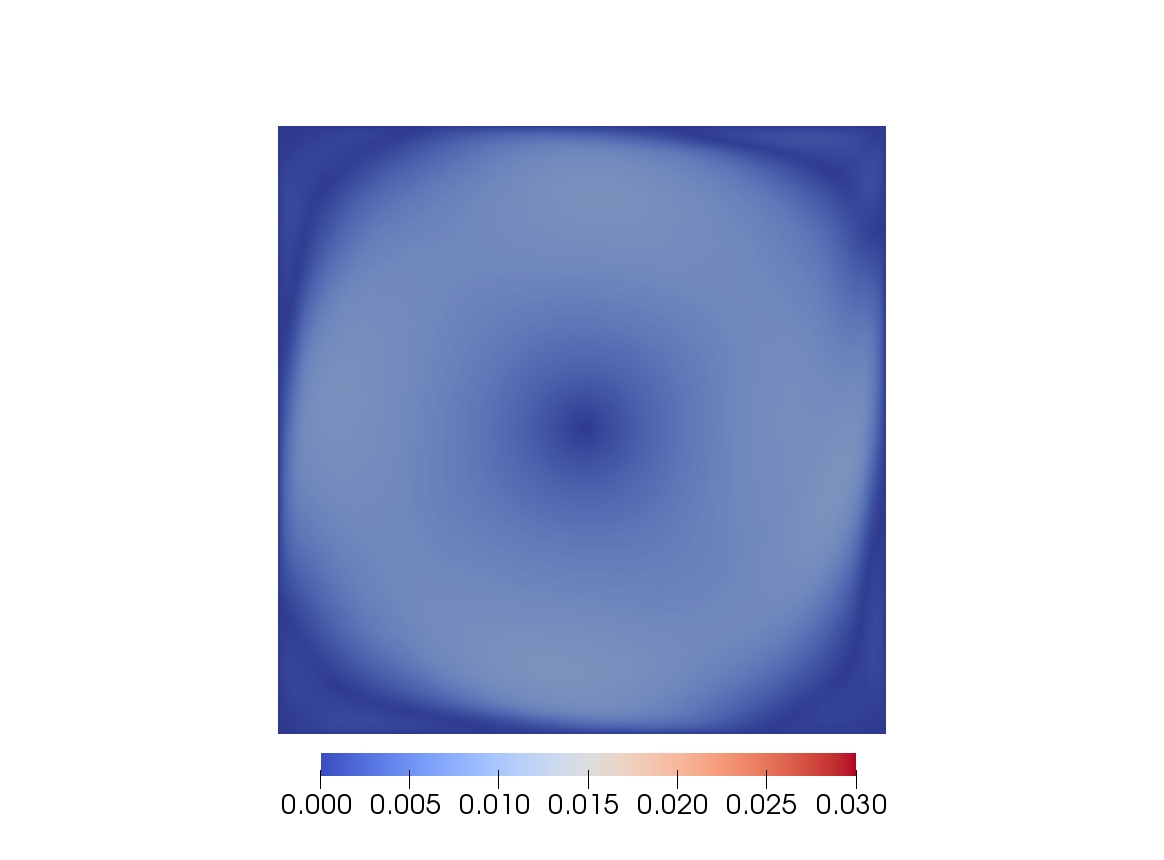}&
		\adjincludegraphics[width=7cm,trim={9cm 1cm 9cm 2cm} ,clip]{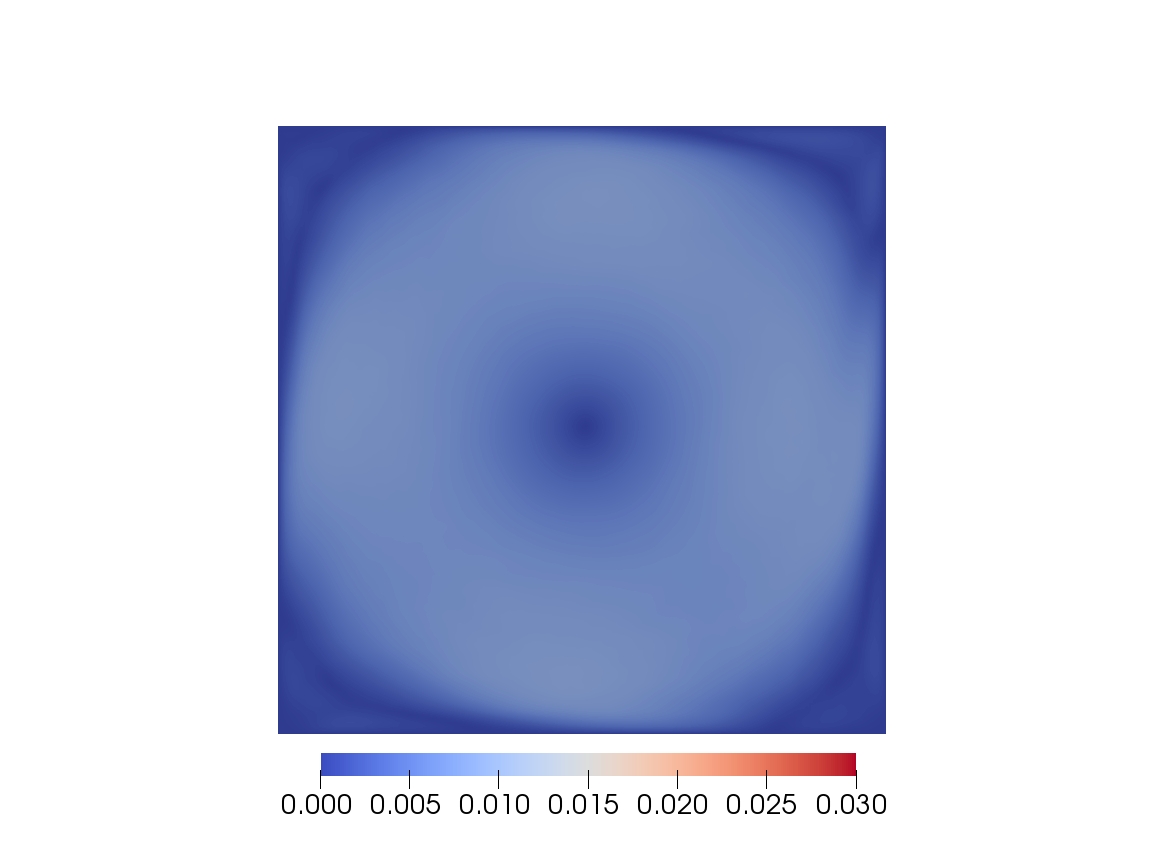}&
		\\
	\end{tabular}
	\caption{First eigenmode for Reynolds number = 100 (top left), 400 (top right), 700 (bottom left), 1000 (bottom right) and $\phi=40\%$.}
	\label{fig:fig4} 
\end{figure*}
\newpage	    

\subsection{50\% Particle Fraction}
\begin{figure*}[h]
	\centering
	\begin{tabular}{lcc}
		\adjincludegraphics[width=7cm,trim={9cm 1cm 9cm 2cm} ,clip]{./images/Re100/phi_1_50pct_rhorat_1000}&
		\adjincludegraphics[width=7cm,trim={9cm 1cm 9cm 2cm} ,clip]{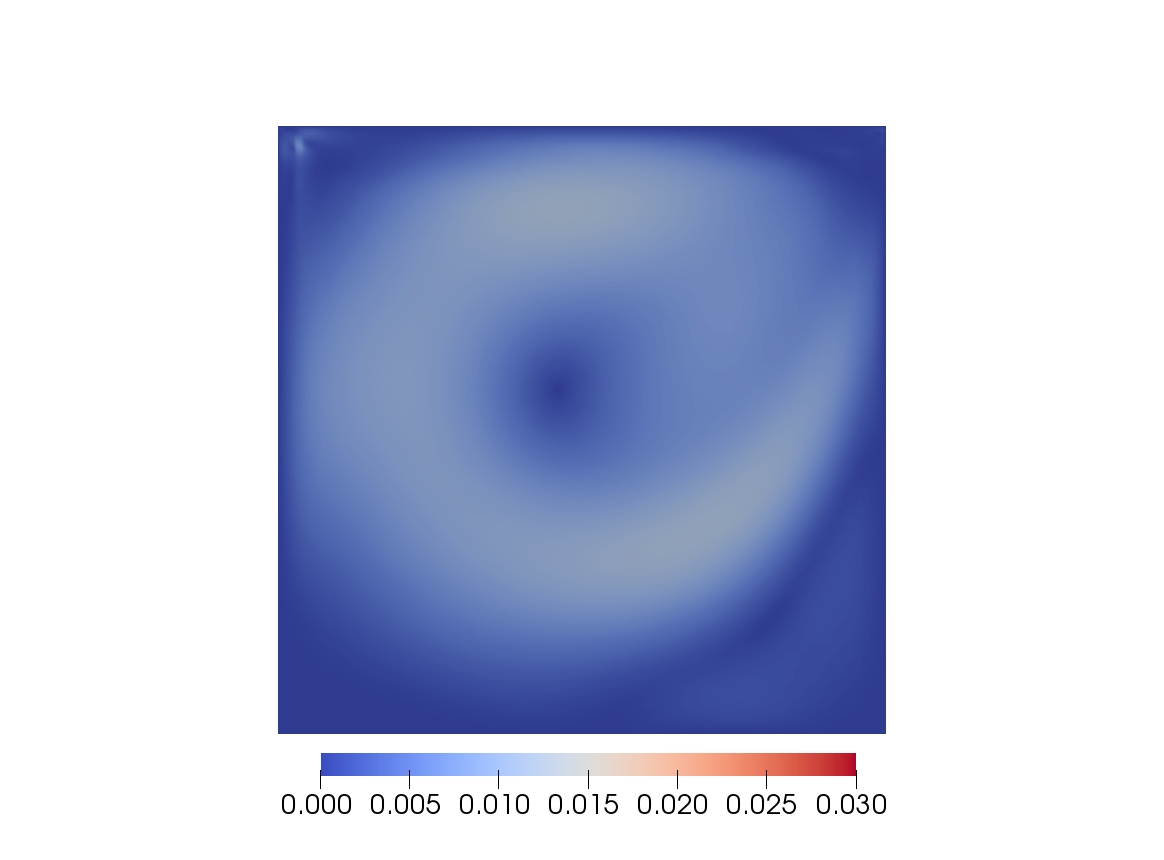}
		\\
		\adjincludegraphics[width=7cm,trim={9cm 1cm 9cm 2cm} ,clip]{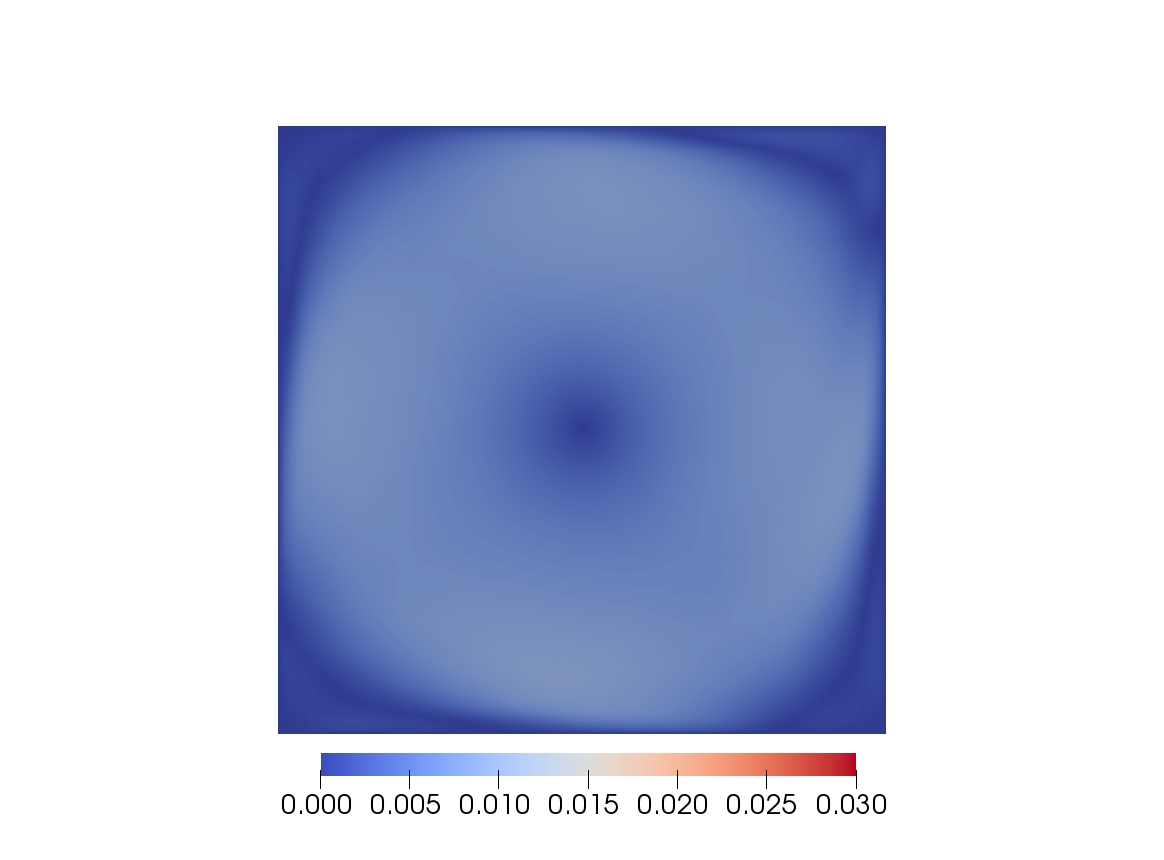}&
		\adjincludegraphics[width=7cm,trim={9cm 1cm 9cm 2cm} ,clip]{./images/Re1000/phi_1_50pct_rhorat_1000}&
		\\
	\end{tabular}
	\caption{First eigenmode for Reynolds number = 100 (top left), 400 (top right), 700 (bottom left), 1000 (bottom right) and $\phi=50\%$.}
	\label{fig:fig4} 
\end{figure*}
\newpage	    

\section{Second Eigenmode: All Particle Fractions}
\subsection{10\% Particle Fraction}
\begin{figure*}[h]
	\centering
	\begin{tabular}{lcc}
		\adjincludegraphics[width=7cm,trim={9cm 1cm 9cm 2cm} ,clip]{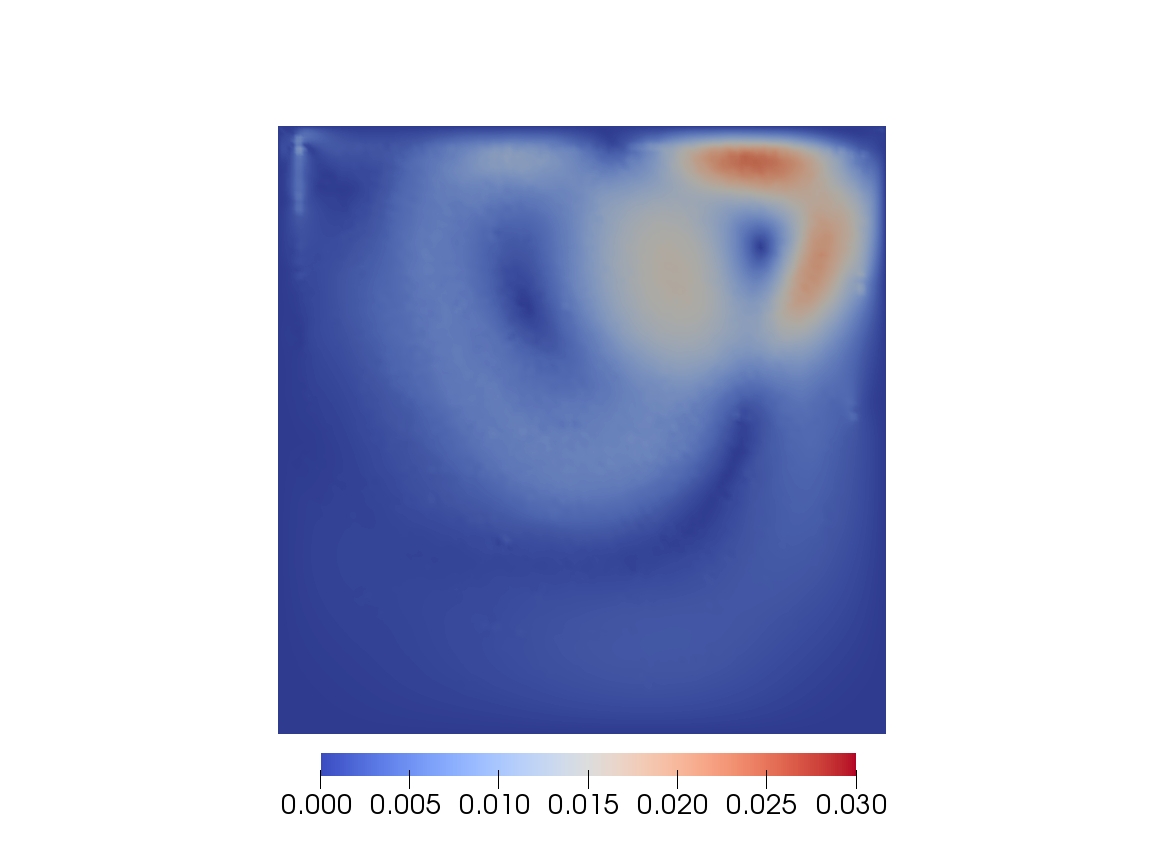}&
		\adjincludegraphics[width=7cm,trim={9cm 1cm 9cm 2cm} ,clip]{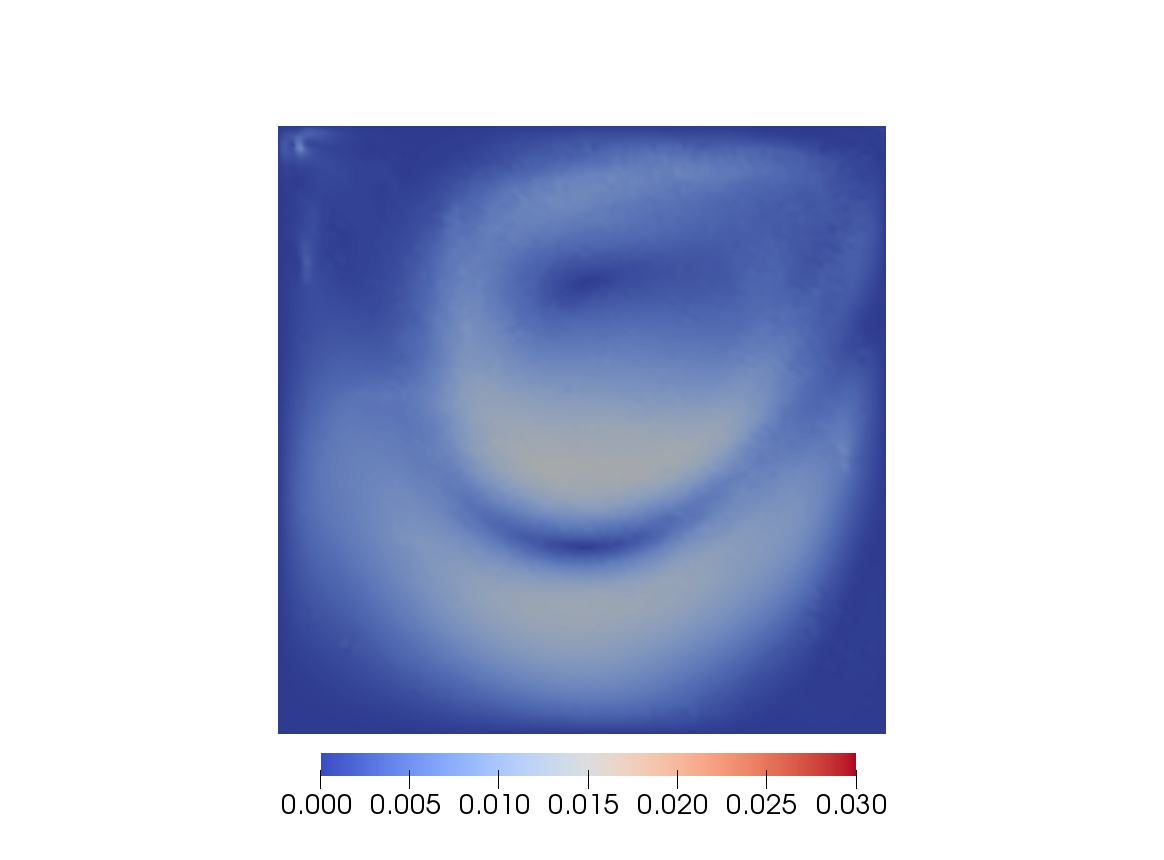}
		\\
		\adjincludegraphics[width=7cm,trim={9cm 1cm 9cm 2cm} ,clip]{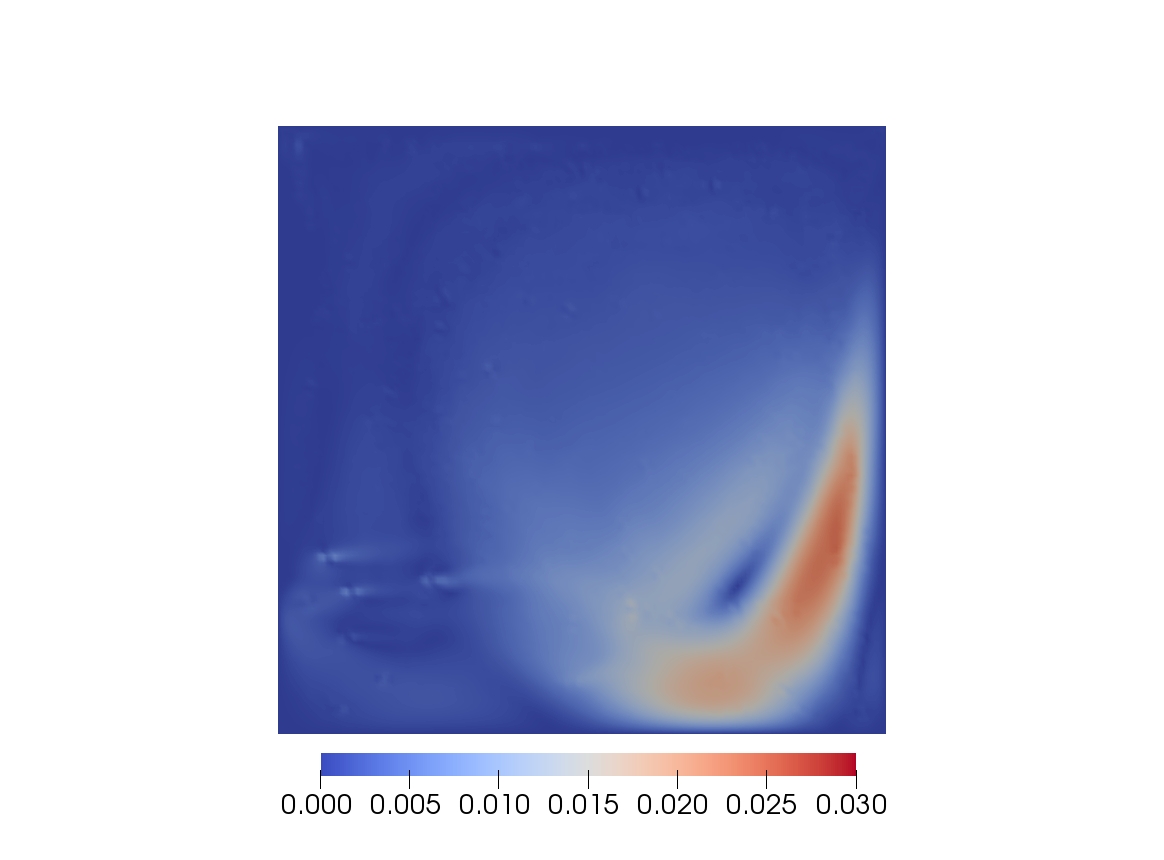}&
		\adjincludegraphics[width=7cm,trim={9cm 1cm 9cm 2cm} ,clip]{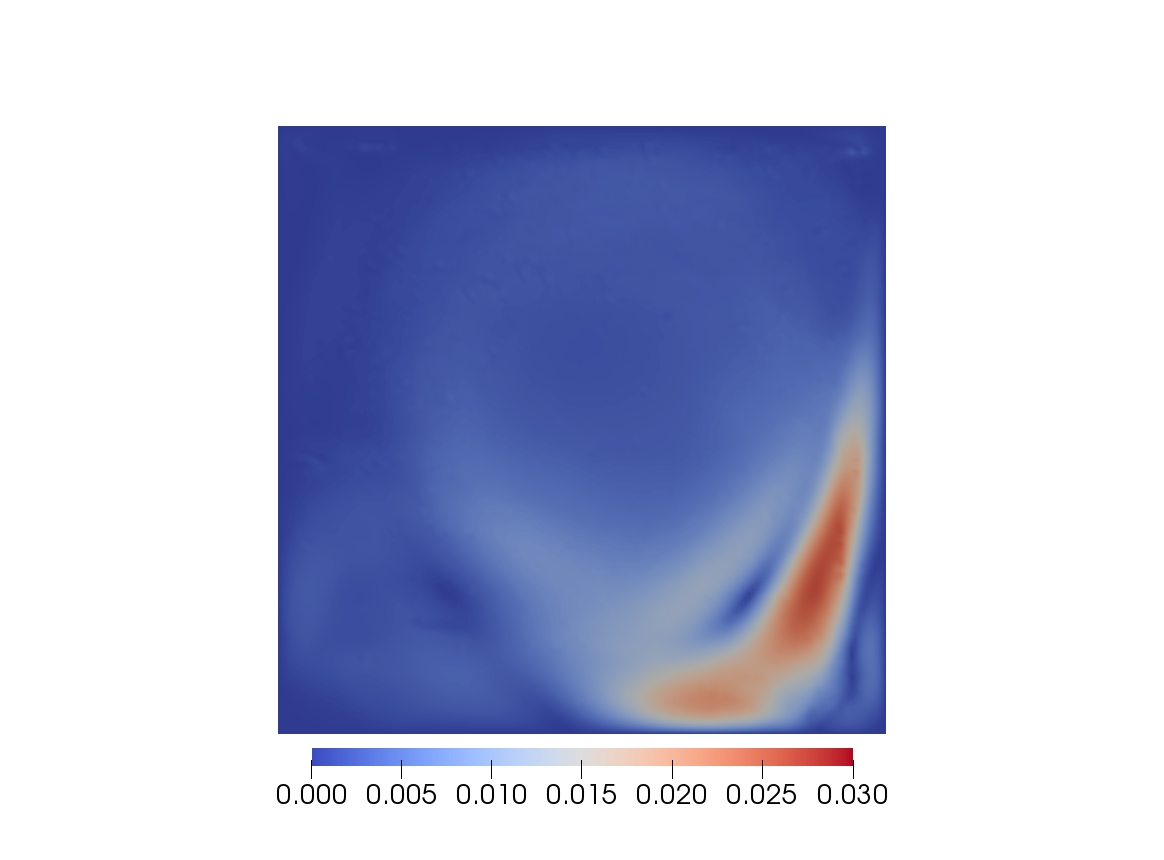}&
		\\
	\end{tabular}
	\caption{Second eigenmode for Reynolds number = 100 (top left), 400 (top right), 700 (bottom left), 1000 (bottom right) and $\phi=10\%$.}
	\label{fig:fig4} 
\end{figure*}
\newpage	    

\subsection{20\% Particle Fraction}
\begin{figure*}[h]
	\centering
	\begin{tabular}{lcc}
		\adjincludegraphics[width=7cm,trim={9cm 1cm 9cm 2cm} ,clip]{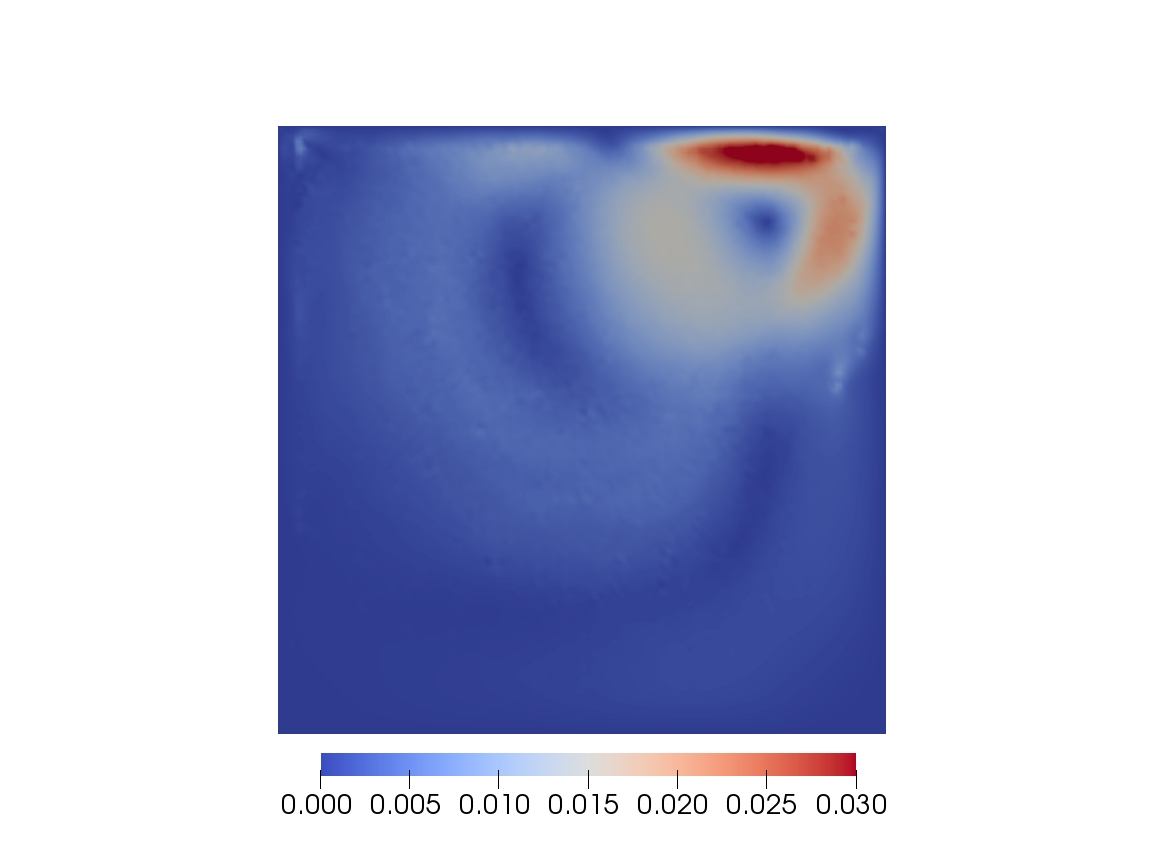}&
		\adjincludegraphics[width=7cm,trim={9cm 1cm 9cm 2cm} ,clip]{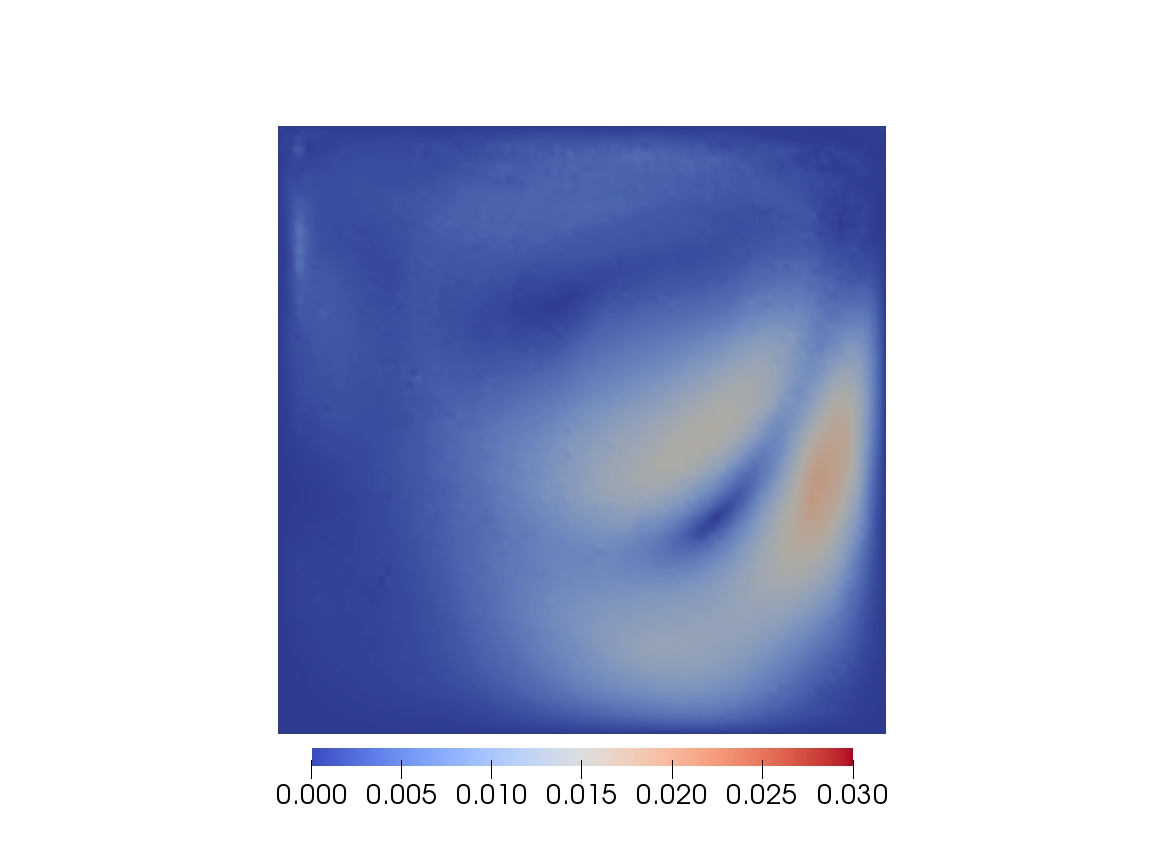}
		\\
		\adjincludegraphics[width=7cm,trim={9cm 1cm 9cm 2cm} ,clip]{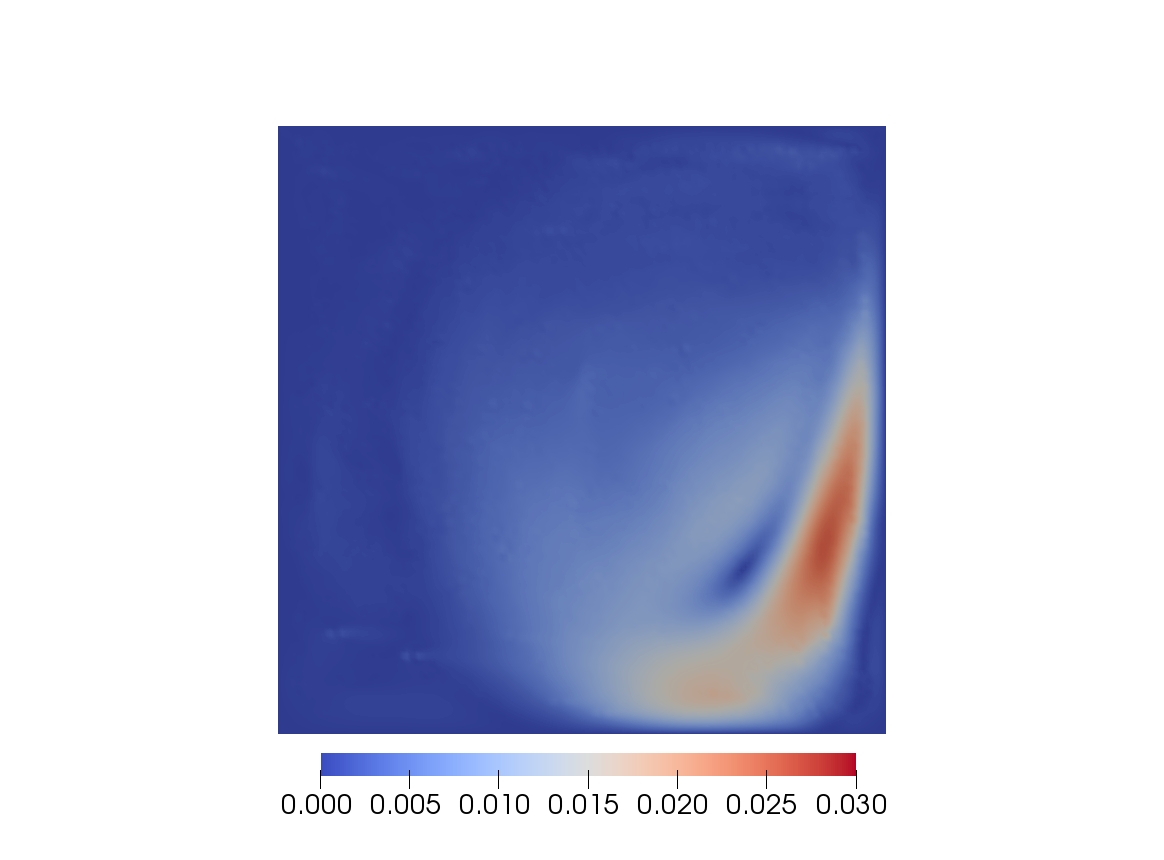}&
		\adjincludegraphics[width=7cm,trim={9cm 1cm 9cm 2cm} ,clip]{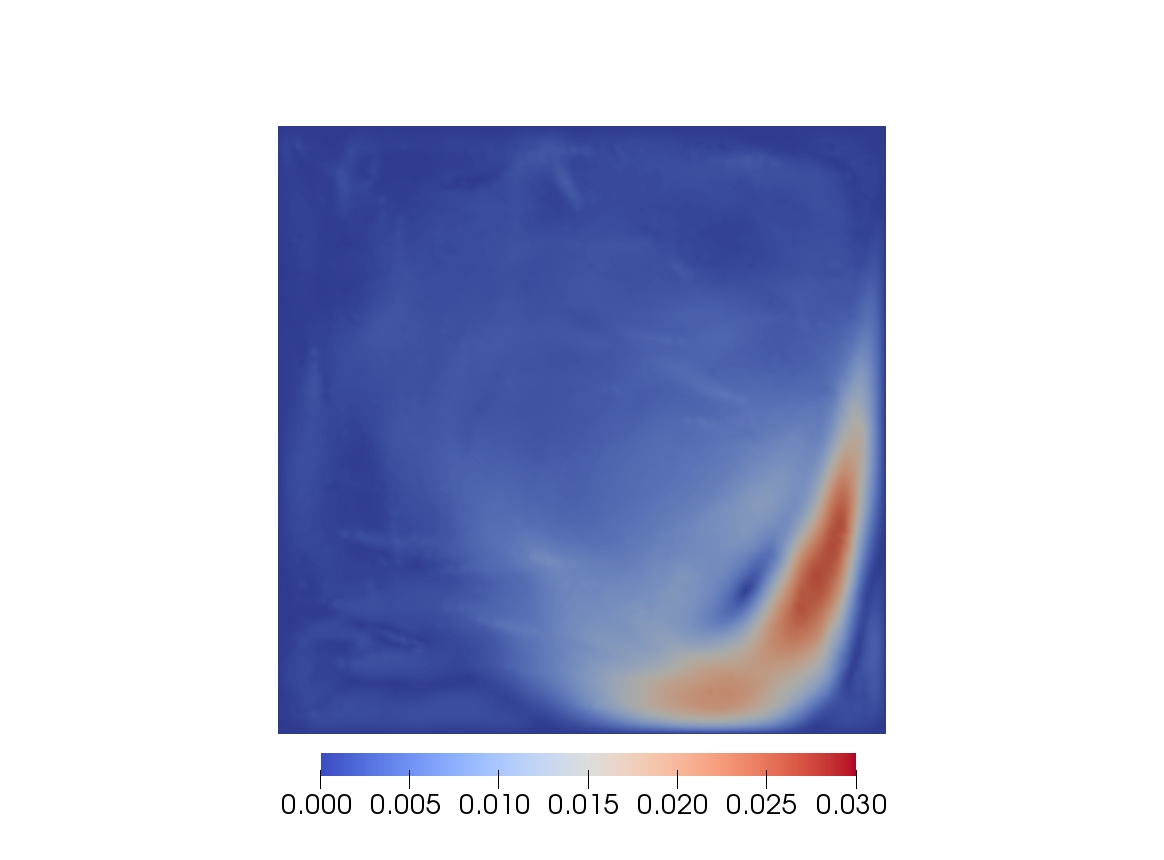}&
		\\
	\end{tabular}
	\caption{Second eigenmode for Reynolds number = 100 (top left), 400 (top right), 700 (bottom left), 1000 (bottom right) and $\phi=20\%$.}
	\label{fig:fig4} 
\end{figure*}
\newpage	    

\subsection{30\% Particle Fraction}
\begin{figure*}[h]
	\centering
	\begin{tabular}{lcc}
		\adjincludegraphics[width=7cm,trim={9cm 1cm 9cm 2cm} ,clip]{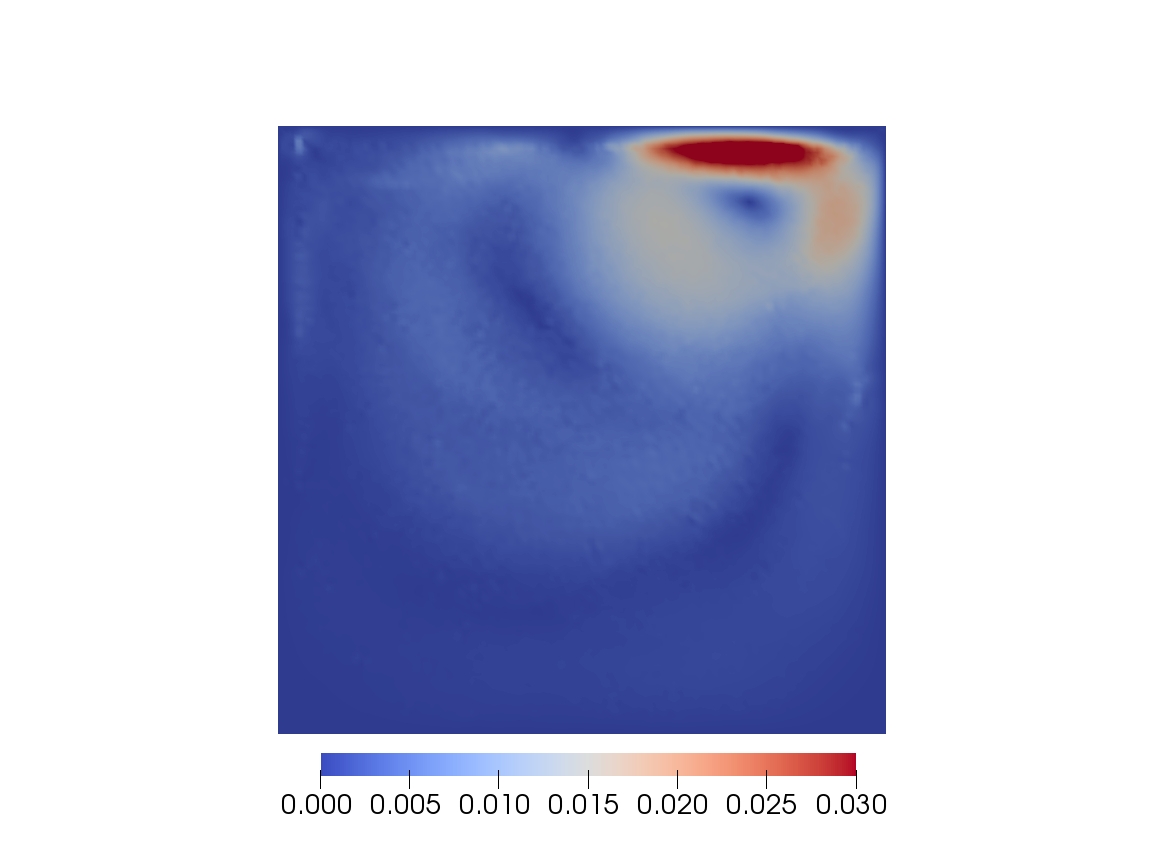}&
		\adjincludegraphics[width=7cm,trim={9cm 1cm 9cm 2cm} ,clip]{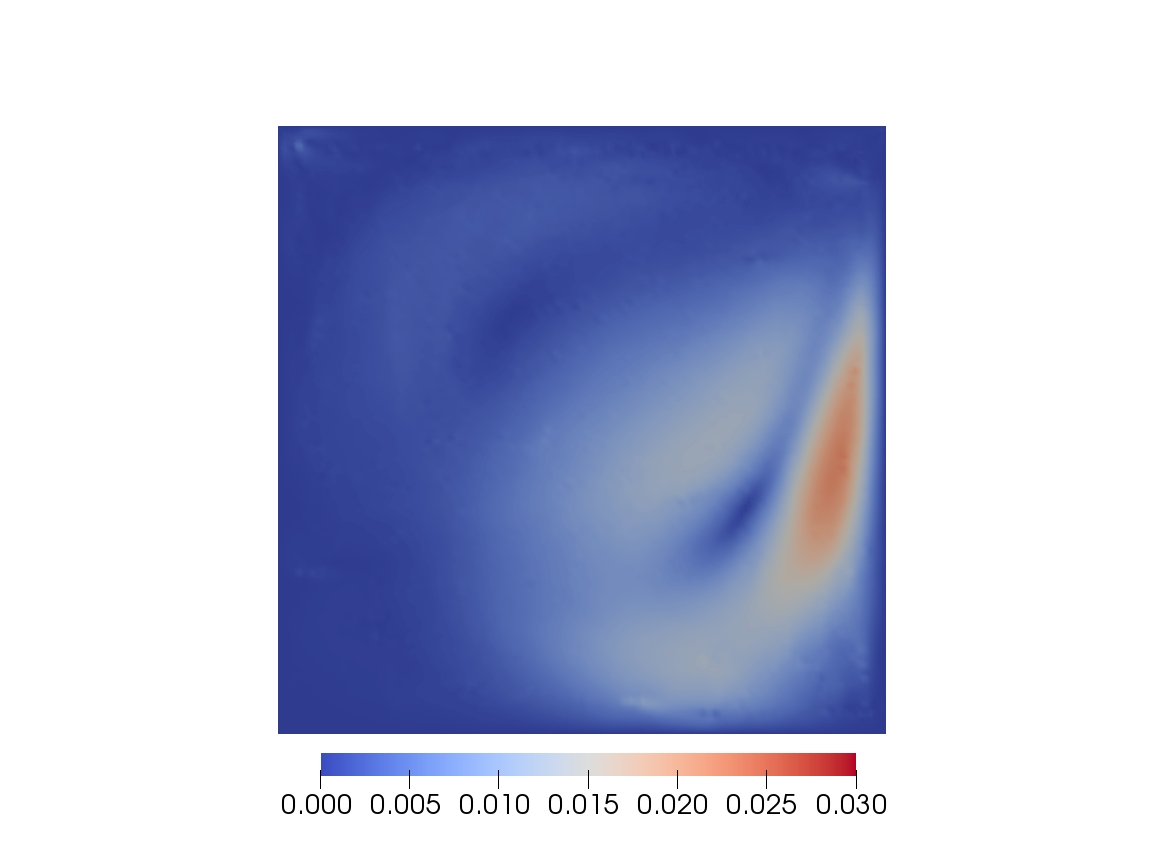}
		\\
		\adjincludegraphics[width=7cm,trim={9cm 1cm 9cm 2cm} ,clip]{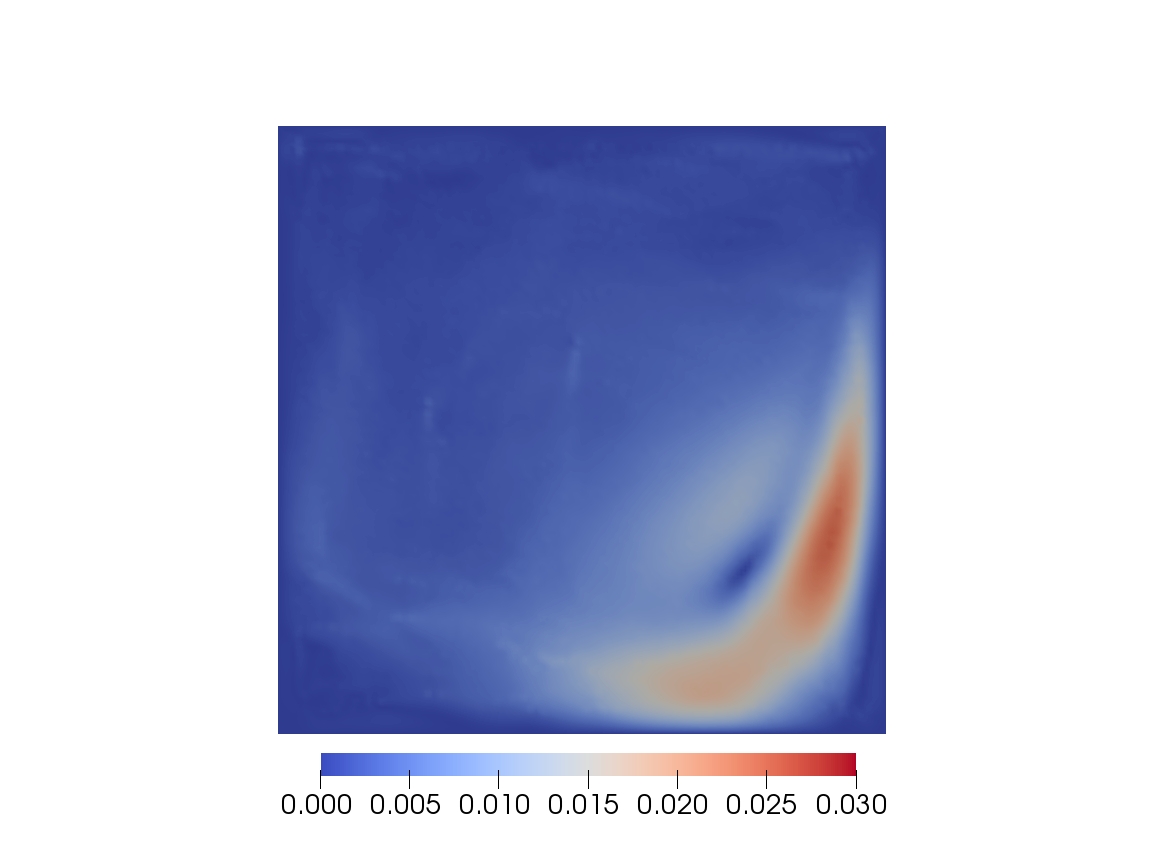}&
		\adjincludegraphics[width=7cm,trim={9cm 1cm 9cm 2cm} ,clip]{./images/Re1000/phi_2_30pct_rhorat_1000}&
		\\
	\end{tabular}
	\caption{Second eigenmode for Reynolds number = 100 (top left), 400 (top right), 700 (bottom left), 1000 (bottom right) and $\phi=30\%$.}
	\label{fig:fig4} 
\end{figure*}
\newpage	    

\subsection{40\% Particle Fraction}
\begin{figure*}[h]
	\centering
	\begin{tabular}{lcc}
		\adjincludegraphics[width=7cm,trim={9cm 1cm 9cm 2cm} ,clip]{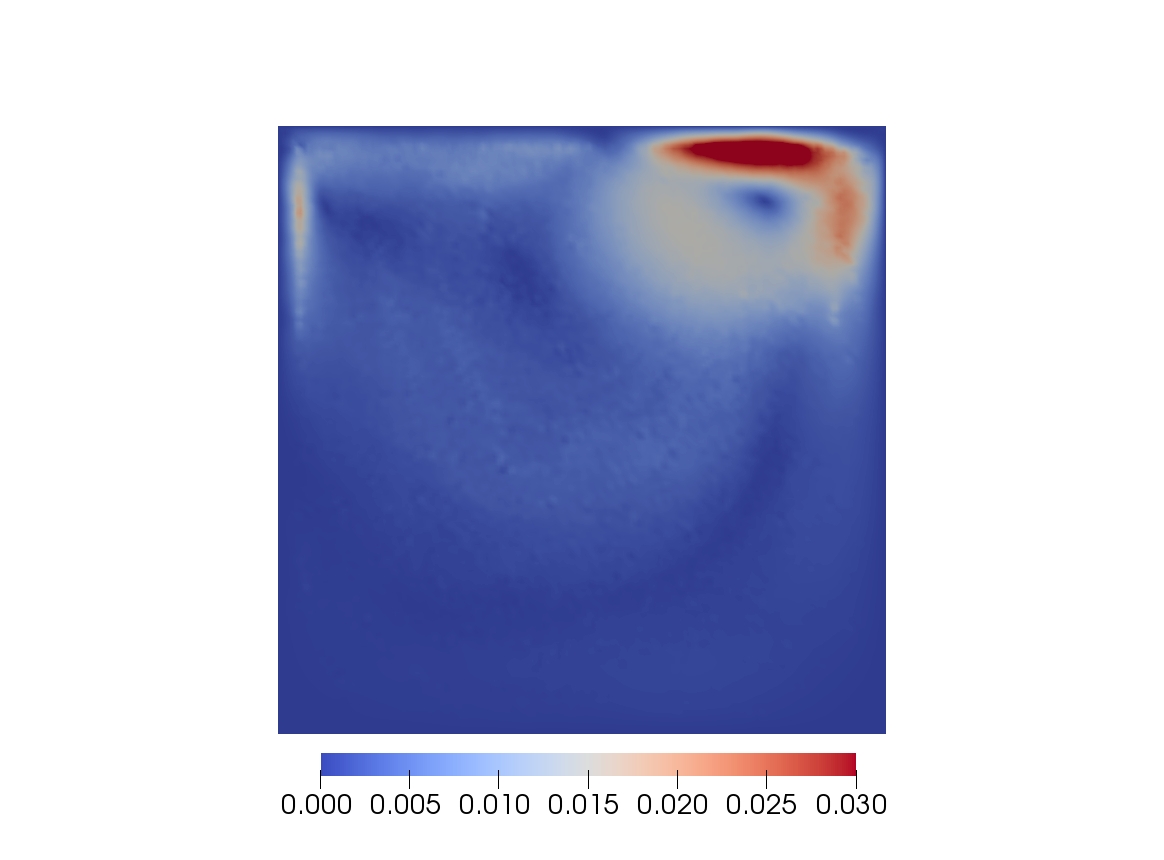}&
		\adjincludegraphics[width=7cm,trim={9cm 1cm 9cm 2cm} ,clip]{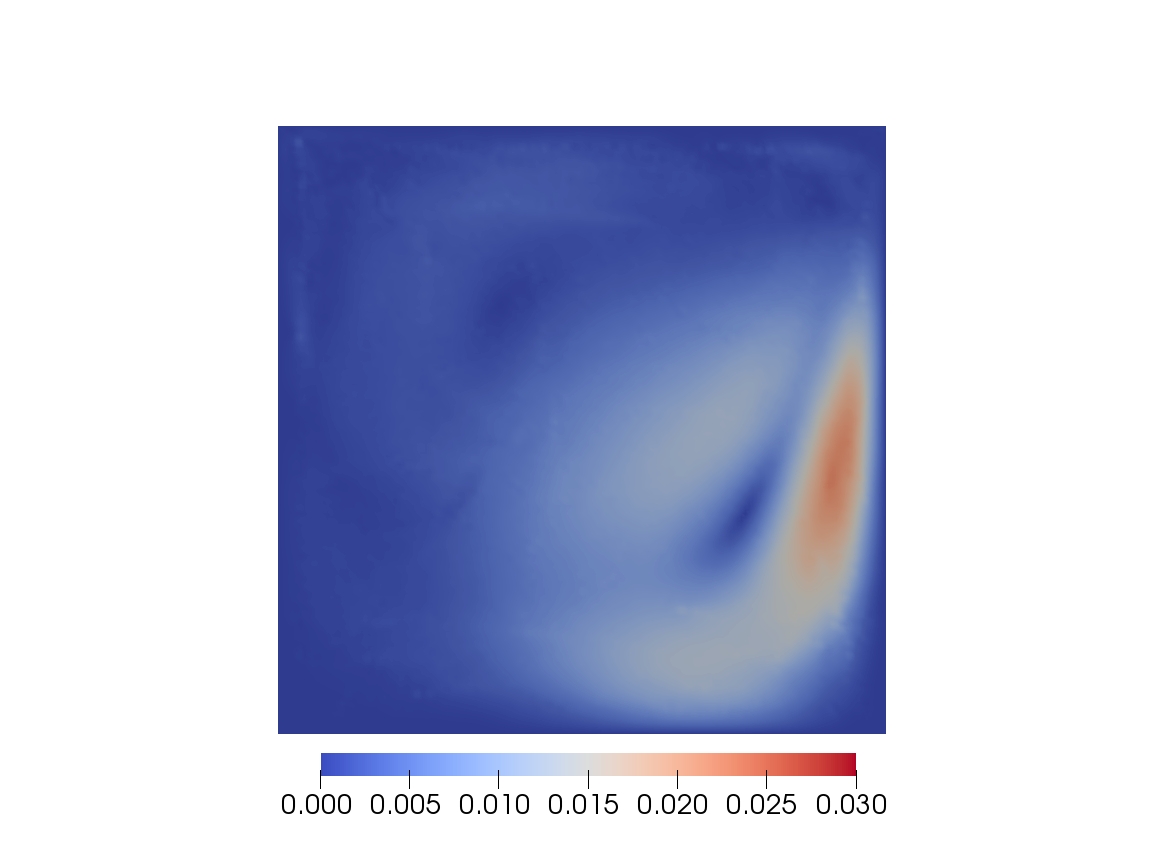}
		\\
		\adjincludegraphics[width=7cm,trim={9cm 1cm 9cm 2cm} ,clip]{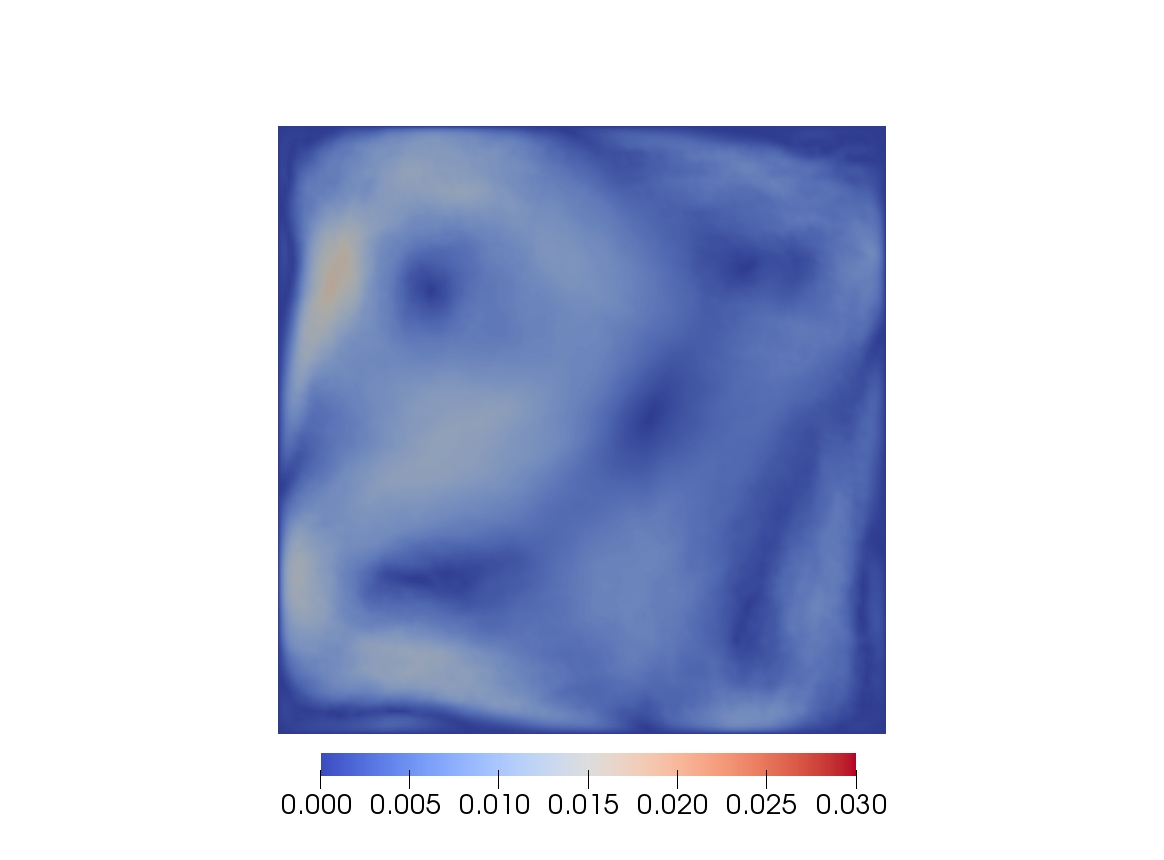}&
		\adjincludegraphics[width=7cm,trim={9cm 1cm 9cm 2cm} ,clip]{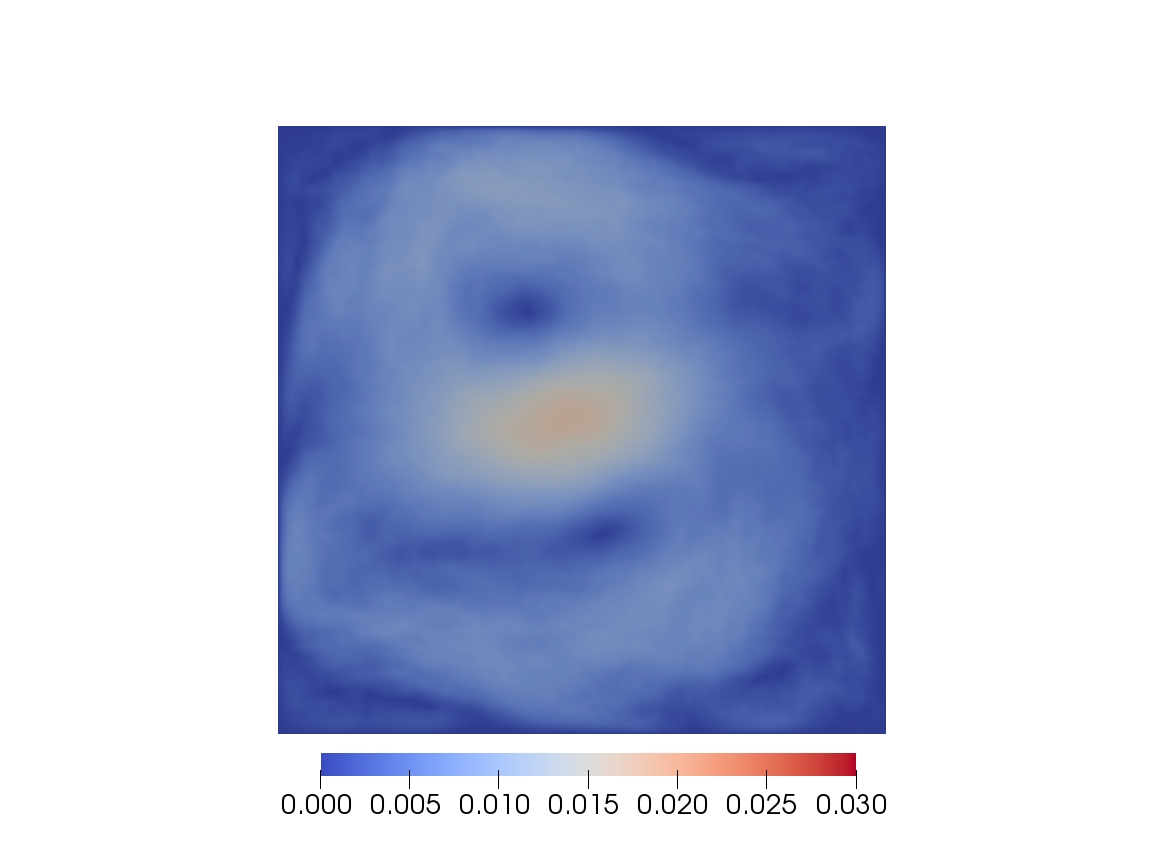}&
		\\
	\end{tabular}
	\caption{Second eigenmode for Reynolds number = 100 (top left), 400 (top right), 700 (bottom left), 1000 (bottom right) and $\phi=40\%$.}
	\label{fig:fig4} 
\end{figure*}
\newpage	    

\subsection{50\% Particle Fraction}
\begin{figure*}[h]
	\centering
	\begin{tabular}{lcc}
		\adjincludegraphics[width=7cm,trim={9cm 1cm 9cm 2cm} ,clip]{./images/Re100/phi_2_50pct_rhorat_1000}&
		\adjincludegraphics[width=7cm,trim={9cm 1cm 9cm 2cm} ,clip]{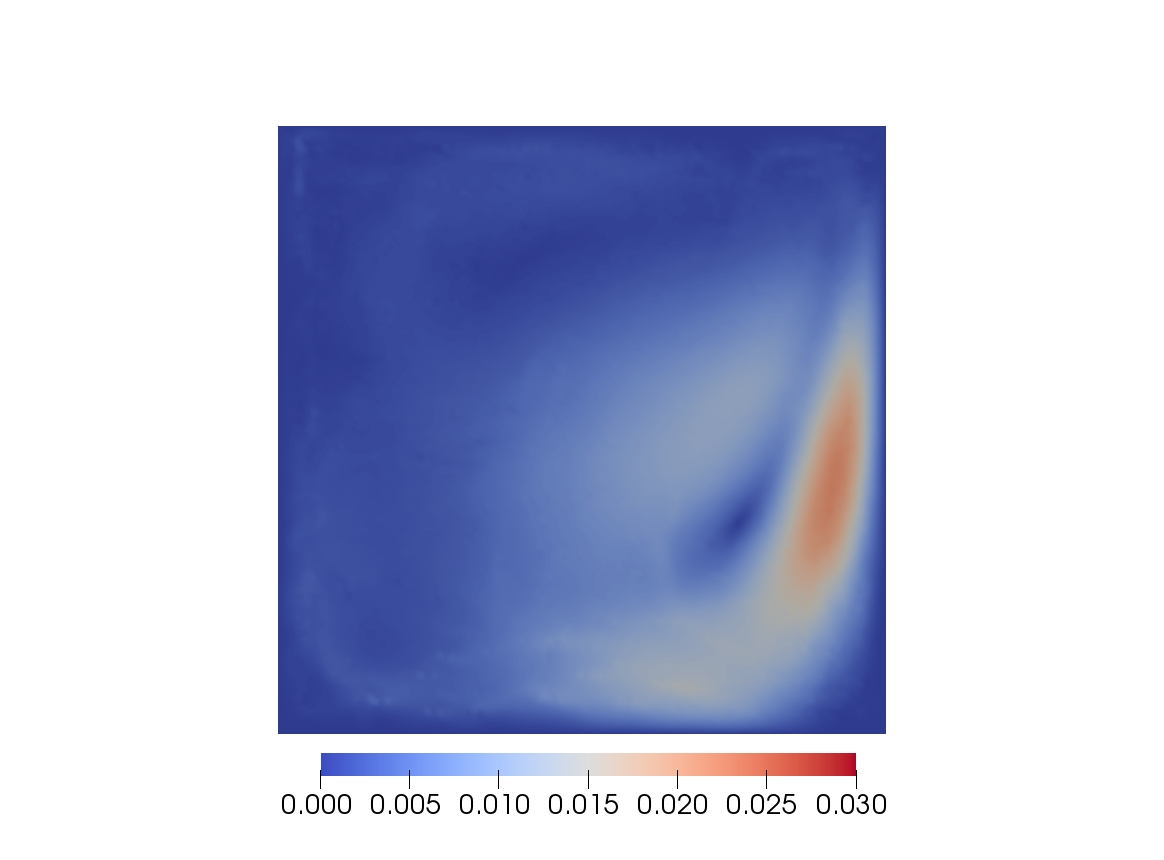}
		\\
		\adjincludegraphics[width=7cm,trim={9cm 1cm 9cm 2cm} ,clip]{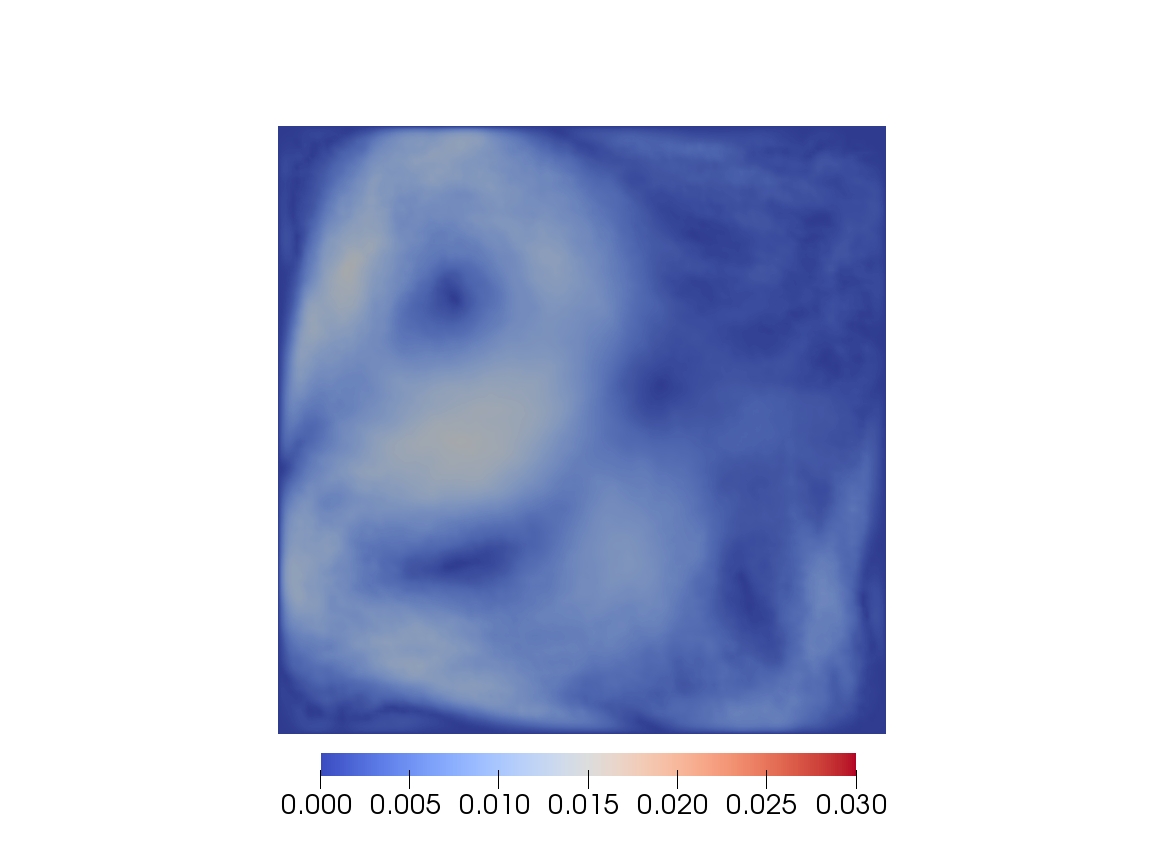}&
		\adjincludegraphics[width=7cm,trim={9cm 1cm 9cm 2cm} ,clip]{./images/Re1000/phi_2_50pct_rhorat_1000}&
		\\
	\end{tabular}
	\caption{Second eigenmode for Reynolds number = 100 (top left), 400 (top right), 700 (bottom left), 1000 (bottom right) and $\phi=50\%$.}
	\label{fig:fig4} 
\end{figure*}
\newpage	    

\section{Third Eigenmode: All Particle Fractions}
\subsection{10\% Particle Fraction}
\begin{figure*}[h]
	\centering
	\begin{tabular}{lcc}
		\adjincludegraphics[width=7cm,trim={9cm 1cm 9cm 2cm} ,clip]{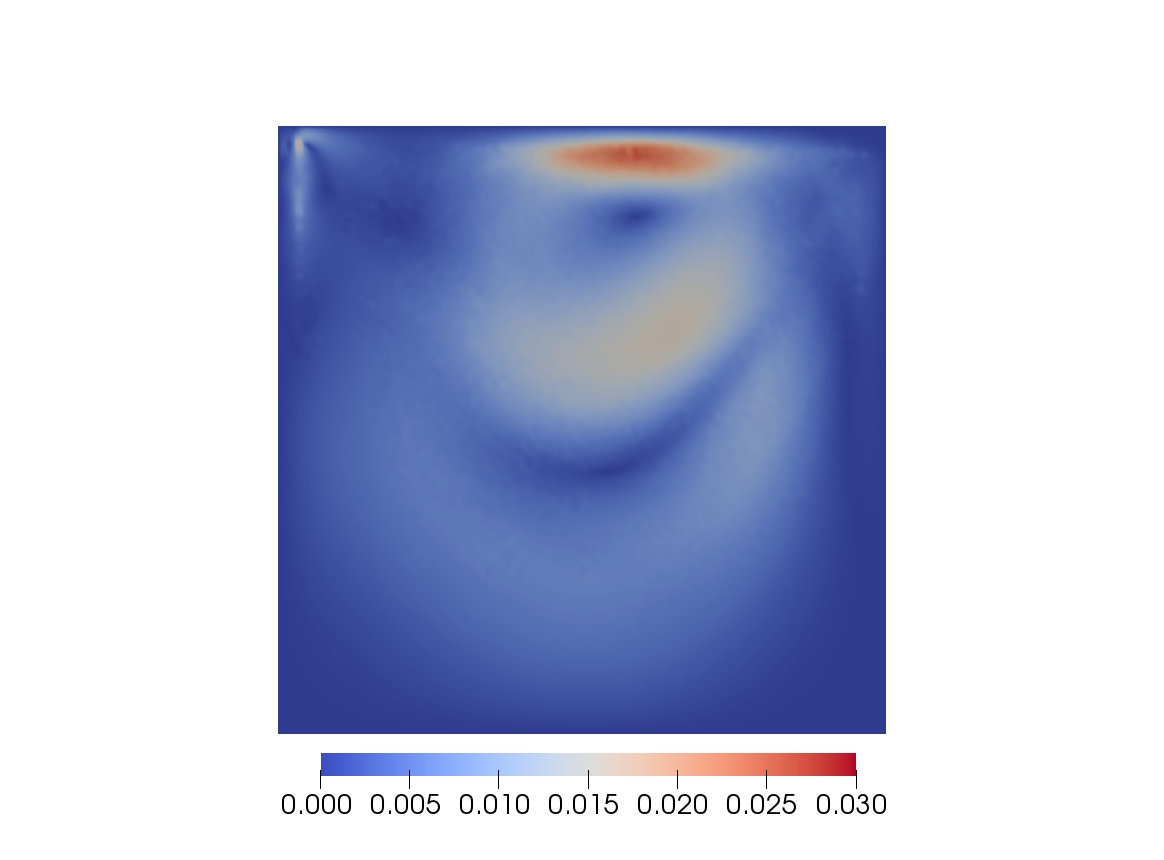}&
		\adjincludegraphics[width=7cm,trim={9cm 1cm 9cm 2cm} ,clip]{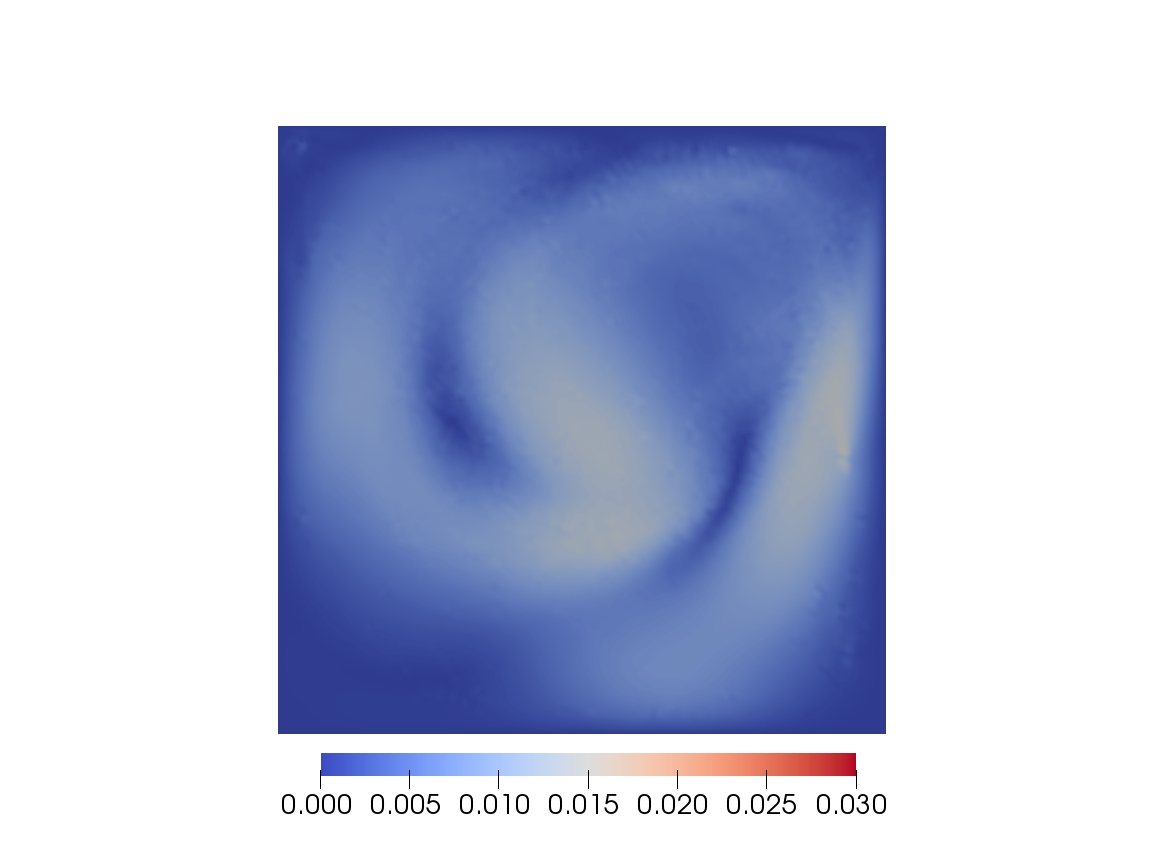}
		\\
		\adjincludegraphics[width=7cm,trim={9cm 1cm 9cm 2cm} ,clip]{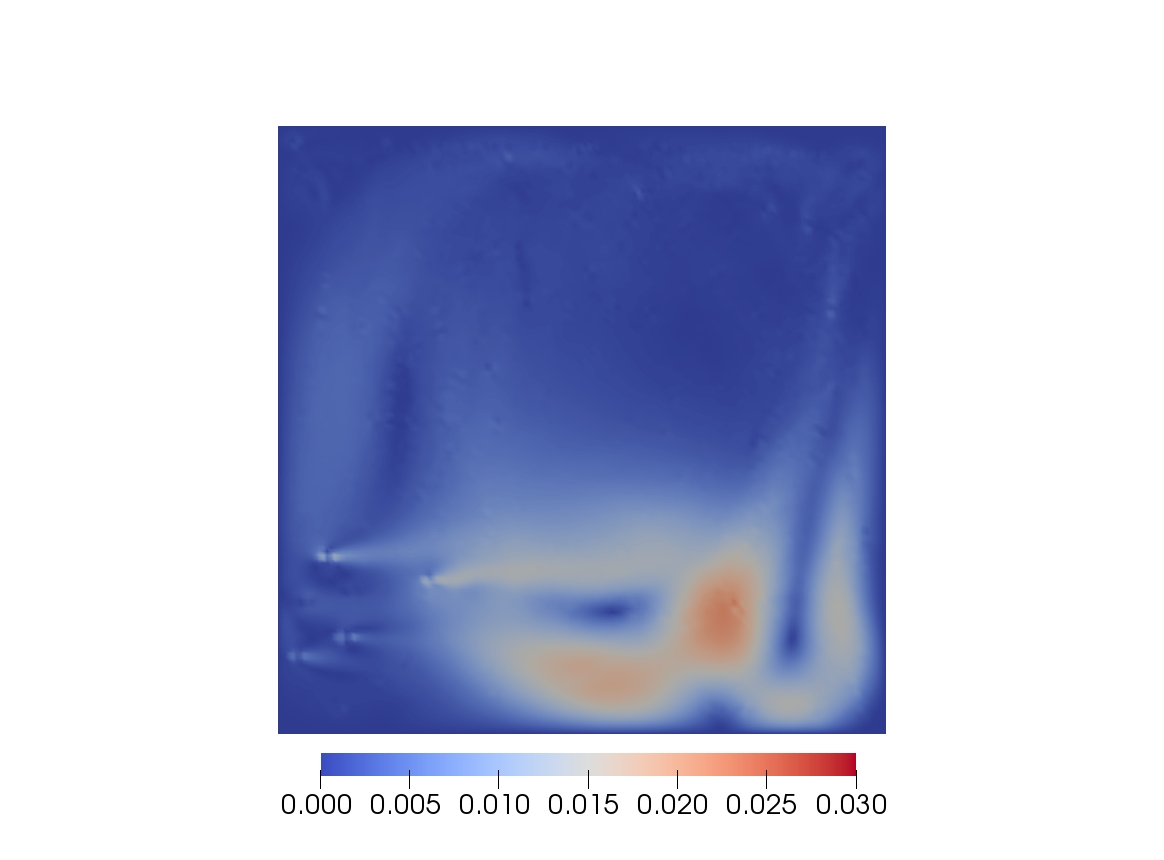}&
		\adjincludegraphics[width=7cm,trim={9cm 1cm 9cm 2cm} ,clip]{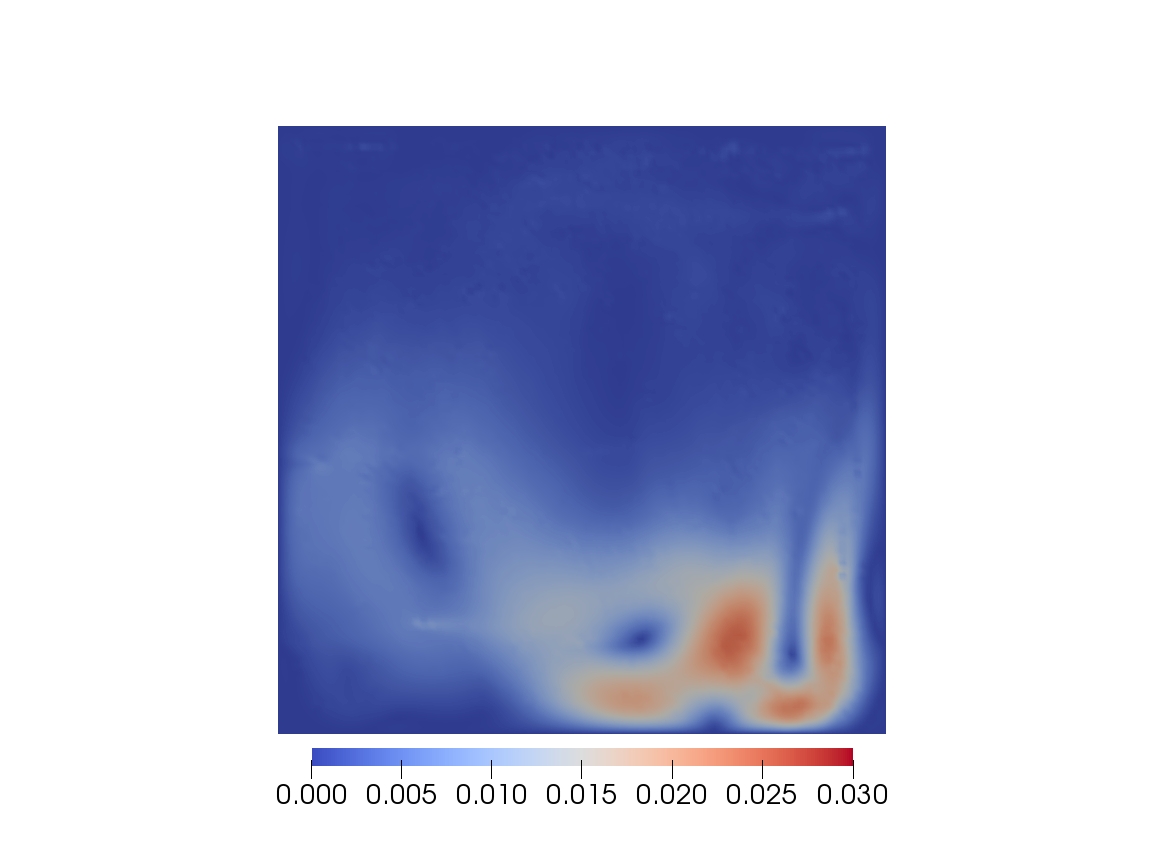}&
		\\
	\end{tabular}
	\caption{Third eigenmode for Reynolds number = 100 (top left), 400 (top right), 700 (bottom left), 1000 (bottom right) and $\phi=10\%$.}
	\label{fig:fig4} 
\end{figure*}
\newpage	    

\subsection{20\% Particle Fraction}
\begin{figure*}[h]
	\centering
	\begin{tabular}{lcc}
		\adjincludegraphics[width=7cm,trim={9cm 1cm 9cm 2cm} ,clip]{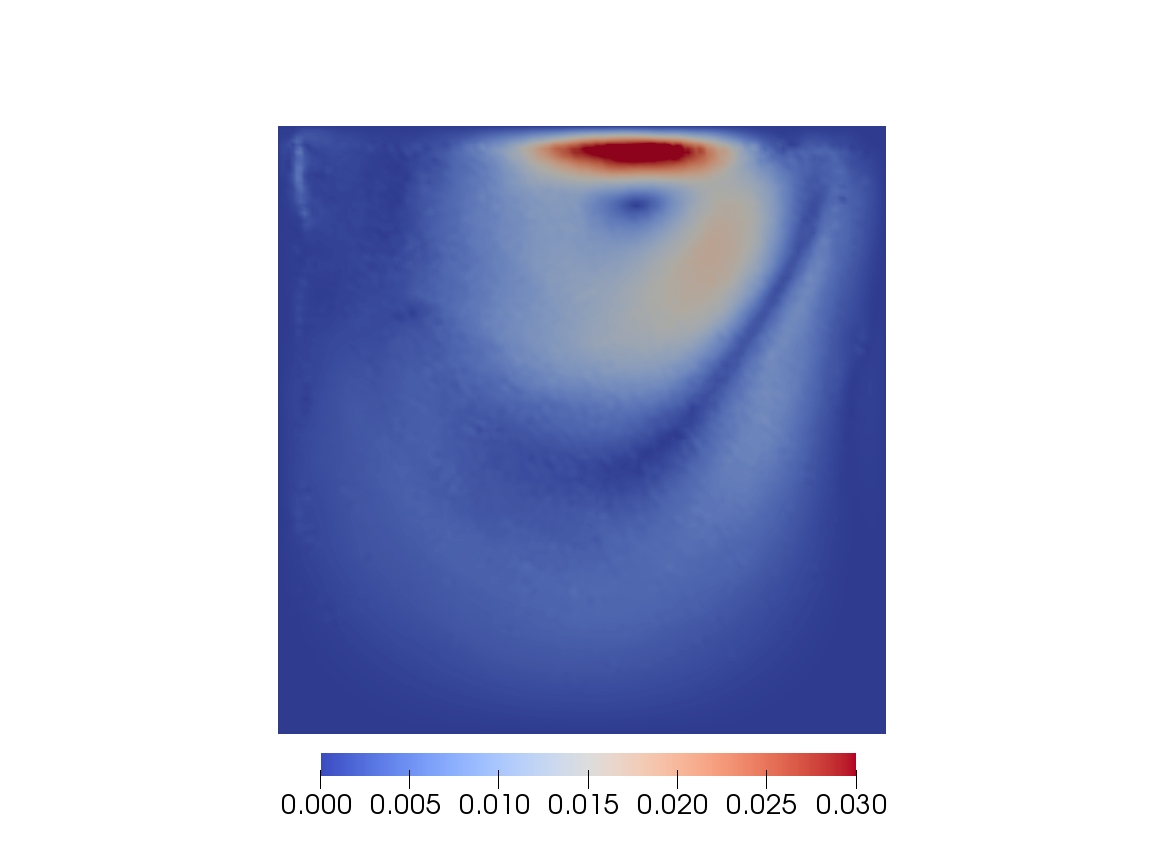}&
		\adjincludegraphics[width=7cm,trim={9cm 1cm 9cm 2cm} ,clip]{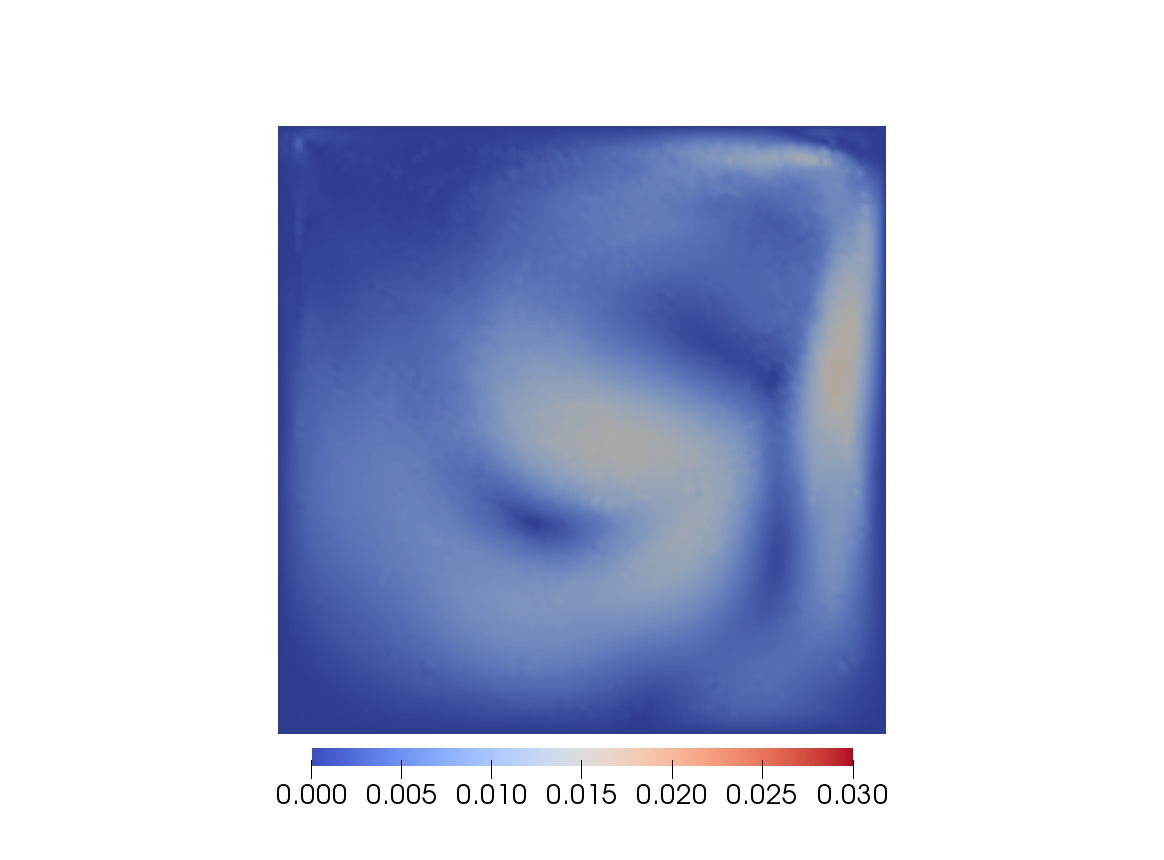}
		\\
		\adjincludegraphics[width=7cm,trim={9cm 1cm 9cm 2cm} ,clip]{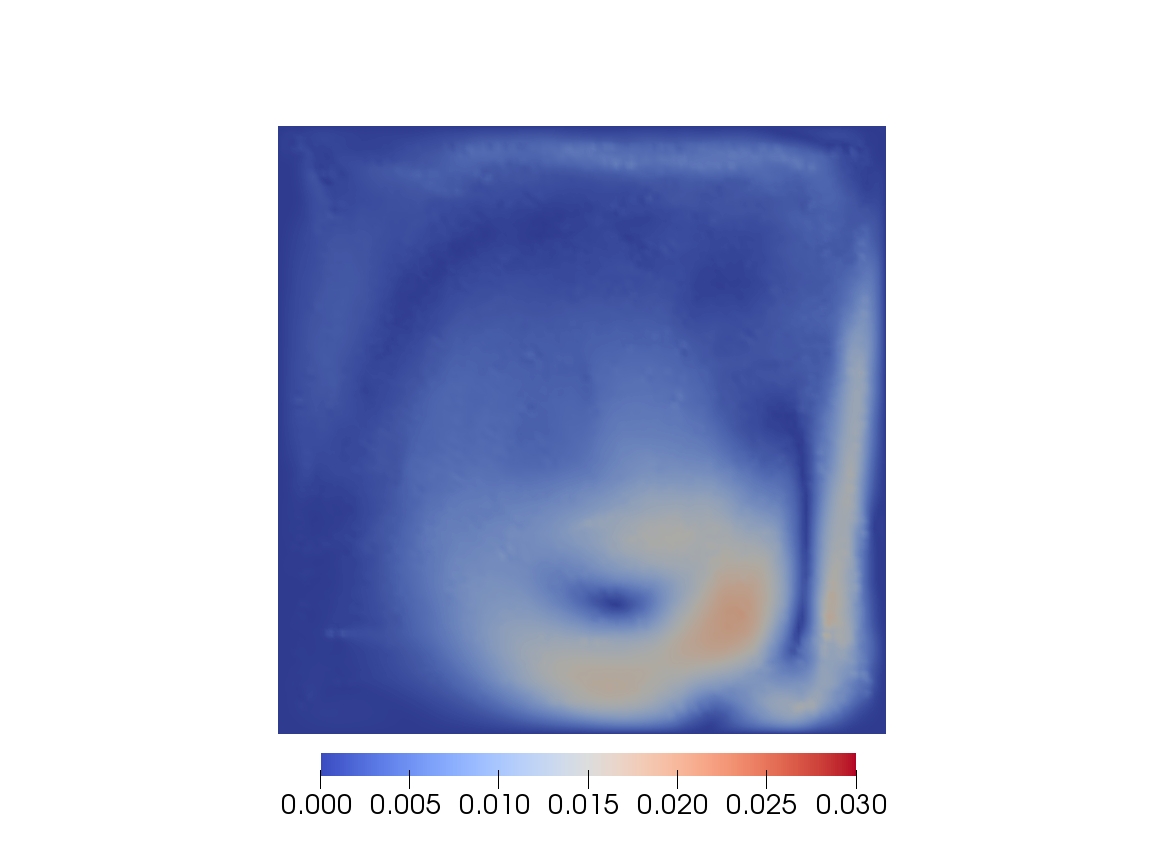}&
		\adjincludegraphics[width=7cm,trim={9cm 1cm 9cm 2cm} ,clip]{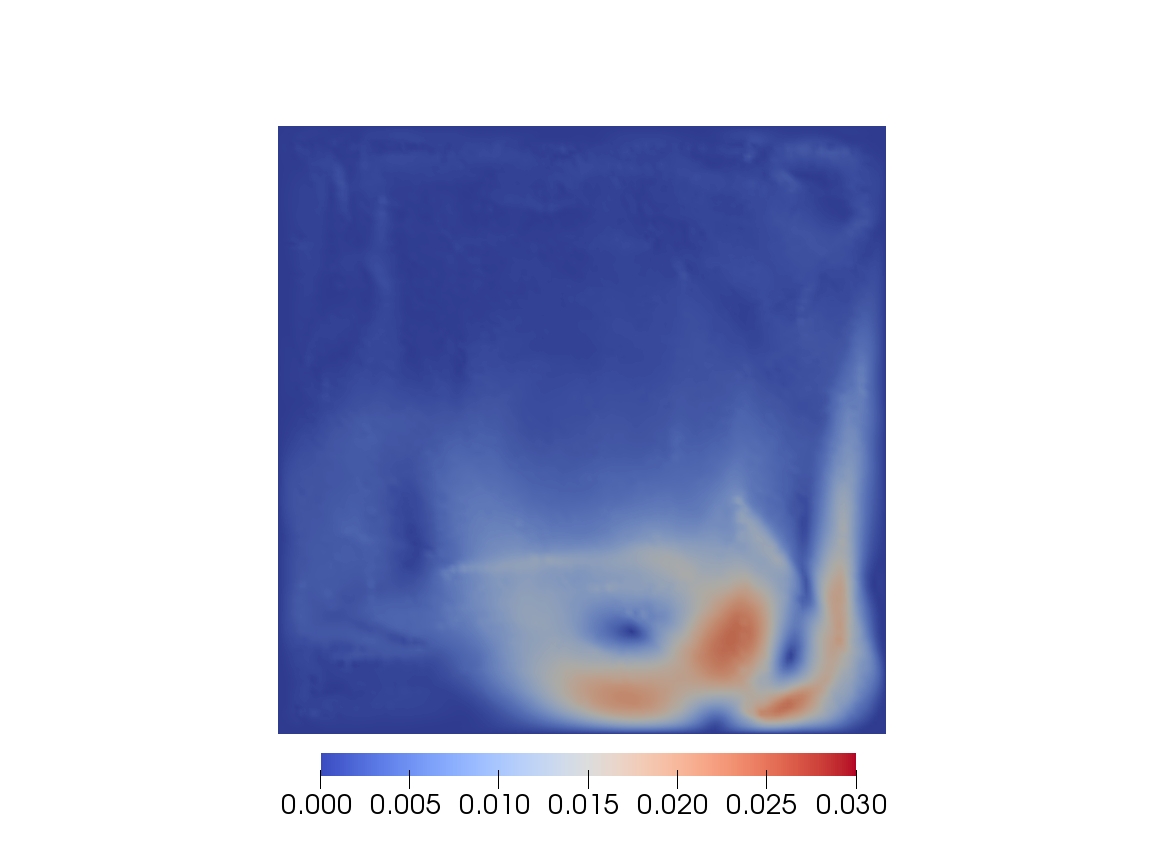}&
		\\
	\end{tabular}
	\caption{Third eigenmode for Reynolds number = 100 (top left), 400 (top right), 700 (bottom left), 1000 (bottom right) and $\phi=20\%$.}
	\label{fig:fig4} 
\end{figure*}
\newpage	    

\subsection{30\% Particle Fraction}
\begin{figure*}[h]
	\centering
	\begin{tabular}{lcc}
		\adjincludegraphics[width=7cm,trim={9cm 1cm 9cm 2cm} ,clip]{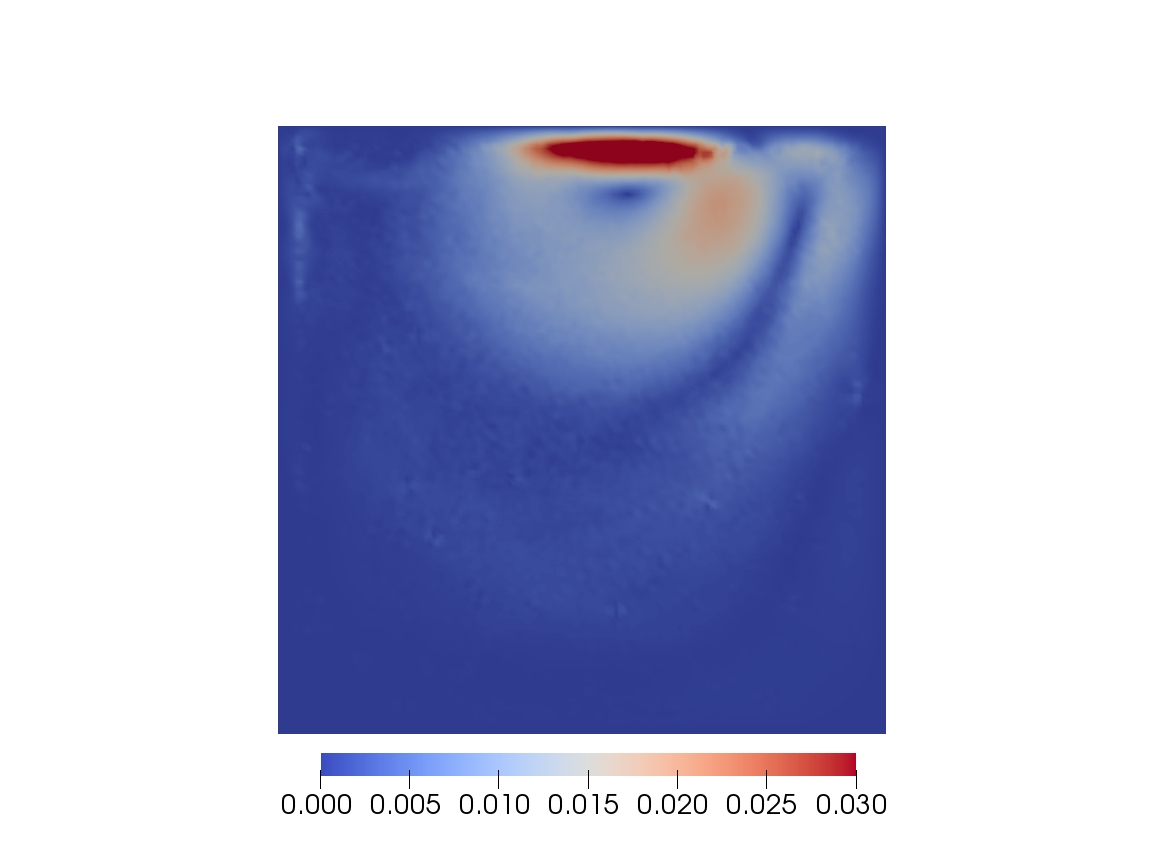}&
		\adjincludegraphics[width=7cm,trim={9cm 1cm 9cm 2cm} ,clip]{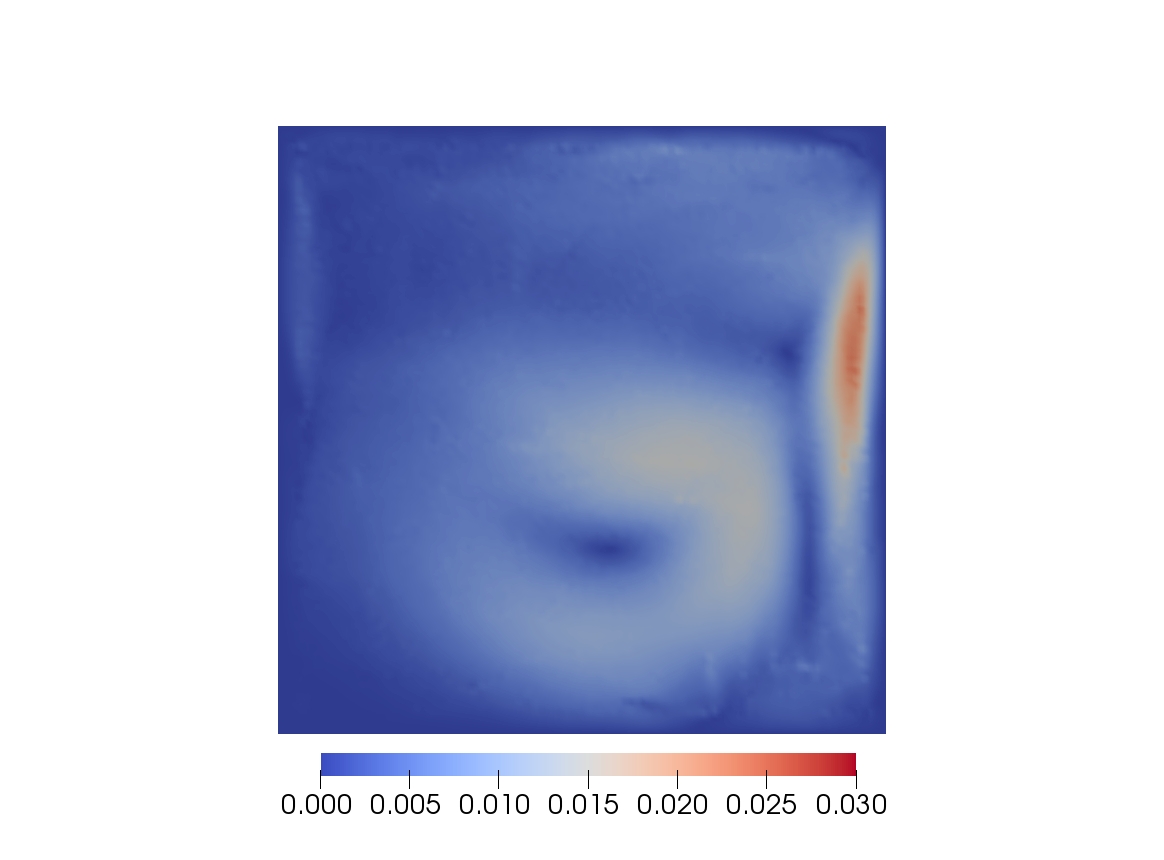}
		\\
		\adjincludegraphics[width=7cm,trim={9cm 1cm 9cm 2cm} ,clip]{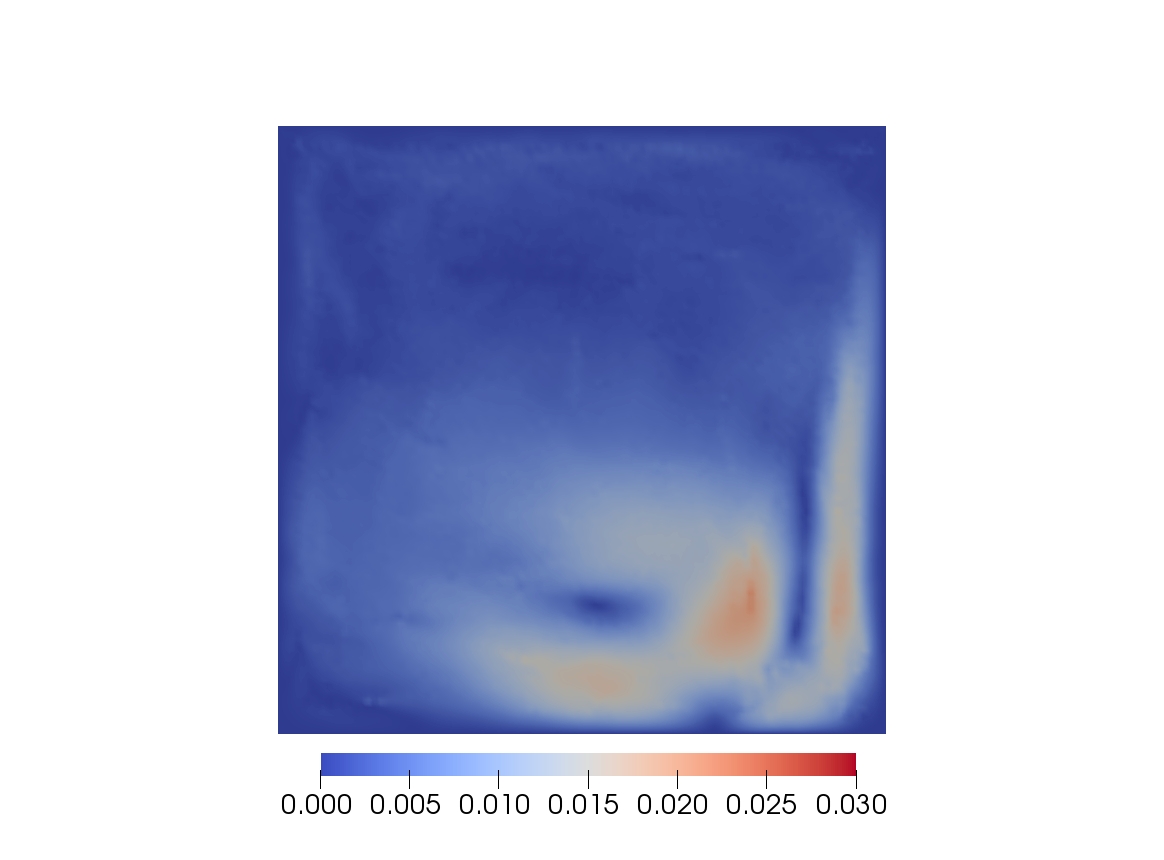}&
		\adjincludegraphics[width=7cm,trim={9cm 1cm 9cm 2cm} ,clip]{./images/Re1000/phi_3_30pct_rhorat_1000}&
		\\
	\end{tabular}
	\caption{Third eigenmode for Reynolds number = 100 (top left), 400 (top right), 700 (bottom left), 1000 (bottom right) and $\phi=30\%$.}
	\label{fig:fig4} 
\end{figure*}
\newpage	    

\subsection{40\% Particle Fraction}
\begin{figure*}[h]
	\centering
	\begin{tabular}{lcc}
		\adjincludegraphics[width=7cm,trim={9cm 1cm 9cm 2cm} ,clip]{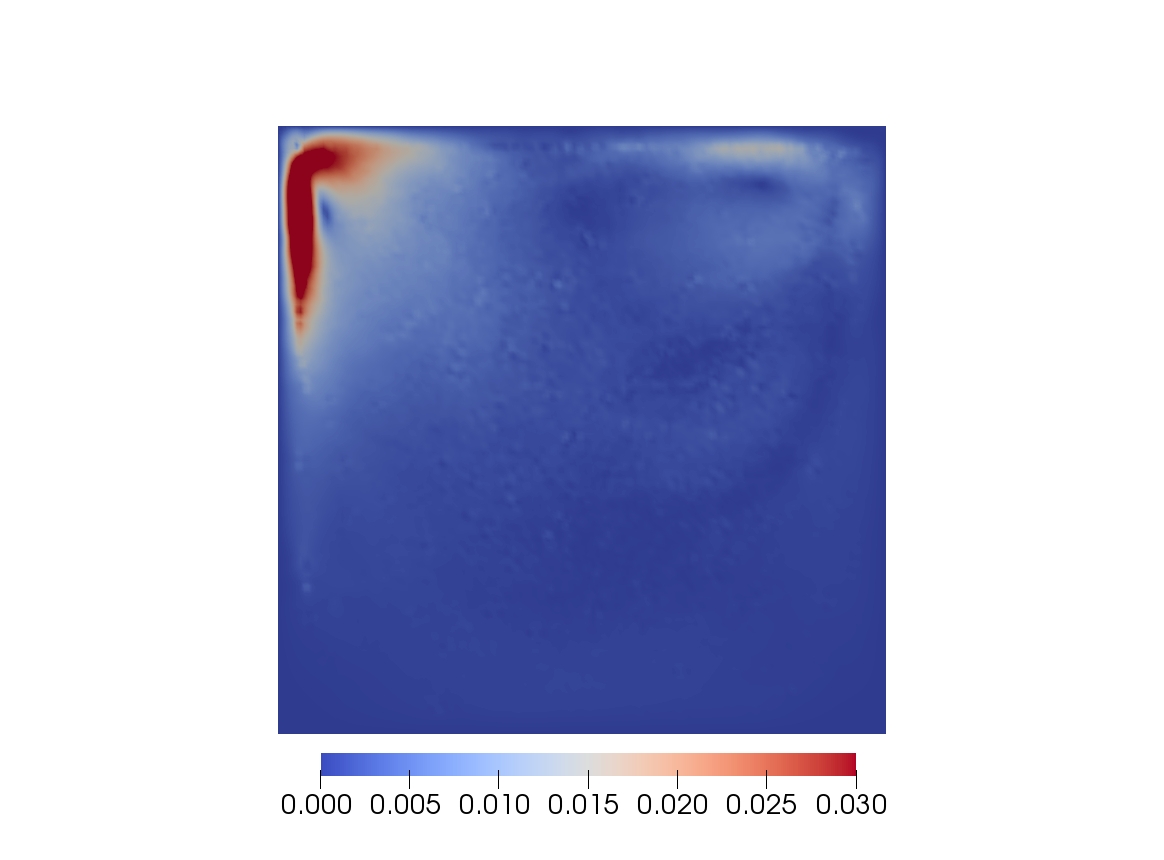}&
		\adjincludegraphics[width=7cm,trim={9cm 1cm 9cm 2cm} ,clip]{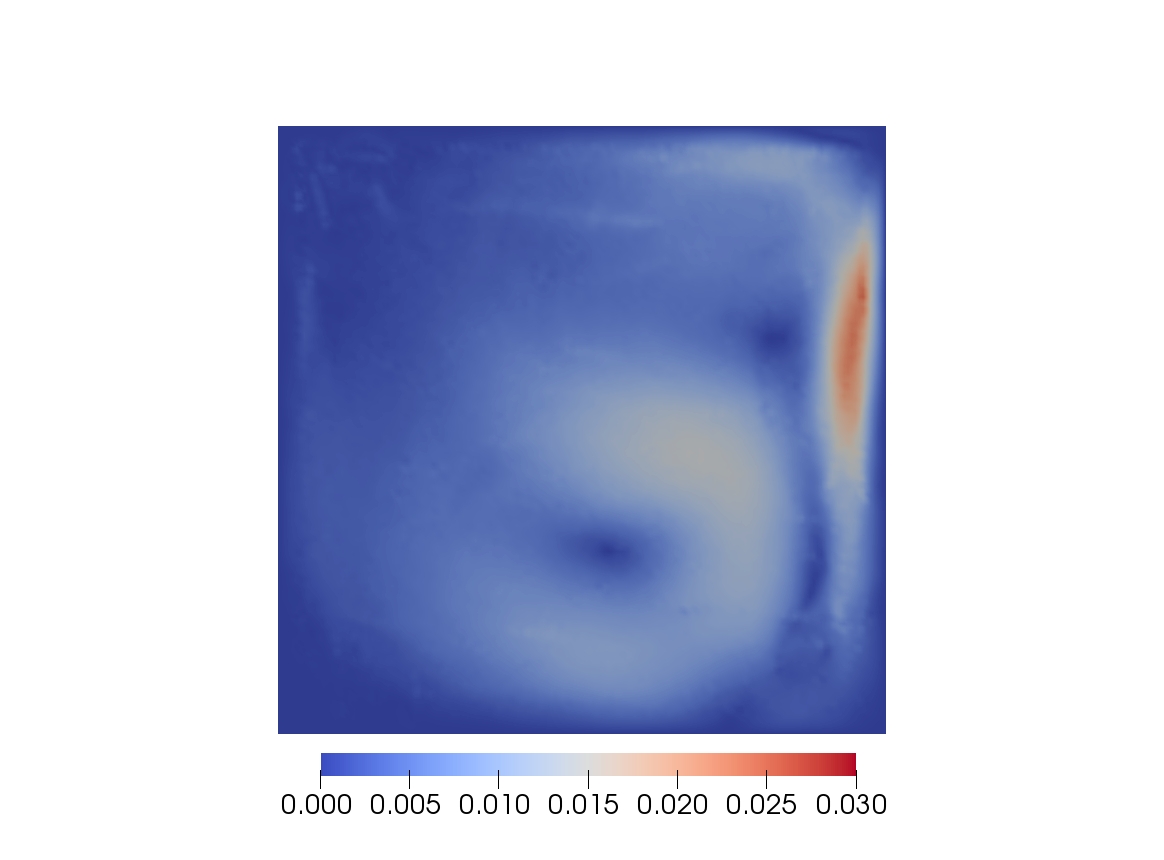}
		\\
		\adjincludegraphics[width=7cm,trim={9cm 1cm 9cm 2cm} ,clip]{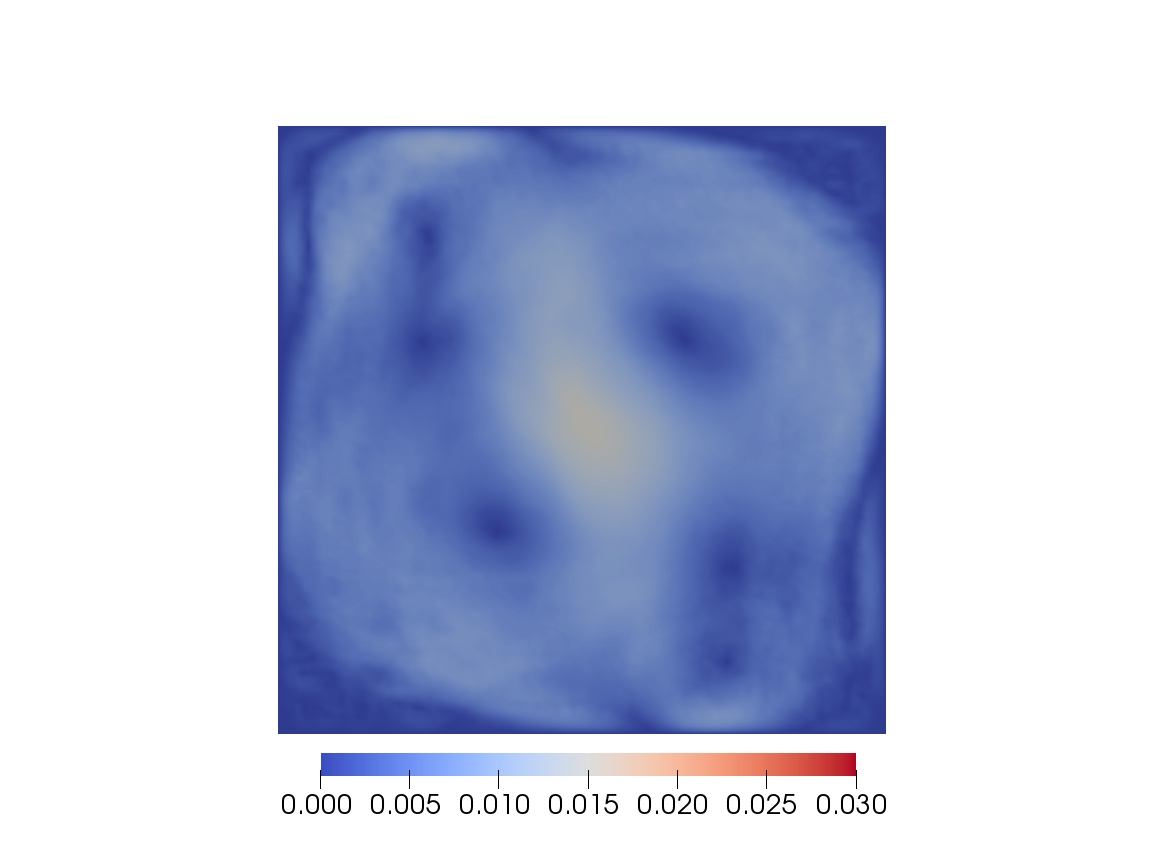}&
		\adjincludegraphics[width=7cm,trim={9cm 1cm 9cm 2cm} ,clip]{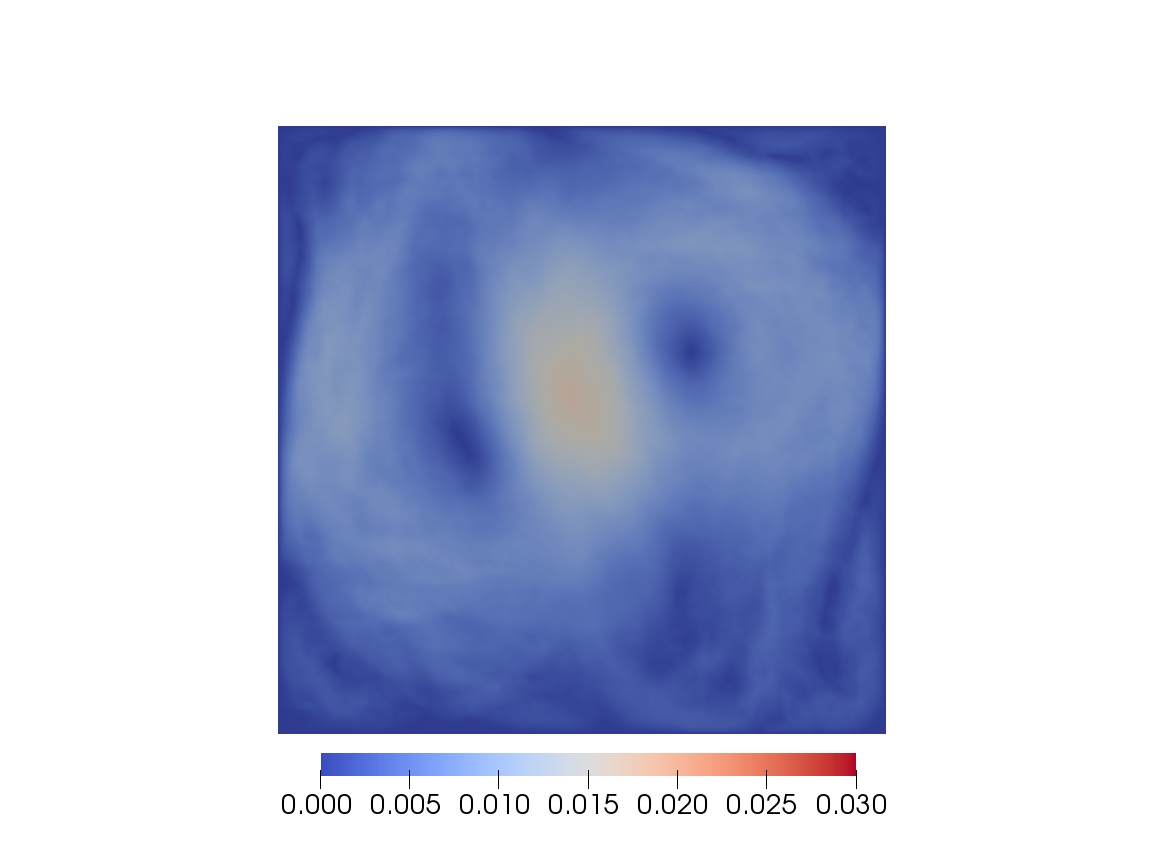}&
		\\
	\end{tabular}
	\caption{Third eigenmode for Reynolds number = 100 (top left), 400 (top right), 700 (bottom left), 1000 (bottom right) and $\phi=40\%$.}
	\label{fig:fig4} 
\end{figure*}
\newpage	    

\subsection{50\% Particle Fraction}
\begin{figure*}[h]
	\centering
	\begin{tabular}{lcc}
		\adjincludegraphics[width=7cm,trim={9cm 1cm 9cm 2cm} ,clip]{./images/Re100/phi_3_50pct_rhorat_1000}&
		\adjincludegraphics[width=7cm,trim={9cm 1cm 9cm 2cm} ,clip]{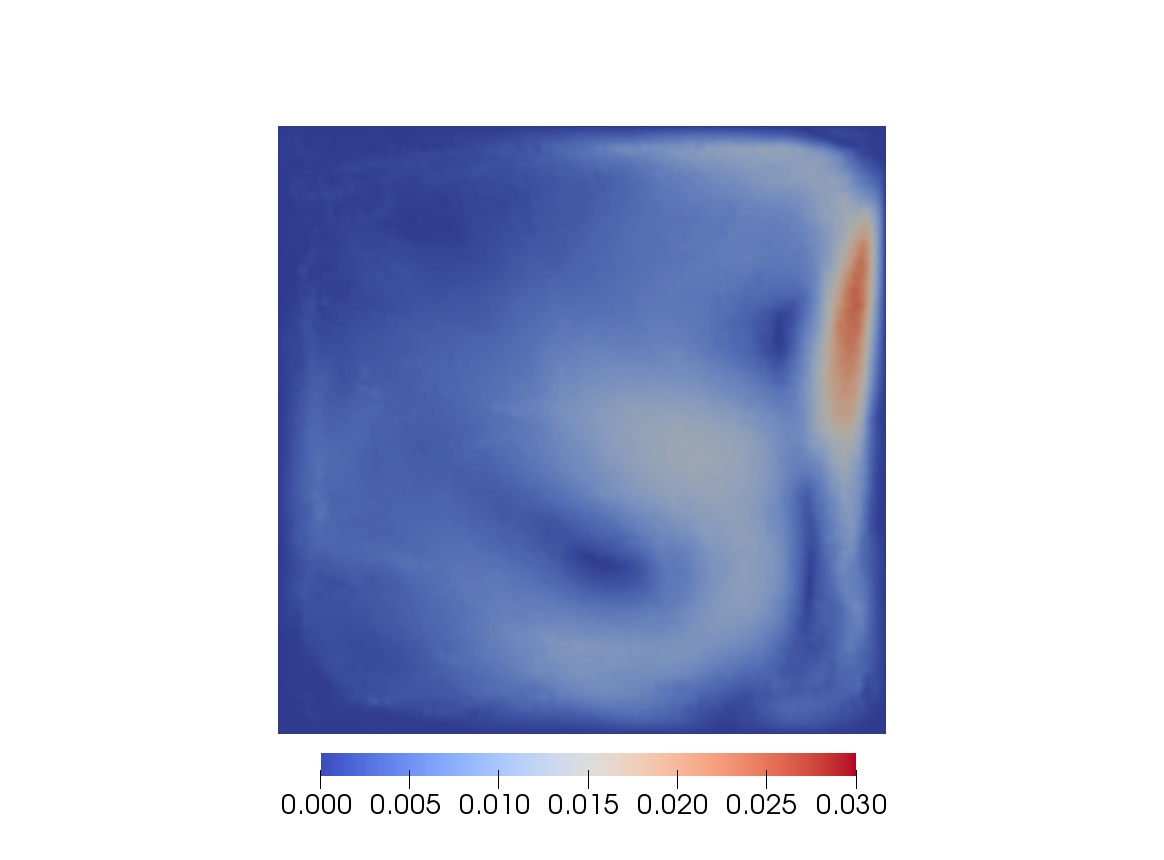}
		\\
		\adjincludegraphics[width=7cm,trim={9cm 1cm 9cm 2cm} ,clip]{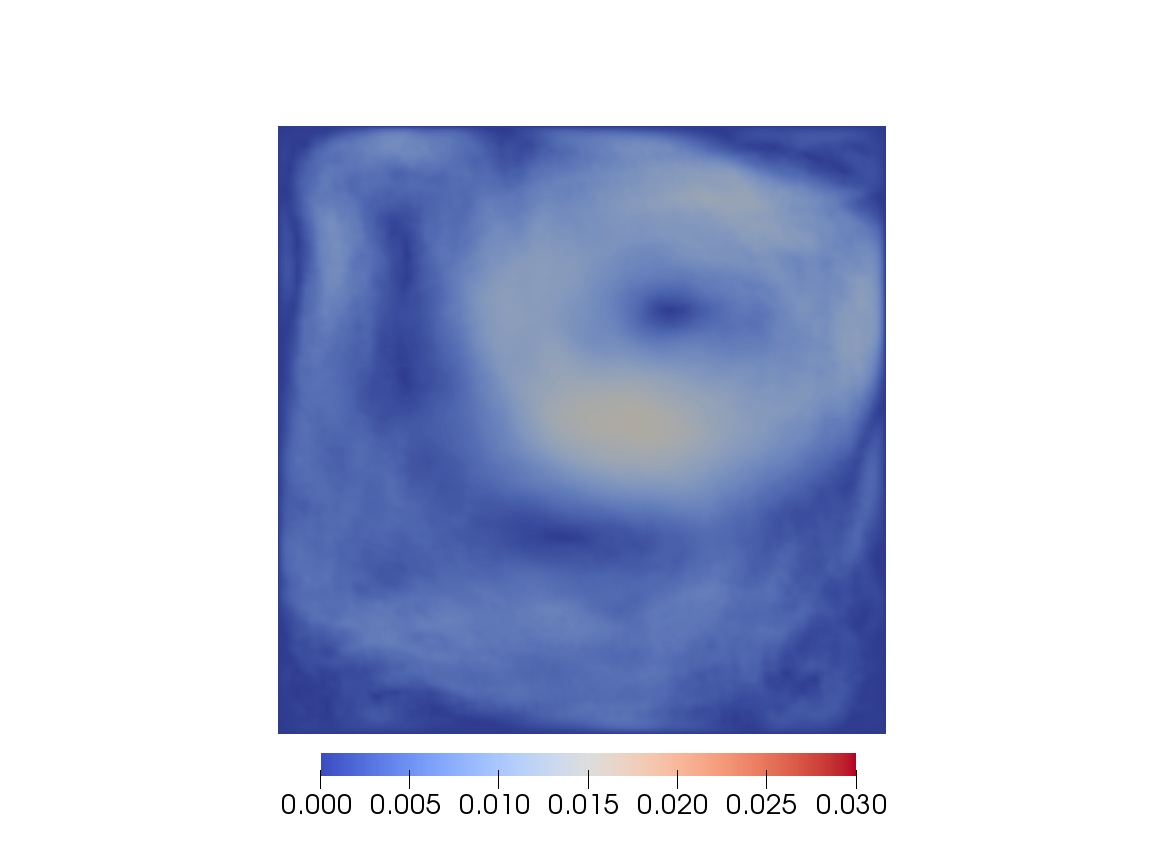}&
		\adjincludegraphics[width=7cm,trim={9cm 1cm 9cm 2cm} ,clip]{./images/Re1000/phi_3_50pct_rhorat_1000}&
		\\
	\end{tabular}
	\caption{Third eigenmode for Reynolds number = 100 (top left), 400 (top right), 700 (bottom left), 1000 (bottom right) and $\phi=50\%$.}
	\label{fig:fig4} 
\end{figure*}
\newpage